\documentclass[12pt, twoside]{report}
\usepackage{mypackages}
\DeclareUnicodeCharacter{0278}{$\phi$}
\pdfoutput=1
\usepackage{fancyhdr} 
\usepackage{titlesec}
\titleformat{\chapter}[hang]
  {\Huge \bfseries}{\thechapter{. }}{0pt}{\Huge}

\title{{\textbf{Cosmic adventures in stringland:}\\ Bosonic and fermionic scattering in gravitating cosmic string spacetimes}\\
\vspace{2.5cm}
{Departamento de Física}\\
{Universidade Federal de Pernambuco}\\}
\vspace{5.0cm}
\author{Marcos Vinicius Santos Silva}
\date{08/11/2021}

\begin{document}
\thispagestyle{plain}
    \begin{center}
        \vspace*{1cm}
            
        \Huge
        \textbf{Cosmic adventures in stringland}
            
        \vspace{0.5cm}
        \LARGE
        Scattering of bosonic and fermionic fields in gravitating cosmic string spacetimes
            
        \vspace{1.5cm}
            
        \textbf{Marcos Vinicius Santos Silva}
            
        \vfill
            
        A thesis presented for the degree of\\
        Master in Physics
            
        \vspace{0.8cm}
            
        \includegraphics[height=0.25\textheight]{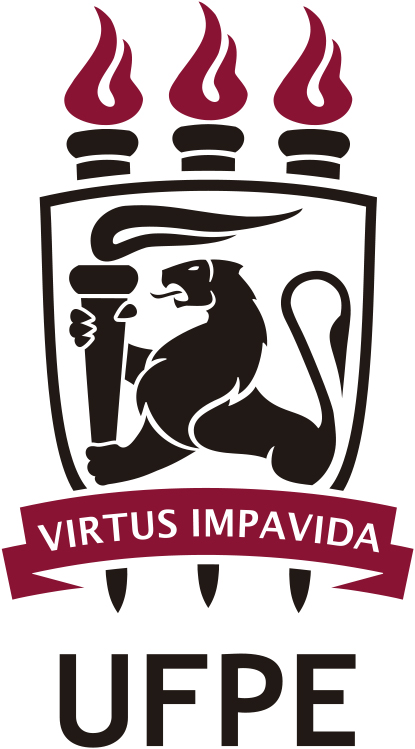}
        
        \Large
        Departamento de Física\\
        Universidade Federal de Pernambuco\\
        Brazil\\
        08/11/2021
            
    \end{center}

\pagebreak

\thispagestyle{plain}
\begin{center}
\hspace{0pt}
\vfill
{\small \textit{This thesis was supervised by}} \\
Azadeh Mohammadi, \\
\vspace{1.5cm}
{\small \textit{and was examined internally by}} \\
Fernando Roberto de Luna Parisio Filho \\
\vspace{1.5cm}
{\small \textit{and externally by}} \\
Ilya Shapiro \\
\vspace{1.5cm}
{\small \textit{on the}}\\
$8^{\text{th}}$ of December, 2021.
\vfill
\hspace{0pt}
\end{center}

\pagebreak

\thispagestyle{plain}

\begin{center}
\textbf{Agradecimentos}
\end{center}
À minha mãe, Eliana, por ter me apoiado em todos os caminhos tortos que decidi seguir na vida. Por ser minha maior inspiração em tudo. E por me permitir tentar. Enfim, por existir.\\
À minha irmã Elisangela, por ter me incentivado a estudar desde cedo e por ter sido minha primeira referência. \\
À Maria Clara por ter me feito ver as coisas com mais cores e detalhes. Por ter me tornado uma pessoa mais aberta, mais alegre e cheia de vida.\\
À Dorinha e Vladimir, que tanto me ajudaram na odisséia que foi morar em Recife. Obrigado pelas ajudas, conversas e comidas.\\
À Azadeh por ter confiado em mim, pela paciência e compreensão na pesquisa e na vida.\\
À Joan e Saullo, que mesmo após tantas idas e vindas da vida ainda me trazem bons momentos. Obrigado pelas conversas, ajudas, sessões de cinema no fim da noite, pelos lanches da madrugada, e por permanecerem comigo por tanto tempo.\\
À Marta Menezes, a matemática do bem, pelas conversas e apoios que têm durado tanto tempo apesar da distância.\\
Aos amigos que Recife me deu: Kivia e Henrique, que sempre estiveram por perto pra tudo.\\
À Julia Medeiros por ter me abrigado num dos momentos mais difíceis desse mestrado insano.\\
Ao grupinho dos Coaches Quanticos: Matheus Valença, Lucas Gabriel, Ricardo Silveira e Hugo Estevão, pelo apoio quase incondicional, pelas reflexões, indicações de músicas, origamis, resoluções de listas, bebedeiras, piadas e trocadilhos. Sou eternamente grato por tudo isso.\\
À Nicolas Pessoa e João Guilherme pela imensa ajuda computacional e por tornar minha caminhada um pouco mais confortável.\\
Aos muitos amigos que o Departamento de Física me trouxe, Ruben, Gustavo, Naudson, Tertius, Pollyanna, Matheus Martins, João Rebouças, e tantos outros. Obrigado pelas conversas, palavras de apoio, brincadeiras, saídas pro bar, e principalmente por tornar o DF-UFPE mais humano.\\

A todos, citados e não-citados, aquele abraço!

\newpage

\thispagestyle{plain}
\begin{center}
    \Large
    \textbf{Cosmic adventures in stringland}
        
    \vspace{0.4cm}
    \large
    Scattering of bosonic and fermionic fields in gravitating cosmic string spacetimes
        
    \vspace{0.4cm}
    \textbf{Marcos Vinicius Santos Silva}
       
    \vspace{0.9cm}
    \textbf{Abstract}
\end{center}

Cosmic strings are line-like topological defects expected to have formed during cosmological phase transitions in the early universe when the Higgs field acquired a non-vanishing expectation value. Because of being thin, on a cosmological scale, their gravitational effects are usually studied using the wire approximation, in which the string has a negligible width. Although the wire approximation seems reasonable in the cosmological context, local gravitational interaction with matter fields can be significant when considering a non-vanishing radius. Generally, the spacetime of a gravitating vortex is flat at the center, has a localized curvature at some finite distance from the core, and is flat and conical far from the center. The conical structure is characterized by an angular variable ranging from 0 to less than $2\pi$, i.e., the space has an angular deficit. This work shows that when considering a scalar field scattering in a gravitating cosmic string spacetime, the standard partial-wave approach's scattering amplitude is singular. In order to avoid the divergence caused by the spacetime asymptotically conical structure, we propose a modification of the asymptotic ansatz in the partial-wave formalism and find the corrections in the phase-shift and total scattering cross-section. We also developed a toy model for the spacetime metric of a cosmic string and showed how local interaction with the vortex gauge field affects the scalar field total cross-section. Then we apply this formalism to a Dirac field and show the explicit formula for the fermionic total cross-section. Finally, we study the scattering of bosonic and fermionic fields in the spacetime of an abelian and a nonabelian gravitating cosmic strings and show that the cross-sections have damped oscillations. In order to understand the origin of this behavior, we used the aforementioned toy model to show that the spacetime particular asymptotical structure causes the observed oscillations.

 \par
\vspace{2cm}

\newpage

\thispagestyle{plain}
\begin{center}
    \Large
    \textbf{Aventuras cósmicas no país das cordas}
        
    \vspace{0.4cm}
    \large
    Espalhamento de campos bosônicos e fermiônicos em espaços-tempo de cordas cósmicas
        
    \vspace{0.4cm}
    \textbf{Marcos Vinicius Santos Silva}
       
    \vspace{0.9cm}
    \textbf{Resumo}
\end{center}

Cordas cósmicas são defeitos topológicos cilindricamente simétricos que espera-se que tenham se formado durante algumas transições de fase no universo primordial quando o campo de Higgs adquiriu um valor não-nulo. Por serem estreitas, numa escala cosmológica, elas são comumente estudadas utilizando a aproximação de fio na qual considera-se que o raio da corda é nulo. Embora essa aproximação faça sentido num contexto cosmológico, a interação gravitacional local desses objetos com campos de matéria pode ser muito relevante quando consideramos um raio não-nulo. De forma geral, o espaço-tempo ao redor de um vórtice é plano no centro, possui curvatura localizada a uma distância finita da origem, e é plano e cônico longe do centro. A estrutura cônica é caracterizada por uma coordenada angular variando de 0 a menos que $2\pi$, isto é, o espaço-tempo possui um déficit angular. Nosso trabalho mostra que quando consideramos o espalhamento de um campo escalar no espaço-tempo de uma corda cósmica, a amplitude de espalhamento encontrada a partir da abordagem usual de ondas parciais é divergente. Para evitar a singularidade causada pela estrutura assintoticamente cônica do espaço-tempo, nós propomos uma modificação do ansatz assintótico no formalismo de ondas parciais e encontramos correções no desvio de fase e na seção de choque total. Também desenvolvemos um modelo simplificado para a métrica de uma corda cósmica e mostramos como a interação com o campo de calibre do vórtice afeta a seção de choque do campo escalar. Após isso aplicamos o formalismo para um campo de Dirac e explicitamente mostramos a fórmula da seção de choque para o caso fermiônico. Por fim, estudamos o espalhamento de campos bosônicos e fermiônicos no espaço-tempo de uma corda cósmica abeliana e não-abeliana e vimos que a seção de choque apresenta oscilações amortecidas. Para entender a origem desse comportamento usamos nosso modelo simplificado para mostrar que a estrutura assintótica do espaço-tempo causa essas oscilações. 

\par
\vspace{2cm}
\textbf{Palavras-chave:} boson; fermion; corda cósmica; gravitação; defeito topologico; transições de fase;

\newpage

\tableofcontents

\chapter{Introduction}

In 1976, Thomas Kibble showed that cosmological phase transitions during the early universe should lead to the formation of topological defects, namely cosmic strings, domain walls and monopoles \cite{kibble1976topology}. Among these, cosmic strings have been showed to be the most promising ones \cite{preskill1979cosmological, vilenkin1981gravitational}. Along the years many attempts have been made in order to develop reasonable observational constraints on the parameters of real cosmic strings \cite{ade2014planck, gott1985gravitational}. Some works suggested cosmic strings as possible sources for gravitational lensing phenomena \cite{schild2004anomalous}; and although some of these claims have been debunked \cite{agol2006hubble}, recent data from 12 years of observation of the NANOGrav collaboration \cite{arzoumanian2020nanograv} suggests that cosmic strings gravitational interaction might be affecting the detection of pulsar blinking \cite{blasi2021has, ellis2021cosmic}. It appears the subject of cosmic strings is about to be revived. However, the attention toward cosmic strings is not exclusively by cosmological reasons. These objects have strong condensed matter counterparts. The U(1) string solution, for instance, is also present in the theory of superconductors \cite{Abrikosov:1956sx, abrikosov2004nobel, feynman1955chapter}. Furthermore, even the gravitational interaction of cosmic strings, in the wire approximation, have condensed matter analogues called wedge disclinations and edge dislocations. Therefore the interest in cosmic strings/vortex solutions, are \emph{not} exclusive for the high-energy physicists.

Many cosmic string solutions have been found in many diverse models, specially during the 20th century, such as the Nielsen-Olesen vortex in 1973 \cite{nielsen1973vortex}, Semilocal strings in 1991 \cite{vachaspati1991semilocal} and Eletroweak strings in 1993 \cite{vachaspati1993electroweak}. Nonetheless their gravitational interaction with nearby matter have been most studied using the wire approximation. This is possibly because it is too hard to study the gravity-coupled system, since one usually has to solve 5 (or more) coupled partial differential equations. Hence one has to resort to numerical methods to have even a basic understanding of the field and metric solutions. Even the flat-spacetime cosmic string solutions have not been analytically solved completely. With a symmetry-breaking model in hands one frequently resorts to proving the existence of such a line-like object without actually finding the solution analytically. Only in some restricted situations the equations of motion are analytically solvable, such that it is of no surprise that the gravity-coupled system is only possible to be studied using numerical methods.

Although Garfinkle \cite{garfinkle1985general} made significant advances in understanding the asymptotical conical limit of an abelian string spacetime, there was still plenty of aspects to be understood. For instance, how do the vortex internal parameters affect the asymptotical metric limit or the curvature profile? Is the conical limit a general property of any cosmic string solution? Do the gravitational description changes significantly when we consider the vortex size to be non-negligible? These are all legitimate questions, but one needs to build a comprehensive understanding of gravitating cosmic string solutions to answer them. The works of \textcite{christensen1999complete} and \textcite{brihaye2000classical} on the abelian-Higgs model appeared as a light in this long cosmic journey. \cite{christensen1999complete} established the default notation in studying gravitationally extended vortex solutions of the abelian-Higgs model and following them \textcite{brihaye2000classical} showed that the metric of the U(1) vortex is asymptotically conical and how the conical parameters depend on the internal ones. Subsequently many gravitating vortex solutions have been found to also be asymptotically conical. This feature, which seems to be general to all gravitating cosmic string solutions, is a key point to the work presented in chapters \ref{chap3} and \ref{chap4} of this thesis.

This thesis is organized as follows. In Chapter \ref{chap1} we present the essential theory of cosmic strings without taking into account their gravitational interaction. We study the Nielsen-Olesen vortex with a reasonable amount of details and present the numerical solution to the scalar and gauge fields as well as the energy density with two different winding numbers. In Chapter \ref{chap2}, we study the gravitational aspects of cosmic strings. We present the wire approximation and two of the local physical effects of a conical spacetime. In the second part of Chapter \ref{chap2} we dive into the world of gravitationally extended vortices. We present the results from \cite{christensen1999complete} and \cite{brihaye2000classical} on the gravitating U(1) string, and the results of \textcite{de2015gravitating} on a non-abelian-Higgs model developed recently. We see that both of these models present conical structure far from the core. In Chapter \ref{chap3} we study the scattering of scalar and fermionic fields in the spacetime of a gravitating cosmic string. We show that the partial-wave formalism presents inconsistencies when applied to the scalar field scattering in this class of spacetimes and propose a modification to solve them. We explicitly show the corrections in the phase-shift, scattering amplitude and total cross-section. In order to apply the formalism we develop a toy-model for the spacetime of a cosmic string and study the scattering of a scalar field in this toy-model. We also take into account the interaction with the gauge field that generates the vortex and compare the cross-section with and without the local gauge-field interaction. In the last part of \ref{chap3} we apply the same formalism to the fermionic field and find the expression of the total scattering cross-section.

In Chapter \ref{chap4} we apply the formalism to the spacetime found by \cite{de2015gravitating}. In the first part we study the scattering of the scalar field and show how the mass of the field affects the total cross-section. We then turn our attention to analyze the reason behind the observed oscillations in the total cross-section. In order to do that we use the aforementioned toy-model to conclude that the oscillations are caused by the spacetime asymptotical conical structure. In the second part of Chapter \ref{chap4} we study the fermionic scattering in the same spacetime. We show the dependence of the cross-section with the fermion mass and also observe oscillations in the total cross-section. Finally we present a crude estimation of why the cross-section is larger when the deficit angle of the asymptotical spacetime is bigger.

Throughout the text we use natural units where $\hbar = c = 1$. Hence mass is measured in Planck mass, $m_p$, and length in Planck length, $l_p$. For future reference, these values are approximately $m_p = 2.1 \times 10^{-8}$ kg and $l_p = 1.6 \times 10^{-35}$ m.

\chapter{Down the cosmic hole: Topological defects and string formation}
\label{chap1}

\section{Topological defects}
\epigraph{\textit{Start with something that everyone knows and understands, people like to hear what they know. Then say something that only the experts understand, lest you be accused of talking trivia. Conclude with something no one, not even you, can understand, just to keep the proper respect for physics.}}{Vicki Weisskopf \cite{freund2007passion}}

In field theory, a topological defect is a field configuration that only depends on the value of the field at the boundaries. Since, for a localized configuration, the boundary field values are also the vacuum ones we can also define topological defects by the non-trivial structure of the vacuum of the theory, which is commonly expressed as a disconnectedness in the vacuum manifold. This motivates a more general definition of topological defects which is also valid outside the field theory context: a topological defect is a discontinuity (defect) in the order parameter\footnote{In field theory the order parameter is the value of field at the vacuum.} that cannot be removed (topological).\\

Furthermore, if the field configuration cannot be continuously transformed to the vacuum keeping the energy finite then it is considered topologically stable since, as we shall see, it is possible to define a conserved charge that depends only on the field values at the boundaries.\\

The non-trivial topology of the vacuum is commonly connected to a form of symmetry breaking, which can be related to phase transitions in a variety of systems, ranging from condensed matter ones, as in liquid crystals, liquid helium, and superconductors, to the formation of structures in the early universe, such as domain walls and cosmic strings.\\
In this chapter, we will see the basic topological defects from a field-theoretical point of view. In the last section, however, we encounter some interesting connections between condensed matter systems and the universe's structure.\\

\subsection{The simplest model: $\phi^4$}
When confronted with a new subject, it is often illuminating to start with a straightforward problem. A simple model that admits a topological defect solution is the 1+1-dimensional $\phi^4$ model, i.e. we are searching for static solutions in 1 spatial dimension.

\begin{equation}
\begin{gathered}
\mathcal{L} = \frac{1}{2}\partial_\mu \phi \, \partial^\mu \phi - V(\phi), \quad \quad
\text{with } \quad V(\phi) = \frac{\lambda}{4}\left(\phi^2 - \eta^2 \right)^2 .
\end{gathered}
\label{phi4_Lagrangian}
\end{equation}
The $\phi^4$ Lagrangian describes a real scalar field with a quartic self-interaction potential. The quadratic term in the field, $\lambda \eta^2 \phi^2/2$, means the mass of the field is $\sqrt{\lambda} \eta$. Also, the minimum of the potential, $V(\phi) = 0$, is achieved when $\phi = \pm \eta$, hence we say $\phi = \pm \eta$ are the vacuum states.\\
The Euler-Lagrange equation yields the solutions
\begin{equation}
\phi(x) = \pm \eta \tanh\left( \sqrt{\frac{\lambda}{2}} \,\eta x \right),
\end{equation}
which are called \emph{kink} (positive sign) and \emph{anti-kink} (negative sign). The kink is a one-dimensional, localized, time-independent, stable solution. We shall see it is also topologically stable, i.e. the boundary conditions enforce stability under field perturbations.
\begin{figure}[H]
\caption{Profiles of the field, potential and energy density of the kink solution.}
\includegraphics[width=1.0\textwidth]{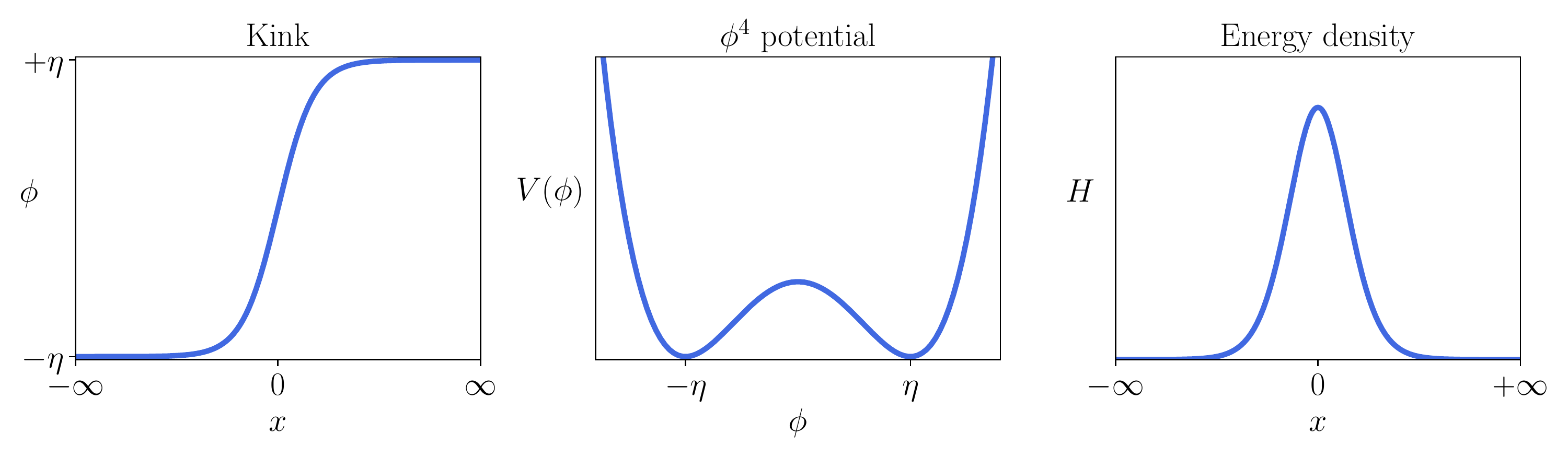}
\caption*{Source: The author (2021).}
\label{fig:kink_plots}
\end{figure}
In Figure \ref{fig:kink_plots} we see that the kink interpolates between both vacuum states, $\pm \eta$, which are at $\pm \infty$. Also, the order parameter is the vacuum expectation value (VEV), $\langle \phi \rangle$, the field's value at the minimum of the potential as it labels the different equilibrium field configurations. \\
The $\phi^4$ model has many applications. For example, it can represent domains walls \cite{zel1974cosmological}, 2-dimensional structures expected to have formed during the early universe, but also non-linear excitations in polyacetylene \cite{rice1979charged}. For a historical overview of the $\phi^4$ model we suggest \cite{campbell2019historical}.

Now, if we look at the vacuum manifold (the space of vacuum states), we see that it is disconnected. It is impossible to go from one vacuum to the other passing only through vacuum states. This means that the order parameter, $\langle \phi \rangle$, has a discontinuity between the vacua; thus, the vacuum manifold is disconnected.  \par

It was mentioned earlier that topological defects are usually connected with symmetry breaking, but which symmetry is broken by the quartic potential? The answer is $Z_2$ parity, which is equivalent to $\phi \to -\phi$. If we make $\eta \to 0$ we see that both vacua coalesce to the same value, namely $\langle \phi \rangle = 0$, and the configuration becomes $Z_2$ invariant. The degenerate minima and the disconnected vacuum manifold only appear because $\eta \neq 0$.\\

\begin{figure}[H]
\caption{Representation of vacuum states interpolated by the kink solution.}
\includegraphics[width=0.5\textwidth]{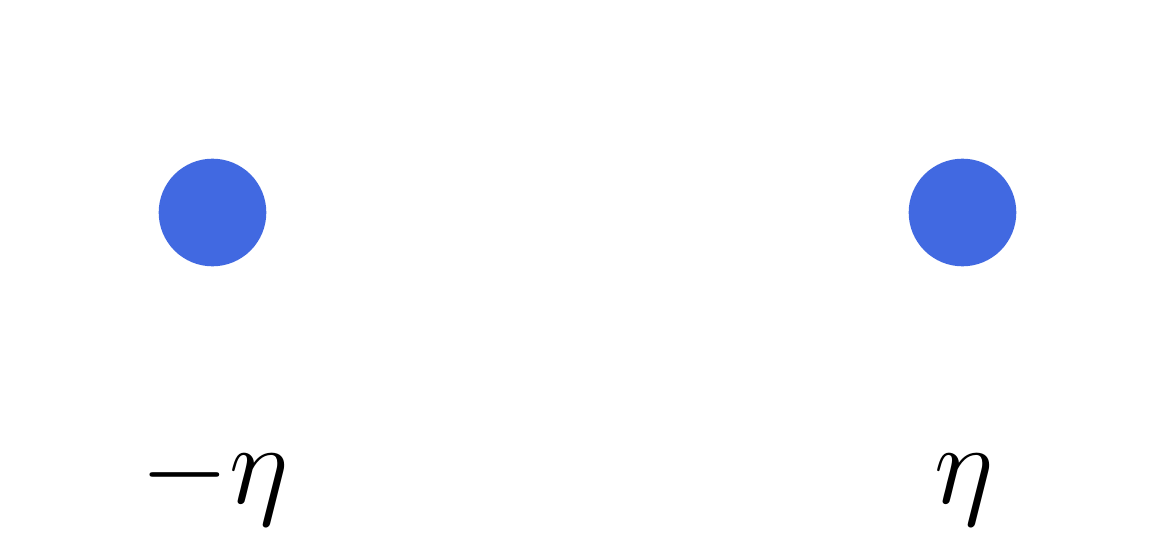}
\caption*{Source: The author (2021).}
\end{figure}
Topological defects are topologically stable. In order to destroy the kink, one would have to take every point of the solution and move to zero, which would require an infinite amount of energy since the solution extends to infinity. This guarantees the stability of the kink. This argument can be tracked down to the disconnectedness of the vacuum, where the vacuum structure ensures stability. 

We can see topological stability more clearly defining the current
\begin{equation}
j^\mu = \epsilon^{\mu \nu} \, \partial_\nu \phi
\label{kink_topol_current}
\end{equation}
where $\epsilon^{\mu \nu}$ is the two-dimensional Levi-Civita antisymmetric tensor. One can easily see that the current (\ref{kink_topol_current}) is conserved, $\partial_\mu j^\mu = 0$, which gives rise to a conserved charge
\begin{equation}
\begin{gathered}
Q = \int_{-\infty}^{\infty}{dx \,j^0} = \phi(\infty) - \phi(-\infty) = 2 \eta.
\label{kink_topol_charge}
\end{gathered}
\end{equation}
The conserved charge derived from current (\ref{kink_topol_current}) only depends on the VEV. Naturally (\ref{kink_topol_current}) is called \emph{topological current}, and (\ref{kink_topol_charge}) the \emph{topological charge}. One can see that \eqref{kink_topol_current} is \emph{not} derived from Noether's theorem, so this is a purely topological conservation law. Once again, the boundary conditions (topology) guarantees stability!

\subsection{Derrick sends in a little bill: Derrick's theorem}

Naturally, one might want to explore solutions of the $\phi^4$ model in more spatial dimensions. However, it is not possible to find such stable static solutions. Derrick's theorem \cite{derrick1964comments} states that
\begin{displayquote}
there are no non-linear scalar field models with stable, time-independent, and localized solutions in more than one dimension
\end{displayquote}
which imposes a big restriction on higher-dimensional solutions to \eqref{phi4_Lagrangian}. Essentially, if one has a non-linear scalar model in more than one spatial dimension it is guaranteed no static and localized\footnote{"Localized" means the energy density vanishes at infinity.} solution is stable under field perturbations. \\
Derrick's argument is based on the fact that any D-dimensional pure scalar field theory, with static solution is $\phi(x)$, has total energy given by

\begin{equation}
   E (\phi(x)) = E = \int{d^D x \left[ \left |\nabla_x  \phi(x) \right|^2 + V(\phi(x)) \right]}.
\end{equation}
where $\nabla_x$ is the gradient with respect to the $x$-coordinates. For further use we define the two quantities

\begin{gather}
\begin{aligned}
    I_1 &= \int{d^Dx  \left | \nabla_x \phi(x) \right|^2}  > 0\, ,\\
    I_2 &= \int{d^Dx V(\phi(x))} > 0 \, .
\end{aligned}
\end{gather}

We can make the rescaling $x \to y = \alpha x$, $\alpha \neq 0$, and define the deformed field $\phi_\alpha = \phi(\alpha x)$. The energy of the new configuration is given by

\begin{align}
    E (\phi_\alpha) = E_\alpha &= \int{d^D x \left[ \left| \nabla_x \phi(\alpha x) \right|^2 + V(\phi(\alpha x)) \right]} \\
    &= \alpha^{-D} \int{d^D y \left[ \alpha^2 \left|\nabla_y \phi(y)\right|^2 + V(\phi(y)) \right]} \\
    &= \alpha^{2 - D} I_1 + \alpha^{-D} I_2.
\end{align}
Now we can analyze how the energy behaves with respect to the rescaling parameter $\alpha$ and the spatial dimensions. The main point here is that if the solution $\phi(x)$ is stable then $\alpha = 1$ is a minimum of the energy. 

Extremizing the energy we find the condition for the solution to be at the equilibrium point:

\begin{equation}
    \left. \frac{dE}{d\alpha} \right |_{\alpha = 1} = 0 \Rightarrow D I_2 = (2 - D) I_1
\end{equation}

Now we need to know if this point is a minimum, $\delta^2 E > 0$, or a maximum, $\delta^2 E < 0$. 

\begin{equation}
    \left. \frac{d^2 E}{d \alpha^2} \right|_{\alpha = 1} = 2 (2 - D) I_1 < 0 \text{for } D < 2, 
\end{equation}

which is greater than zero only when $D < 2$. This means that for $D \geq 2$ a deformation of $\phi(x)$ shall decrease the energy of the configuration, hence the solution is not stable. \\

The alternatives to construct stable non-linear scalar field solutions in higher dimensions are split into two categories: 1) extending the model by including more fields, like a gauge or spinor field, and b) relaxing the time-independence or localized condition. Higher-dimensional topological defects can be obtained following any of these two paths. 
Before diving into any of these routes, we need to understand a straightforward generalization of (\ref{phi4_Lagrangian}).\\

\section{Spontaneous Symmetry Breaking}
\epigraph{\textit{Come forth into the light of things, let nature be your teacher.}}{William Wordsworth}

Knowing the symmetries of the Lagrangian is often regarded as the first step in getting a comprehensive description of a system. If a symmetry is continuous, Noether's theorem assures a related conserved quantity, making some calculations a lot easier. Nevertheless, sometimes a symmetry of the Lagrangian might not be the symmetry of the vacuum. In such cases, we say the symmetry is spontaneously broken.

\subsection{A simple model: Goldstone}
We can illustrate the main features of spontaneous symmetry breaking using the Goldstone model
\begin{equation}
\begin{gathered}
\mathcal{L} = \partial_\mu \phi^{*} \, \partial^\mu \phi - V(\phi), \quad \quad
\text{with } \quad
V(\phi) = \frac{\lambda}{4} \left( |\phi|^2 - \eta^2 \right)^2 ,
\end{gathered}
\label{goldstone_Lagrangian}
\end{equation}
which looks similar to the $\phi^4$ one, but now $\phi$ is a complex scalar field instead of a real one. Although, according to Derrick's theorem, model \eqref{goldstone_Lagrangian} in more than one dimension is unstable, it is also pedagogically useful to illustrate spontaneous symmetry breaking. 
We see that (\ref{goldstone_Lagrangian}) is invariant under the action of the global $U(1)$ group, i.e., (\ref{goldstone_Lagrangian}) does not change if we multiply $\phi$ by a constant phase $\alpha$, $\phi \rightarrow e^{i \alpha} \phi$. The vacuum state, $\phi = \eta e^{i\theta}$, however, is \emph{not} invariant under $U(1)$. The global $U(1)$ symmetry is spontaneously broken.\\

\begin{figure}[H]
\caption{Potential of the Goldstone model, also called mexican hat or Higgs potential.}
\includegraphics[width=0.45\textwidth]{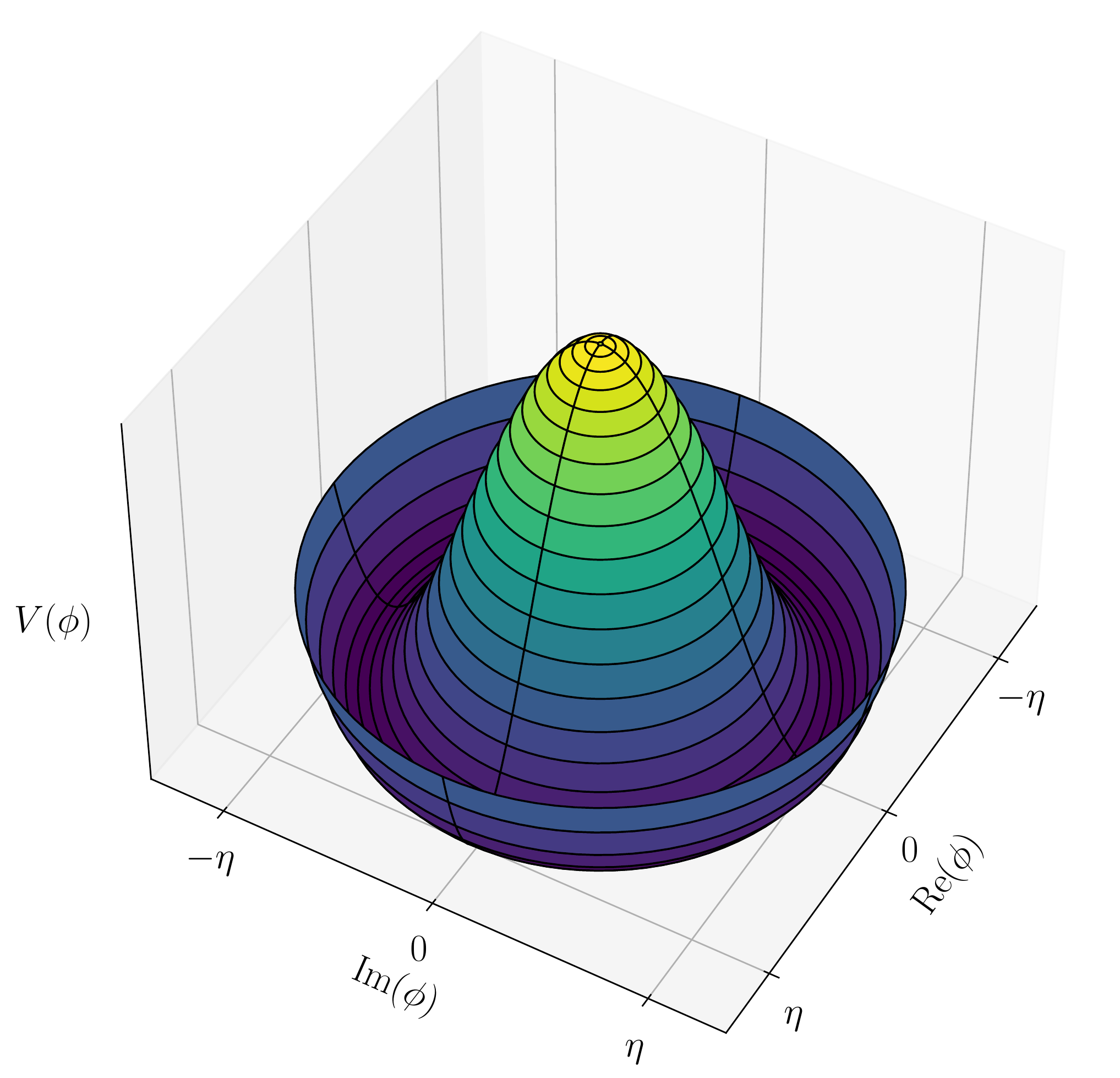}
\caption*{Source: the author (2021)}
\label{fig:mexican_hat}
\end{figure}
Since the coupling constant $\lambda$ is considered to be small, we can use the term \emph{vacuum expectation value} referring to the ground state, since when $\lambda \rightarrow 0$, the classical and quantum values become the same.
For small $\eta$, the potential $V(\phi)$ can be approximated by
\begin{equation}
V(\phi) \approx |\phi|^4 - 2 \lambda \eta^2 |\phi|^2,
\end{equation}
which has a maximum $V_{max} = \lambda \eta^2$ at $\phi = 0$ and a minimum $V_{min} = 0$ at $|\phi| = \eta$. In order to keep track of field excitations, we expand $\phi$ around the specific ground state $\phi = \eta$,

\begin{equation}
\phi = \eta + \frac{1}{\sqrt{2}} (\phi_1 + i \phi_2),
\label{goldstone_gs_state_exp}
\end{equation}
where $\phi_1$ and $\phi_2$ are treated as two independent real scalar fields. Each term of \eqref{goldstone_Lagrangian} becomes
\begin{gather}
\begin{aligned}
\partial_\mu \phi^* \partial^\mu \phi &= \frac{1}{2} \left[\partial_\mu \phi_1  \partial^\mu \phi_1 + \partial_\mu \phi_2 \partial^\mu \phi_2 \right] ,\\
V(\phi) &= \frac{\lambda}{4} \left[\frac{1}{2} (\phi_{1}^2 - \phi_{2}^2) + \sqrt{2}\eta\phi_{1} \right]^2 
\\ &= \frac{\lambda}{4} \left[ \frac{1}{4} ( \phi_{1}^4 + \phi_{2}^4 - 2 \phi_{1}^2 \phi_{2}^2) + 2 \eta^2 \phi_{1}^2 + \sqrt{2} \eta \phi_1 (\phi_{1}^2 - \phi_{2}^2) \right].\\
\end{aligned}
\label{goldstone_gs_exp_terms}
\end{gather}
Now plugging \eqref{goldstone_gs_exp_terms} into \eqref{goldstone_Lagrangian} yields
\begin{equation}
\mathcal{L}_{exc} = \underbrace{\frac{1}{2} (\partial_\mu \phi_1)^2 - \frac{1}{2} \eta^2 \lambda \phi_{1}^2 + \frac{1}{2} (\partial_\mu \phi_2)^2}_{\text{free part}} + \mathcal{L}_{int},
\end{equation}
where $\mathcal{L}_{exc}$ refers to the excitation of the scalar field and $\mathcal{L}_{int}$ includes all the interaction terms. What is important to notice here is that the free part describes a massive scalar field $\phi_1$, with mass $\eta \sqrt{\lambda/2}$, and a massless scalar field $\phi_2$. This suggests the symmetry breaking, which can be tracked down to the inclusion of a complex scalar field, generates a massless excitation of the Higgs field, $\phi_2$, called \emph{Goldstone boson}. In 1961 Jeffrey Goldstone conjectured this to be a general feature of models with symmetry breaking. The Goldstone conjecture \cite{goldstone1961field} states that
``every spontaneous breaking of a continuous symmetry generates a massless scalar particle''. It was demonstrated, becoming a theorem, in 1962 by Goldstone, Salam and Weinberg \cite{goldstone1962broken}.\par 

We can give meaning to the Goldstone boson by looking at Figure \ref{fig:mexican_hat}. In order to use the expansion \eqref{goldstone_gs_state_exp} suppose that the system is at the vacuum with $\theta = 0$, i.e. at Im$(\phi) = 0$ and Re$(\phi) = \eta$. In this situation, a change in $\phi_1$ (the real part of the field) is energetically disfavoured, since the state would have to climb the potential hill, while a change in $\phi_2$ takes the field to an energetically equivalent state, only changing the representative vacuum (value of $\theta$). The Goldstone boson encodes the energetic symmetry of the vacua. Naturally, any interpretation to $\phi_1$ or $\phi_2$, as defined in \eqref{goldstone_gs_state_exp}, \emph{have} to consider them as perturbations around $\phi = \eta$.
There is, however, a loophole to the Goldstone theorem. If we consider a broken \emph{local} symmetry, we can make the Goldstone boson disappear. One might take a moment and try to figure out why. 

\subsection{A hard one: abelian-Higgs}
Now we are going to see what happens when we impose a local symmetry in the Goldstone model. First of all, local symmetry plus Lorentz-invariance forces the existence of a gauge field communicating the symmetry. Consider the abelian-Higgs model with the following Lagrangian
\begin{equation}
\begin{gathered}
\mathcal{L}_{AH} = (D_\mu \phi)^* (D^\mu \phi) - \frac{1}{4}F_{\mu \nu} F^{\mu \nu} - V(\phi), \quad \quad
\text{with } \quad V(\phi) = \frac{\lambda}{4} (|\phi|^2 - \eta^2)^2 ,
\end{gathered}
\label{AH_Lagrangian}
\end{equation}
where $D_\mu = \partial_\mu -ieA_\mu$ is the covariant derivative in field space, $e$ is the coupling to the gauge field (usually referred to as charge), $A_\mu$ is the gauge field, and $F_{\mu \nu} = \partial_\mu A_\nu - \partial_\nu A_\mu$ is the Faraday tensor. Because we are dealing with a gauge field, there is a subtlety. The Lagrangian \eqref{AH_Lagrangian} have $U(1)$ \emph{local} symmetry, which means invariance under \emph{gauge transformations}

\begin{gather}
\left \{
\begin{aligned}
\phi & \rightarrow \phi^\prime = e^{i \alpha(x)} \phi \\
\phi^* & \rightarrow {\phi^*}^\prime = e^{-i\alpha(x)} \phi^* \\
A_\mu & \rightarrow A_{\mu}^\prime = A_\mu + \frac{1}{e}\partial_\mu \alpha(x), \\
\end{aligned}
\label{AH_gaugetransform}
\right.
\end{gather}
for any arbitrary coordinate dependent function $\alpha(x)$. Transformations \eqref{AH_gaugetransform} tell us that for every choice of the four-vector $A_\mu$, the vector $A_{\mu}^\prime$ is equivalent. This is called \emph{gauge freedom}. It means that we have the freedom of choosing a function $\alpha$, the \emph{gauge}, that better suits our intentions. 
In this case we can choose a gauge that annihilates the imaginary part of $\phi$, the Goldstone boson. Expanding the field around vacuum
\begin{equation}
\phi = \eta + \frac{1}{\sqrt{2}} \phi_1,
\end{equation}
results in the Lagrangian
\begin{equation}
\mathcal{L}_{exc} = \frac{1}{2} (\partial_\mu \phi_1)^2 + \frac{\lambda}{2} \eta^2 \phi_1^2 + e^2 \eta^2 A_\mu A^\mu + \frac{1}{4} F_{\mu \nu} F^{\mu \nu} + \mathcal{L}_{int}.
\label{AH_Lagrangian_exc}
\end{equation}
One might notice that now we do not have any massless scalar boson in \eqref{AH_Lagrangian_exc}, the gauge boson "ate" the Goldstone boson.
In addition, the field $A_\mu$ has a self-interaction term, $e^2 \eta^2 A_\mu A^\mu$, meaning that the gauge boson, initially massless, acquires a mass $m_A = \sqrt{2} e \eta$. This result can be generalized to any number of non-abelian gauge fields. It can be shown that the corresponding gauge boson acquires a mass proportional to the vacuum energy for every broken symmetry.

\subsection{Global strings} 
Now consider the Goldstone potential
\begin{equation}
V(\phi) = \frac{\lambda}{4} (|\phi|^2 - \eta^2)^2,
\label{goldstone_potential}
\end{equation}
and assume the field $\phi$ is near the minima, $\phi = \phi_0 e^{i\theta}$, where $\theta$ is the phase in the field space. We can map circles, in field space, near the minima of the potential. which we denote by the letter $L$, to loops in physical space, which we denote by $S$, since both are periodic closed curves. Once we travel around a loop $S$, in physical space, we know the angular coordinate, $\varphi$, must change by $2\pi$ while the phase of $L$, the image of $S$ in field space, $\theta$, must change by an integer multiple of $2 \pi$, $\Delta\theta = 2\pi n$. Now imagine one continuously deform the loop $S$, in physical space, to a point $p$. Since the field has to be uniquely determined at $p$ there are two options: $\phi_0(p) = 0$ or $\phi_0 (p) \neq 0$. If $\phi_0(p) = 0$ then we know there is an energy excitation inside $S$.\par
\begin{figure}[H]
\caption{$|\phi|$ has to vanish at some point when contracted from a minimum of the potential $V(\phi)$ to a single point $p$.}
	\begin{subfigure}[b]{0.65\textwidth}
	\includegraphics[width=\textwidth]{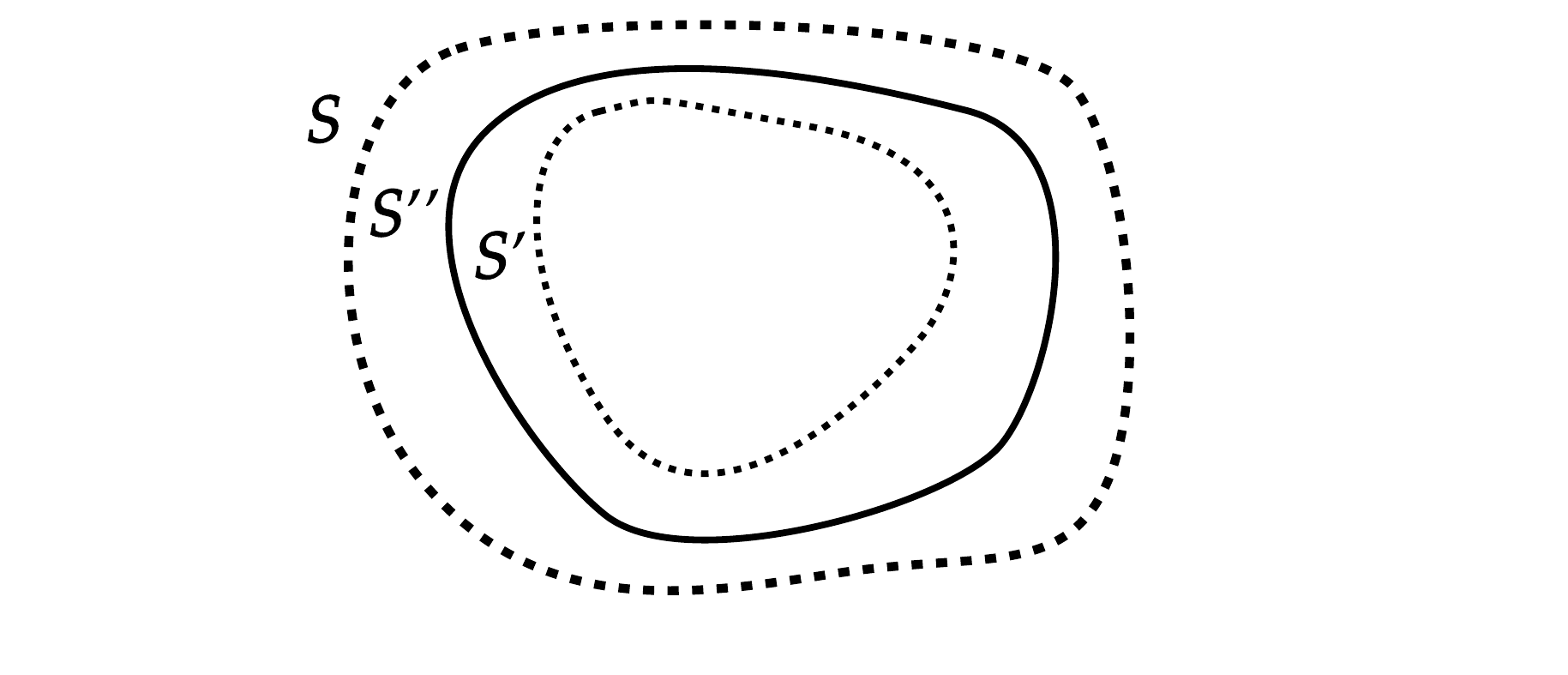}
	\caption{Loops in physical space.}
	\label{fig:deformation_argument1}
	\end{subfigure}
	\begin{subfigure}[b]{0.75\textwidth}
	\includegraphics[width=\textwidth]{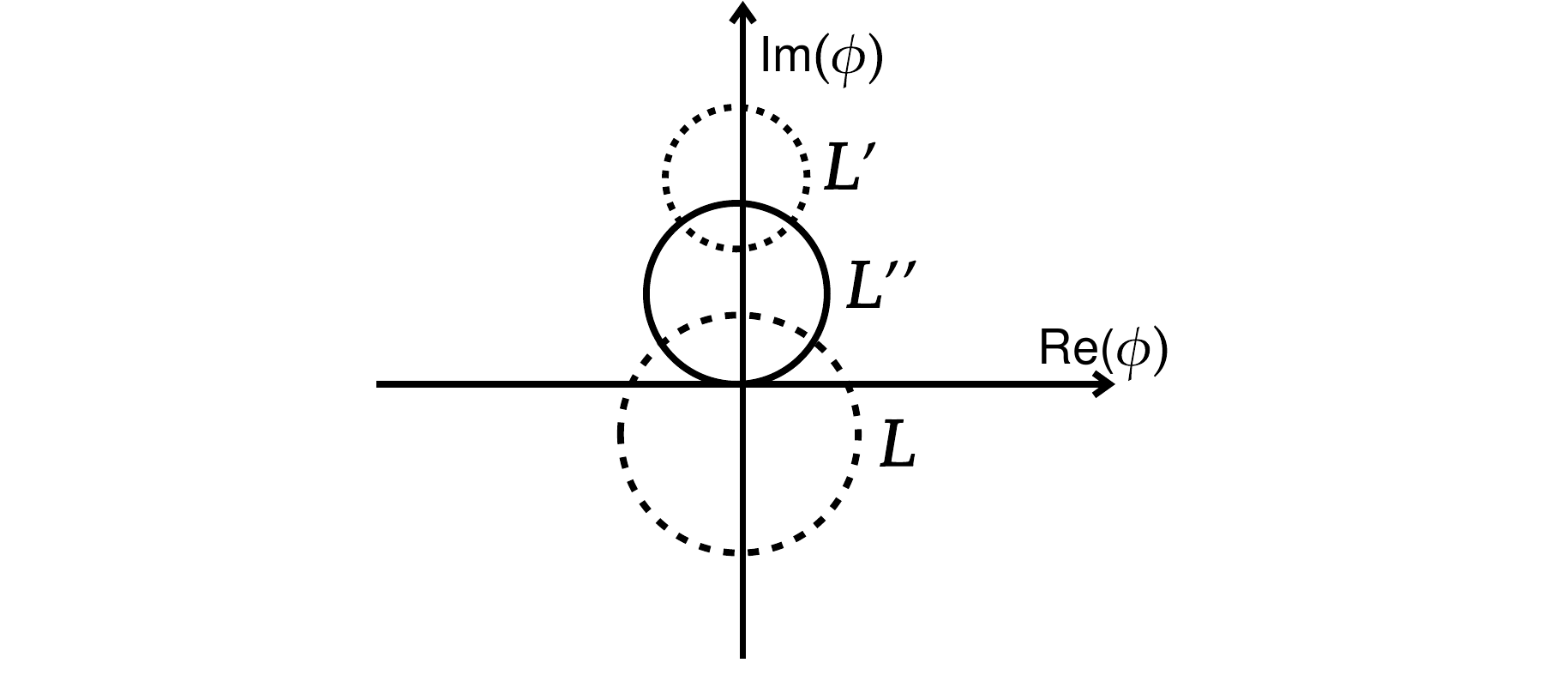}
	\caption{Loops in physical space mapped onto field space.}
	\label{fig:deformation_argument2}
	\end{subfigure}
\caption*{Source: The author (2021).}
\label{fig:deformation_argument}
\end{figure}

If $\phi_0(p) \neq 0$, because $\phi$ is single-valued at $p$ then $\Delta\theta = 0$ at some intermediate loop $L'$, which is mapped to $S'$ between $S$ and $p$ in physical space, but since $\Delta\theta$ has to be an integer multiple of $2 \pi$, $\Delta\theta$ has to jump from $2 \pi$ to $0$ at $L'$, which contradicts the hypothesis that $\phi$ is a smooth function of the coordinates. Hence $\phi_0 = 0$ at least at one point of a loop $L''$ located between $L$ and $L'$. The excitation confined by the loop $S''$, the image of $L''$ in physical space, is called a \emph{cosmic string}.
This reasoning can be better visualized in Figure \ref{fig:deformation_argument} and works for any model that possesses a symmetry breaking potential of the form (\eqref{goldstone_potential}).\\

We can show the phase difference around the loop $L''$, or any other above it, in Figure \ref{fig:deformation_argument2} is indeed zero. The phase $\theta$ of any point in $L''$ is

\begin{equation}
\theta = \tan^{-1}\left( \frac{\text{Im}(\phi)}{\text{Re}(\phi)} \right).
\end{equation}
So, starting at point $(\text{Re}(\phi), \text{Im}(\phi)) = (0, 0)$ and going anticlockwise on the loop $L''$ we see that $\theta$ increases in the first quarter of the circle, and so it does on the second quarter. However, on the third quarter, when $\text{Re}(\phi) < 0$, the phase difference is negative, and so it is on the fourth quarter, yielding $\Delta\theta = 0$.\par

The deformation argument outlined above assures that there must be an excitation of the Higgs field at some point of the surface bounded by $L$, and it is clear that this excitation should have cylindrical symmetry since $L$ is on a bounded 2-surface. Moreover, the fact that $\Delta\theta$ should be an integer multiple of $2 \pi$ means that we can make the correspondence $\theta = n \varphi$, where $n$ is called the \emph{winding number}. Finally, the vacuum manifold is a circle, which is \emph{not} simply connected because a loop on the circle cannot be continuously shrunk to a point on the circle \cite{nakahara2018geometry}.\\

\begin{figure}[H]
\caption{The vacuum manifold of the Goldstone model is a circle.}
\includegraphics[width=0.18\textwidth]{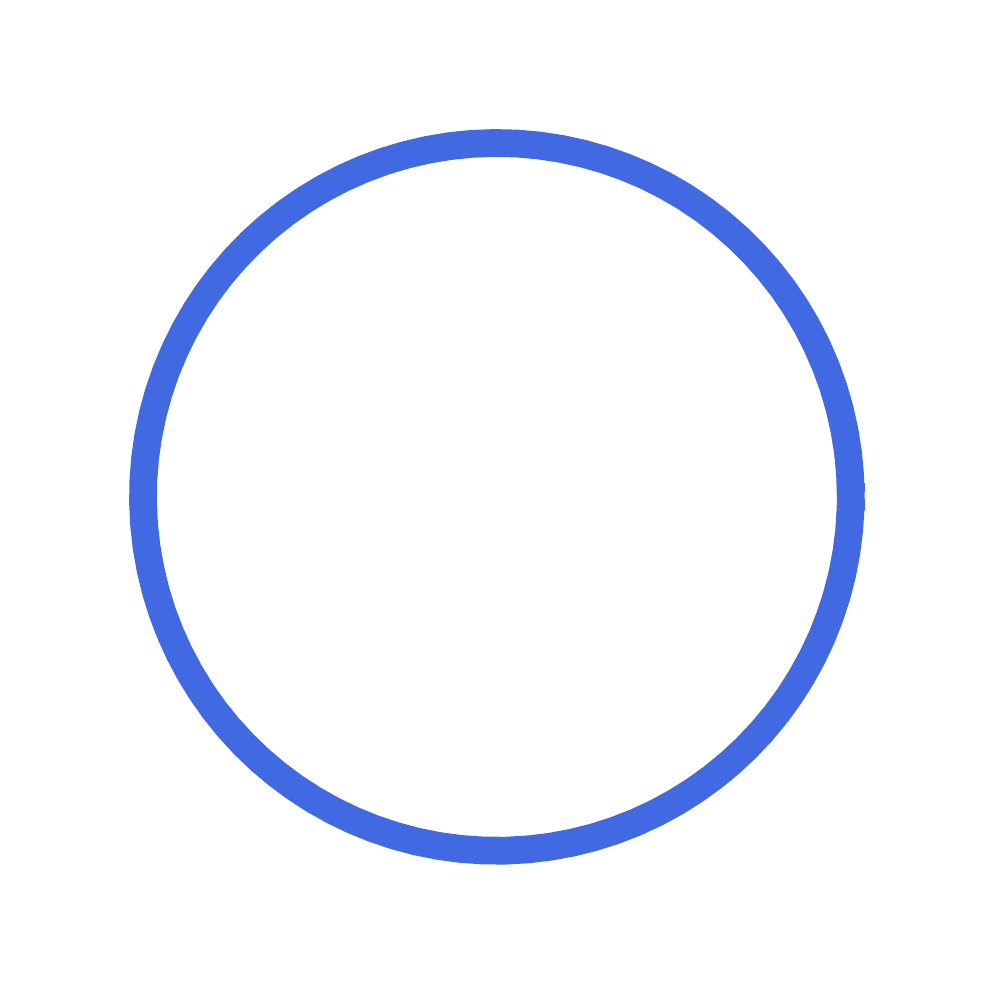}
\caption*{Source: the author (2020)}
\end{figure}

Now we turn our attention to the simplest vortex solution, the global string. This solution arises from the Goldstone model (\ref{goldstone_Lagrangian}), which breaks global $U(1)$ symmetry. Consider a localized configuration, $V(\phi) \xrightarrow{r \to \infty} 0$, which means that far from the core of the string the scalar field is approximately\ $\phi = \eta e^{i n \varphi}$. The energy density is given by\footnote{$V(\phi)$ being localized implies that the vacuum value of $|\phi|$ is a constant, which in turn means that the energy density $H(\phi)$ is localized.}
\begin{equation}
H = |\nabla \phi|^2 + V(\phi) , \, H(r \rightarrow \infty) \approx |\nabla \phi|^2,
\label{energydensity_globalstring}
\end{equation}
which gives the total energy (excluding the contribution from the core)
\begin{gather}
\begin{aligned}
E &= \int{|\nabla \phi|^2 d^2x} = \int{\eta^2 n^2 \frac{1}{r^2} r dr d\varphi} = 2\pi n^2 \eta^2 \int_{\delta}^{R}{\frac{1}{r} dr} \\
& = 2 \pi n^2 \eta^2 \, \ln \left(\frac{R}{\delta} \right).
\end{aligned}
\label{globalstring_energy}
\end{gather}
In the above expression, $\delta$ is the width of the string\footnote{We define the width of the string as the point at which the contribution from the potential energy is irrelevant.}, $R$ is the cutoff radius, and the winding number $n$ labels the kinetic energy of the string. We see that \eqref{globalstring_energy} diverges as $R \to \infty$,  suggesting the isolated global string is unstable, which is expected from Derrick's theorem since this a static solution of a non-linear pure scalar model. We could anticipate the instability based on Derrick's theorem.

\subsection{Local strings}
A local string is a vortex solution with local gauge symmetry breaking. Here we consider the abelian-Higgs model (\ref{AH_Lagrangian}), which is the Goldstone model coupled with the $U(1)$ gauge field. Notice that, because we are coupling the scalar field with a gauge field, Derrick's theorem no longer predicts instability. The actual vacuum is the same as in the pure Goldstone case, so the scalar field cannot be affected by the gauge field far from the core, i.e., $|\phi| \xrightarrow{r \to \infty} \eta$. The effect of the gauge field should be concentrated on the string core since we are searching for a localized solution.\par

The Euler-Lagrange equations for the abelian-Higgs model are
\begin{subequations}\label{AH_eom1}
\begin{align}
\partial_\mu \partial^\mu \phi - 2 i e A^\mu \partial_\mu \phi + [\lambda \, (|\phi|^2 - \eta^2) - e^2 A_\mu A^\mu] \phi = 0, \label{AH_eom11} \\
-\frac{1}{2} i e \left(\phi \partial^\mu \phi^* - \phi^* \partial^\mu \phi \right) + \frac{1}{2} e^2 A^\mu \phi^* \phi \propto \partial_\nu F^{\nu \mu}. \label{AH_eom12}
\end{align}
\end{subequations}
Now plugging the field far from the core $\phi = \eta e^{i n \varphi}$ in \eqref{AH_eom12}, we get
\begin{equation}
A_\varphi = -\frac{n}{e},
\end{equation}
which indicates a magnetic flux, $\phi_B$, on the string
\begin{equation}
\phi_B = \int_{S}{\vec{B} .\vec{dS}} 
= \int_{S}{(\nabla \times \vec{A}) \, . \vec{dS}} 
= \oint_{\partial S}{\vec{A} \cdot \vec{dl}} 
= \int_{\partial S}{A^\varphi r d\varphi} 
= \int_{0}^{2\pi}{\frac{n}{er}\, r d\varphi} 
= \frac{2\pi}{e}n.
\end{equation}
The additional minus sign in $A^\varphi$ appears because the metric has signature -2.
We have seen that the winding number $n$ labels the kinetic energy of the string, and now we see that it also labels the magnetic flux, which is quantized!\par

Denoting the field covariant derivative as $D_\mu = \partial_\mu - ieA_\mu$, we know that far from the core
\begin{gather}
\begin{aligned}
D_\mu \phi (r \rightarrow \infty) & \approx 0, \\
F_{\mu \nu} (r \rightarrow \infty) & \approx 0 ,
\end{aligned}
\end{gather}
which makes the energy depend purely on $V(\phi)$, $H \approx V(\phi)$, which has finite integral. It can be shown that the mass per unit length $\mu$ of the local $U(1)$ string is given by \cite{vilenkin1994cosmic}

\begin{equation}
\mu \approx 2 \pi \eta^2 \ln\left( \frac{m_\phi}{m_A} \right),
\end{equation}

where $m_\phi = \sqrt{\lambda} \eta$ and $m_A = \sqrt{2} e \eta$. We have dodged Derrick's theorem by employing a gauge-field coupling.\par

In 1973 Nielsen and Olesen \cite{nielsen1973vortex} found explicitly the existence of a static cylindrically symmetric solution to the field equations \eqref{AH_eom1}. For this reason, the $U(1)$ local string, or abelian-Higgs string, is usually called the Nielsen-Olesen vortex. In the next section, we show the derivation of the Nielsen-Olesen solution and discuss some of its properties. Also, from now on, we use the words \emph{string} and \emph{vortex} interchangeably.

\section{Nielsen-Olesen vortex}
\epigraph{\textit{The scientist does not study nature because it is useful to do so. He studies it because he takes pleasure in it, and he takes pleasure in it because it is beautiful. If nature were not beautiful it would not be worth knowing, and life would not be worth living. I am not speaking, of course, of the beauty which strikes the senses, of the beauty of qualities and appearances. I am far from despising this, but it has nothing to do with science. What I mean is that more intimate beauty which comes from the harmonious order of its parts, and which a pure intelligence can grasp.}}{Henri Poincaré}

The equations of motion of the abelian-Higgs model can be expressed in the form
\begin{subequations}\label{AH_eom2}
\begin{gather}
(\partial_\mu + ieA_\mu)(\partial^\mu + ieA^\mu)\phi + \frac{\lambda}{2}( |\phi|^2 - \eta^2)\phi = 0, \label{AH_eom21} \\
\partial_\mu F^{\mu \nu} = j^\nu, \\ \label{AH_eom22}
j^\nu = 2e \, Im[\phi^*(\partial^\nu - ieA^\nu)\phi].
\end{gather}
\end{subequations}
For convenience we rescale the quantities, $(\phi, A_\mu, \partial_\mu) \rightarrow \frac{1}{\eta} \, (\phi, A_\mu, \partial_\mu)$, and choose the Lorentz gauge $\partial_\mu A^\mu = 0$. Besides that, we take the following ansatz
\begin{gather}
\begin{aligned}
\phi(\vec{r}) &= e^{in\varphi} f(r), \\
A^\varphi (\vec{r}) &= -\frac{n}{er} \alpha(r),
\end{aligned}
\label{NO_ansatz}
\end{gather}
where $f(r)$ and $\alpha(r)$ are unknown functions that satisfy the boundary conditions
\begin{gather}
\begin{aligned}
r &\to \infty: f(r) = \alpha(r) = 1, \\
r &\to 0: f(0) = \alpha(0) = 0.
\end{aligned}
\end{gather}

Notice that far from the string we recover $A^\varphi = \frac{n}{er}$, $\phi = \eta e^{in\varphi}$. Also, the magnetic flux is in fact given by $2\pi n/e$ since $\alpha(r\to\infty) = 1$. Now we can determine $f(r)$ and $\alpha(r)$ by plugging the ansatz \eqref{NO_ansatz} in the equations of motion \eqref{AH_eom2}. One of the terms is
\begin{gather}
\begin{aligned}
(\partial_\mu + ie A_\mu)(\partial^\mu + ieA^\mu)\phi &= \partial_\mu \partial^\mu \phi - ie A_\mu \partial^\mu \phi - ie \partial_\mu(A^\mu \phi) - e^2 A_\mu A^\mu \phi \\
&= -\left(\nabla^2 \phi - ie \vec{A} \cdot \vec{\nabla} \phi - \nabla \cdot (\vec{A} \phi) - e^2 |A|^2 \phi \right)\\
&= -\left(\nabla^2 \phi - 2 ie \vec{A} \cdot \vec{\nabla} \phi - \phi \underbrace{\nabla \cdot \vec{A}}_{= \,0} - e^2 |A|^2 \phi \right)\\
&= -e^{in\varphi} \Bigg[ \frac{1}{r} \frac{df}{dr} + \frac{d^2r}{dr^2} + \underbrace{ \frac{2n^2}{r^2}f \alpha - \frac{n^2}{r^2} \alpha^2 - \frac{n^2}{r^2} f }_{ \frac{n^2f}{r^2}(\alpha - 1)^2} \Bigg] ,
\end{aligned}
\end{gather}
giving rise to
\begin{equation}
f'' + \frac{1}{r} f' - \frac{n^2}{r^2} (\alpha - 1)^2 f - \frac{\lambda}{2} (f^2 - 1)f = 0
\label{NO_eom1}
\end{equation}
for \eqref{AH_eom21}. Moreover, we obtain \eqref{AH_eom22}
\begin{gather}
\begin{aligned}
\partial_\mu F^{\mu \nu} &= \partial_\mu(\partial^\mu A^\nu) - \partial_\mu(\partial^\nu A^\mu), \\
&= -\nabla^2 \vec{A} = -\left[ \nabla^2(A^\varphi) - \frac{A^\varphi}{r^2} \right] \hat{\varphi}, \\
&=-\left[\frac{1}{r} \frac{\partial A^\theta}{\partial} + \frac{\partial^2 A^\theta}{\partial r^2} - \frac{A^\varphi}{r^2} \right] \hat{\varphi}, \\
&= \frac{n}{er} \left( \alpha'' - \frac{1}{r} \alpha' \right) \hat{\varphi}, \\
\end{aligned}
\end{gather}
and
\begin{gather}
\begin{aligned}
\phi^* (\partial^\nu +ieA^\nu)\phi &= -e^{-in\varphi} f \left[\frac{in}{r} f e^{in\varphi}  - \frac{in}{r} \alpha f e^{in\varphi} \right], \\
&= -i\frac{n}{r} f^2 (1 - \alpha) \hat{\varphi}. \\
\end{aligned}
\end{gather}
Finally, the equations of motion take the form
\begin{subequations}\label{NO_eom2}
\begin{gather}
f''  + \frac{1}{r}f' - \frac{n^2}{r^2} f (\alpha - 1)^2 - \frac{\lambda}{2} f (f^2 - 1) = 0, \label{NO_eom2_1} \\
\alpha'' - \frac{1}{r}\alpha' - 2e^2 f^2 (\alpha - 1) = 0.
\label{NO_eom2_2}
\end{gather}
\end{subequations}
An analitical solution to \eqref{NO_eom2} can only be obtained in some restricted situations.\\
In the regime $r \to \infty$ we have $f(r \to \infty) = 1$, which simplifies \eqref{NO_eom2_2}. By rescaling the radius, $\tilde{r} = \sqrt{2}e r$, and changing the variable, $\alpha(r) - 1 = \tilde{r}\xi(r)$, where $\xi(r)$ is to be determined, we arrive at a new form of \eqref{NO_eom2_2} in the asymptotic regime,
\begin{equation}
\tilde{r}^2 \xi'' + \tilde{r} \xi' - (\tilde{r}^2 + 1) \xi = 0,
\label{NOeq_asymptotic}
\end{equation}
where prime denotes differentiation with respect to $\tilde{r}$. Equation \eqref{NOeq_asymptotic} is the modified Bessel equation, and since we want the solution to be finite at large $r$, it shall be proportional to the modified Bessel function of the second kind of order 1, $K_1(\tilde{r})$. The asymptotic behavior of $\alpha(r)$ is given by
\begin{equation}
\xi (\tilde{r}) = \pm K_1(\tilde{r}) \to \alpha(r) = 1 \pm \sqrt{2}e r K_1 \left(\sqrt{2}e r \right) ,
\end{equation}
where the integration constant is set to $\pm 1$. \par
\begin{figure}[H]
\caption{Dependence of the scalar and gauge fields on the parameter $\beta$. Here we used $n = 2$.}
\includegraphics[width=0.85\textwidth]{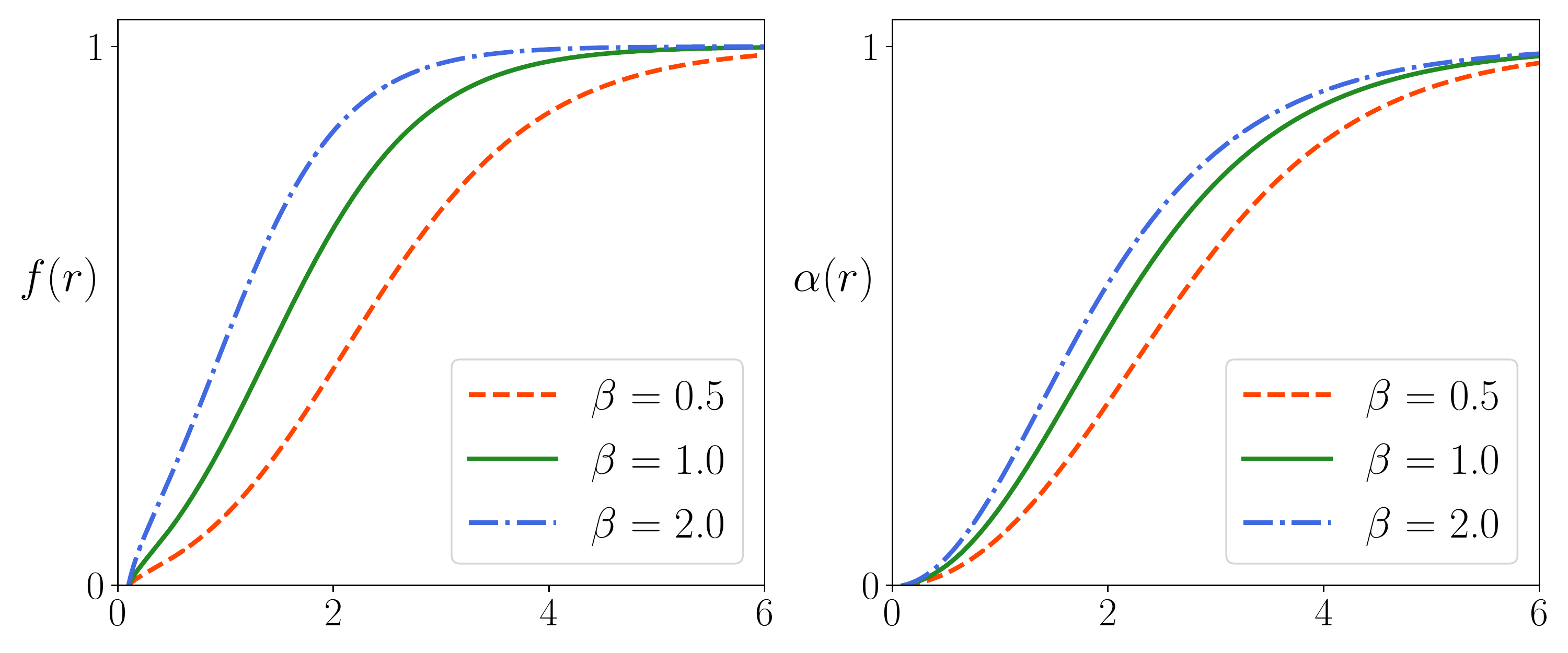}
\caption*{Source: The author (2021).}
\label{fig:NO_fieldprofile}
\end{figure}

If we rescale the coordinates $x \to x/e$ in \eqref{AH_eom2} we can see that the solution to the field equations are only dependent on the parameter $\beta = (m_{\phi}/m_A)^2$, which measures the relative strength between scalar and gauge fields. In Figure \ref{fig:NO_fieldprofile} we show the numerical solution of $f(r)$, $\alpha(r)$ and their dependence on $\beta$. We shall see that the parameter $\beta$ measures the ratio between disruptive and confining forces.

\begin{figure}[H]
\caption{The width of the Nielsen-Olesen vortex is proportional to the winding number $n$.}
\includegraphics[width=0.8\textwidth]{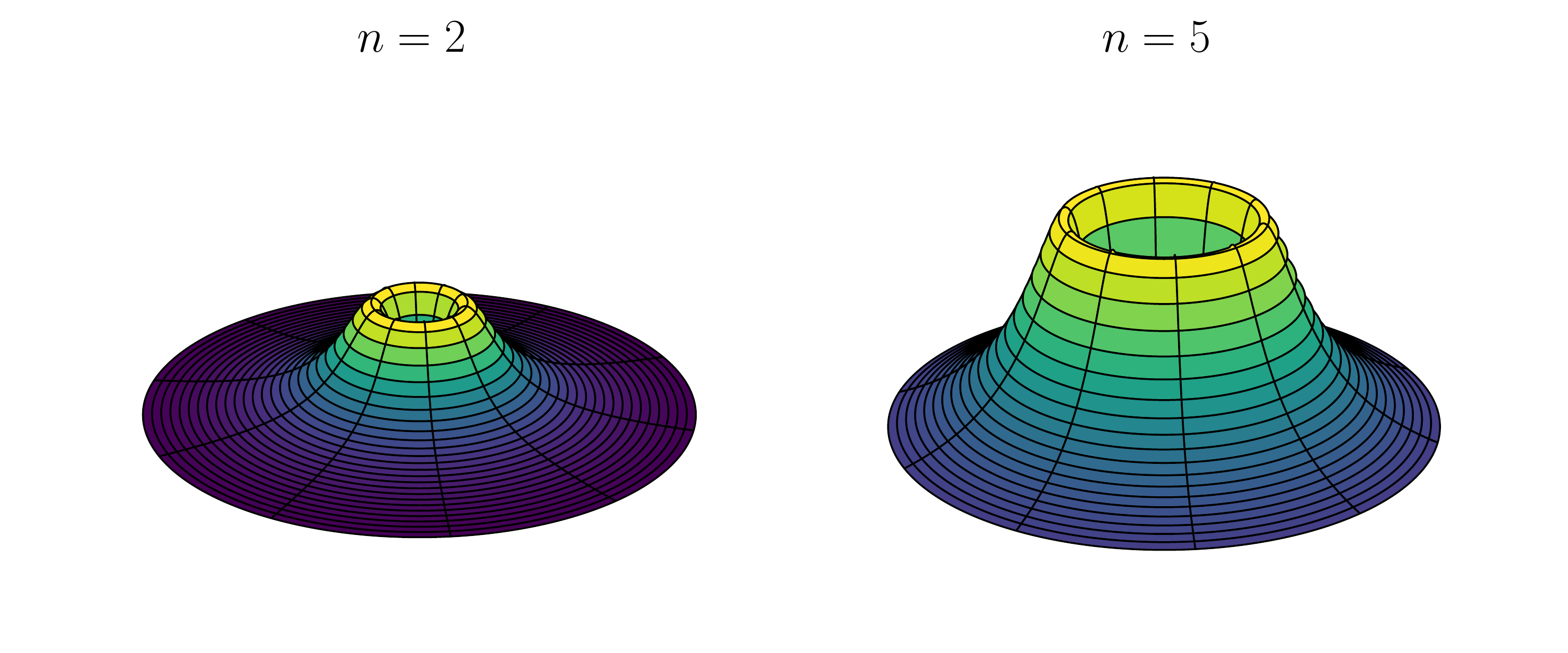}
\caption*{Source: The author (2021).}
\label{fig:NO_energydensity}
\end{figure}

In Figure \ref{fig:NO_energydensity} we see how the vortex energy density is affected by the winding number. In general, the string has zero energy density at the axis and is surrounded by a cylindrical energy density that goes to zero at infinity. This feature will be important when coupling strings with gravity. \par
Moreover, we notice that the width of the vortex is proportional to the winding number $n$. In fact, if we consider the magnetic flux uniformly distributed over the vortex, the characteristic width is proportional to $\sqrt{n}$. The reason for this is that the vortex width should increase with $n$ to accommodate the increasing magnetic flux. Recently Alexander Penin and Quinten Weller, in \cite{penin2020becomes} and \cite{penin2021theory}, used an asymptotic expansion of the field functions in inverse powers of $n$ to find a piece-wise perturbative analytical solution of \eqref{NO_eom2} in the limit $n \to \infty$.

\subsection{Properties of the Nielsen-Olesen vortex}
Equipped with the solution of the Nielsen-Olesen vortex, we can start analyzing some of its properties.

\subsubsection{Energy-momentum tensor}

Given an action $S$ defined in a spacetime with metric $g_{\mu \nu}$ with signature -2, the energy-momentum tensor $\tensor{T}{_\mu_\nu}$ is given by
\begin{equation}
\tensor{T}{_\mu _\nu} = \frac{2}{\sqrt{-g}} \frac{\delta S}{\delta g^{\mu \nu}},
\label{EMtensor_general1}
\end{equation}
where $g$ is the determinant of the metric. For a Lagrangian $\mathcal{L}$, \eqref{EMtensor_general1} yields
\begin{equation}
\tensor{T}{_\mu _\nu} = 2 \frac{\partial \mathcal{L}}{\partial g^{\mu \nu}} - g_{\mu \nu} \mathcal{L}.
\label{EMtensor_general2}
\end{equation}
Now if we apply \eqref{EMtensor_general2} to the Lagrangian \eqref{AH_Lagrangian} we get
\begin{equation}
\tensor{T}{_\mu _\nu} = \partial_\mu \phi \partial_\nu \phi^{*} + ie (A_\mu \phi \partial_\nu \phi^* - A_\nu \phi^* \partial_\mu \phi) + e^2 A_\mu A_\nu |\phi|^2 + \tensor{F}{_\mu _\sigma} \tensor{F}{^\sigma _\nu} - g_{\mu \nu} \mathcal{L},
 \end{equation}
which for the ansatz \eqref{NO_ansatz} results in
\begin{subequations} \label{NO_EM}
\begin{align}
T_{rr} &= \frac{n^2}{2 e^2 r^2} \left(\frac{d\alpha}{dr}\right)^2 + \frac{1}{2} \left(\frac{df}{dr} \right)^2 - \frac{f^2}{2 r^2} n^2(1 + \alpha)^2 + \frac{\lambda \eta^2}{2} - \frac{\lambda}{4}f^4  - \frac{\lambda \eta^4}{4},
\label{NO_EMrr}\\
T_{\varphi \varphi} &= \frac{n^2}{2e^2} \left(\frac{d\alpha}{dr}\right)^2 + \frac{1}{2} f^2 n^2(1 + \alpha)^2 + r^2 \left[ \frac{\lambda \eta^2}{2} f^2 -  \frac{\lambda}{4}f^4 - \frac{1}{2} \left(\frac{df}{dr} \right)^2 - \frac{\lambda \eta^4}{4} \right], 
\label{NO_EMphiphi} \\
T_{tt} &= -T_{zz} = \frac{n^2}{2 e^2 r^2} \left(\frac{d\alpha}{dr}\right)^2 + \frac{1}{2} \left[ \left(\frac{df}{df} \right)^2 + \frac{f^2}{r^2} n^2(1 + \alpha)^2 \right] - \frac{\lambda \eta^2}{2} + \frac{\lambda}{4} f^4 + \frac{\lambda \eta^4}{4} .
\label{NO_EMtt} 
\end{align}
\end{subequations}
Equations \eqref{NO_EM} tell us that besides the energy density, $\varepsilon = \tensor{T}{^t _t}$, and the longitudinal tension, $p^z = \tensor{T}{^z _z}$, there is also radial flux of momentum, $p^r = \tensor{T}{^r _r}$, and azimutal momentum $p^\varphi = \tensor{T}{^\varphi _\varphi}$. Although in some restricted situations we can neglect $p^r$ and $p^\varphi$, they are essential to analyze the vortex internal dynamics which are non-vanishing in general. \par 

\subsubsection{Stability}
One interesting feature of the Nielsen-Olesen solution is that if the winding number $n$ is larger than 1, the vortex can \emph{unwind}, i.e: the string decays into $n$ strings each with magnetic flux equal to $2 \pi/e$. In 1976 Eugene Bogomol'nyi \cite{bogomol1976stability} showed that Nielsen-Olesen strings are unstable to such unwinding if $\beta = \left( \frac{m_{\phi}}{m_A} \right)^2 > 1$ and stable if $\beta < 1$. This can be explained by the following reasoning. The magnetic lines repel each other while the Higgs field act to confine the vortex. Since the field strength is inversely proportional to the mass, when $m_A > m_{\phi}$ ($\beta < 1$) the Higgs field is stronger than the repulsion of the magnetic lines, while when $m_A < m_{\phi}$ ($\beta > 1$) the repulsion overcomes the confinement and the string unwind. Hence we can see $\beta$ as the ratio between disruptive and confining forces.
At this point, we can see the connection between cosmic strings and vortex lines in superconductors. Type-II superconductors ($\beta < 1$) at certain values of the applied magnetic field present stable vortex lines with quantized magnetic flux, while in type-I superconductors ($\beta > 1$), the quantized vortex lines are unstable and immediately unwind due to Meissner effect. In 1976, Vega and Schaposnik \cite{de1976classical} showed that in the critical coupling, $\beta = 1$, the equations of motion decouple, and it is possible to construct an analytical power-series solution to the field equations. In this situation $T_{rr} = 0$\footnote{It implies $T_{\varphi \varphi} = 0$, via energy-momentum conservation $\partial_\mu \tensor{T}{^\mu _\nu} = 0$.} suggesting an equilibrium in the radial direction, i.e: there is no radial net flux.\par
Here we finish our discussion on the Nielsen-Olesen solution. In the next section, we will see a simple mechanism for the cosmological creation of cosmic strings.

\section{Kibble and the cosmic tale}
\epigraph{\textit{...so many out-of-way things had happened lately, that Alice begun to think very few things indeed were really impossible.}}{From "Alice's adventures in Wonderland", by Lewis Carrol.}
The discussion of the Goldstone potential up to now has been simplified for pedagogical purposes. When the temperature is above 0K, the Goldstone potential has a different form caused by the interaction of the matter with the environment energy. 
Furthermore, according to the standard model of cosmology, the universe, with all its fields, started in a highly dense, hot state and cooled through a series of epochs, during which matter interaction drastically changed. In what follows, we will see how the temperature affects the Higgs field and how it is related to the formation of topological defects in the early universe. 
\subsection{The role of temperature}
Before getting to the Kibble mechanism in the next section, we sketch the essential temperature-dependence of the Goldstone potential. The discussion of this section is summarized and for further details and calculations we refer the reader to \cite{vilenkin1994cosmic}
The Goldstone potential at non-zero temperature can be approximated by \cite{vilenkin1994cosmic}  \begin{equation}
V(\phi, T) = m^2(T) |\phi|^2 + \frac{\lambda}{4} |\phi|^4 - \frac{2 \pi^2}{45} T^4,
\label{goldstone_potential_temp}
\end{equation}
via perturbative corrections, where $m(T)$ is the effective temperature-dependent mass of the field
\begin{equation}
m^2(T) = \frac{\lambda}{12} \left( T^2 - 6 \eta^2 \right). \\
\end{equation}
For $T > T_c = \sqrt{6}\eta$, $m^2(T)$ is positive and the VEV is $\langle \phi \rangle = 0$. As the universe evolved, the temperature dropped and, for $T < T_c$, $m^2(T) < 0$. When  $m^2(T) < 0$ the state $\langle \phi \rangle = 0$ is not stable anymore and the scalar field acquires a non-zero VEV, leading to $U(1)$ symmetry breaking. Minimizing $V(\phi, T)$ results in
\begin{equation}
|\phi| = \frac{1}{\sqrt{6}} \sqrt{T_{c}^2 - T^2},
\end{equation}
which reduces to $\eta$ by setting $T = 0$. In summary, decreasing temperature drives symmetry breaking.\par
Finally, when $T \approx T_c$ the approximation is not valid anymore because this regime represents a second-order phase transition and we need to include further corrections to \eqref{goldstone_potential_temp}.

\subsection{Kibble mechanism}
Looking at a piece of iron at high temperatures, we see that the magnetization direction varies from point to point; magnetization does not have a preferred direction. This is the most symmetric state of the system since it possesses rotational symmetry, i.e., we can rotate the iron, and magnetization will look the same. Now suppose we cool it; the molecules have less kinetic energy and, if the cooling is uniform throughout the material, the magnetization in each region ends up in a common direction: the iron becomes a magnet! The magnetized metal is the less symmetric state of the system since we no longer have rotational symmetry. But if things happen this way, why do not all pieces of iron we encounter are magnets?\par

The loophole of the reasoning is the assumption of uniform cooling. In nature, such uniform cooling never happens. We certainly can achieve uniformity in a laboratory with ever-increasing levels of precision, but this is not favored spontaneously. In nature or most simple industrial processes, the non-uniform cooling leads to different regions choosing different magnetization directions, leading to a non-magnetic material. In the early universe, things were quite similar.
As the universe cooled down, the Higgs field went from a symmetric state, $\langle \phi \rangle = 0$, to an asymmetric one $\langle \phi \rangle \neq 0$. Nevertheless, the phase of the field, i.e., the specific representative vacuum it will be in, is not \emph{a priori} defined. Random fluctuations of $\phi$ will determine it. As one can anticipate, this settling process is not uniform through the universe. Causally disconnected regions may choose different vacuum states, and the way these regions organize themselves must give rise to topological defects.\par
Consider the $\phi^4$ model. Two nearby regions in space might end up in opposite minima of the potential. Between them, there can emerge a kink interpolating both vacua. The emerging 2-dimensional (though the dynamics is restricted to 1 dimension) kink-like structure is called \emph{domain wall}.\par
Now consider the Higgs potential. We can imagine a region of space where the field is in the symmetric state $\langle \phi \rangle = 0$ surrounded by patches where the field is in the asymmetric state $| \langle \phi \rangle | \neq \eta$. This gives birth to 1-dimensional defects, called cosmic strings. This mechanism for the cosmological formation of topological defects is called \emph{Kibble mechanism}, after Thomas Kibble, who proposed it in 1976 \cite{kibble1976topology}.

\begin{figure}[H]
\caption{Casually disconnected regions in space can end up in different vacua, creating topological defects.}
	\begin{subfigure}[b]{0.35\textwidth}
	\includegraphics[width=\textwidth]{../images/domainwall_form.pdf}
	\caption{}
	\end{subfigure}
	\quad
	\begin{subfigure}[b]{0.35\textwidth}
	\includegraphics[width=\textwidth]{../images/cosmicstring_form.pdf}
	\caption{}
	\end{subfigure}
\caption*{Source: The author (2021).}
\label{fig:kibble_mech}
\end{figure}

\chapter{Gravitating vortices}
\label{chap2}

\section{The ideal string}
\epigraph{\textit{When it comes to the world around us, is there any choice but to explore?}}{Lisa Randall, in "Warped Passages".}
In this section, we will see how a zero-thickness cosmic string affects the geometry of spacetime. To this end, we use the wire approximation, which treats the vortex as a thin pipe with negligible radius and constant energy density. This is a good starting point since some physical properties can be worked out analytically, and it gives us some valuable physical intuition when studying more complicated, non-analytical scenarios. We start with the consequences of the wire approximation on the energy-momentum tensor and then proceed to its gravitational effects.

\subsection{Wire approximation}
By treating the vortex as a thin wire we can average the energy-momentum (EM) tensor $\tensor{T}{^\mu _\nu}$ over the string cross-section and consider all the energy to be localized at one point in the x-y plane,
\begin{equation}
\tensor{\tilde{T}}{^\mu _\nu} = \delta(x) \delta(y) \int{\tensor{T}{^\mu _\nu} \,dx dy},
\label{EM_wirestring}
\end{equation}
This is called the \emph{wire approximation} which is specially important when considering gravitational effects of cosmic strings. Because the string is supposed to be invariant under boosts along the z-axis we must have $\tensor{T}{^0 _0} = \tensor{T}{^3 _3}$, such that when we apply a Lorentz transformation to \eqref{EM_wirestring} it does not mix these components. In addition, we assume the string does not have any internal shear forces, $\tensor{T}{^3 _i} = 0$ for i = 1,2 or energy flux in any direction $\tensor{T}{^0 _i} = 0$ for i = 1, 2, 3. These assumptions guarantee the string does not have any considerable internal dynamics. In addition, because the divergence of the EM tensor vanishes, $\partial_\mu \tensor{T}{^\mu _\nu} = 0$, we have
\begin{equation}
0 = \int{x \left( \partial_x \tensor{T}{^x _\nu} + \partial_y \tensor{T}{^y _\nu} \right) \, dxdy} = \int{ (x \tensor{T}{^x_\nu}) \Big{|}_{x = -\infty}^{x = \infty} dy} + \int{(x \tensor{T}{^y _\nu}) \Big{|}_{y = -\infty}^{y = \infty} dx} - \int{\tensor{T}{^x _\nu} \, dx dy},
\end{equation}
where we used integration by parts. Assuming that the components of the EM tensor fall exponentially when $r \rightarrow \infty$, one can say $\tensor{T}{^x _\nu} = 0$, and, by a similar reasoning, also $\tensor{T}{^y _\nu} = 0$ for $\nu = 1, 2$. Hence $\tensor{T}{^i _j} = 0$ for i, j = 1, 2. The only non-vanishing components of $\tensor{T}{^\mu _\nu}$ are $\tensor{T}{^0 _0}$ and $\tensor{T}{^3 _3}$. Denoting the mass per unit length by $\mu$ and knowing $\tensor{T}{^0 _0} = \varepsilon(r)$ is the energy density, we end up with
\begin{equation}
\begin{gathered}
\tensor{T}{^\mu _\nu} = \mu \delta(x) \delta(y) \, diag(1, 0, 0, 1), \\
\mu = 2\pi \int_{0}^{\infty}{\varepsilon(r) r dr}.
\end{gathered}
\end{equation}
Notice that the only conditions we used were that the EM should fall exponentially at $r \rightarrow \infty$ and that it is invariant under boosts along the $z$-axis, so these results are valid for every localized narrow \footnote{The vortex width should be small compared to the other parameters of the system.} string configuration. The string solution with these simplifications is an \emph{ideal string}. Alternatively, one could use the lorentz-invariance, localizability of the energy, the no-shear and no disruptive force conditions to show that any static infinite straight string should have an EM tensor proportional to diag(1, 0, 0, 1).\par

We have seen that the Nielsen-Olesen vortex does have momentum in $r$ and $\varphi$ directions, so the wire approximation could only represent this solution when $\beta = 1$, which is also called supersymmetric limit. However, it could be useful when considering large-scale gravitational effects, when the internal dynamic of the string is not important.

\subsection{Gravitating ideal string}
In this section, we shall see how an infinite straight wire affects spacetime around itself. We will follow the approach independently done by Richard Gott \cite{gott1985gravitational} and William Hiscock \cite{hiscock1985exact}. The general procedure is straightforward. We consider a cylindrically symmetric energy density $\varepsilon(r)$ with finite radius $r_0$, find the metric in $r < r_0$, and match it with the exterior, $r > r_0$, vacuum solution of the Einstein field equations (EFEs). In the end, we take the limit $r_0 \to 0$ in the exterior metric, which yields the metric of a zero-thickness infinite straight string.\par
\begin{figure}[H]
\caption{The figure approximates a cosmic string as an energy density confined in a tube of radius $r_0$.}
\includegraphics[width=0.75\textwidth]{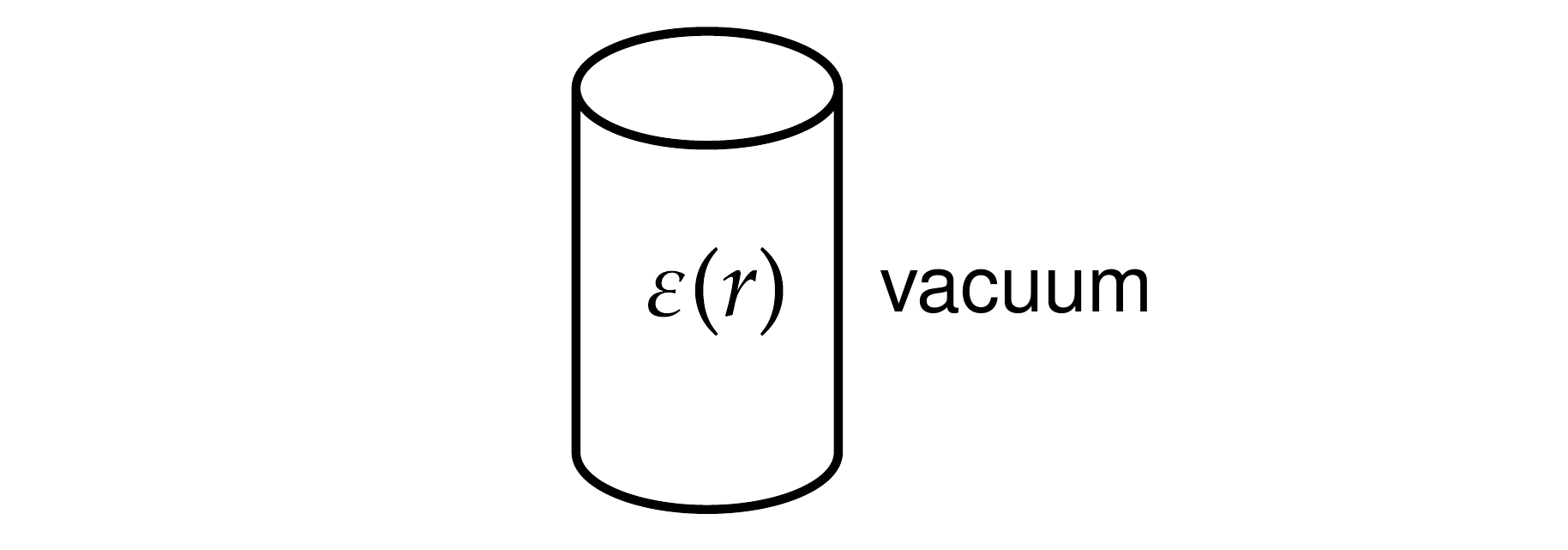}
\caption*{Source: The author (2021).}
\label{fig:cylinder}
\end{figure}
We need to solve EFEs
\begin{equation}
\tensor{G}{^\mu _\nu} = \tensor{R}{^\mu _\nu} - \frac{1}{2}R \delta^{\mu}_{\nu} = \tensor{T}{^\mu _\nu},
\label{wire_einsteq}
\end{equation}
with an energy-momentum tensor inside the tube given by
\begin{equation}
\tensor{T}{^t _t} = \tensor{T}{^z _z} = \varepsilon(r) \quad \text{ and } \quad \tensor{T}{^r _r} = \tensor{T}{^\varphi _\varphi} = 0,
\label{wire_EM}
\end{equation}
where $\varepsilon$ is considered to be any $r$-dependent function. Though Gott and Hiscock used constant energy density we will stick to a general approach and consider a constant energy density $\varepsilon(r) = \varepsilon_0$, only when needed. The ansatz for the line-element proposed by \textcite{gott1985gravitational, hiscock1985exact} is
\begin{equation}
ds^2 = e^{2\chi} dt^2 - e^{2\psi} \left(dr^2 + dz^2 \right) - e^{2\omega} d\phi^2,
\label{wire_metric_ansatz}
\end{equation}
where $\chi(r), \psi(r), \omega(r)$ are functions of $r$ only. We use exponentials just to ensure the metric components are positive. Equation \eqref{wire_metric_ansatz} yields the following non-zero components of the Einstein tensor $\tensor{G}{^\mu _\nu}$
\begin{subequations}\label{wire_einstensor}
\begin{align}
\tensor{G}{^t _t} &= - \left[ \left(\omega^\prime \right)^2 + \omega'' + \psi'' \right] e^{-2 \psi} \label{wire_einstensor1}\\
\tensor{G}{^r _r} &= - \left[ \chi' \omega' + \chi'\psi' + \omega' \psi' \right] e^{-2 \psi} \label{wire_einstensor2} \\
\tensor{G}{^\varphi _\varphi} &= -\left[ \left(\chi'\right)^2 + \chi'' + \psi'' \right] e^{-2 \psi} \label{wire_einstensor3}\\
\tensor{G}{^z _z} &= -\left[ \left(\chi'\right)^2 + \chi'\omega' - \chi'\omega' + \chi'\psi' + \chi'' + \left(\omega'\right)^2 - \omega'\psi' + \omega'' \right] e^{-2\psi}, \label{wire_einstensor4}
\end{align}
\end{subequations}
where prime denotes derivative with respect to $r$. Conservation of energy-momentum, $\nabla_\mu \tensor{T}{^\mu _\nu} = 0$, where $\nabla$ denotes the covariant derivative, gives one more constraint on the metric functions
\begin{equation}
\nabla_\mu \tensor{T}{^\mu _r} = - \left(\chi' + \psi' \right) \varepsilon = 0,
\end{equation}
which implies
\begin{equation}
\chi' + \psi' = 0 \quad \text{and} \quad \chi'' + \psi'' = 0.
\label{wire_metricfunc_cond1}
\end{equation}
Using \eqref{wire_metricfunc_cond1} in (\ref{wire_einstensor3}) we get $(\chi')^2 e^{-2 \psi} = \tensor{T}{^\varphi _\varphi} = 0$ which means $\chi' = \psi' = 0$, hence $\chi$ and  $\psi$ are constants. From now on we denote $e^{\chi} = \chi_0$ and $e^{\psi} = \psi_0$. With these results we notice that only the three last terms of \eqref{wire_einstensor4} do not vanish, resulting in the equation
\begin{equation}
(\omega')^2 + \omega''   = -8 \pi  \bar{\varepsilon},
\label{wire_eom_omega1}
\end{equation}
where $\bar{\varepsilon} = \psi_{0}^2 \varepsilon$. Equation \eqref{wire_eom_omega1} is equivalent to
\begin{equation}
(e^\omega)'' = \bar{\varepsilon} e^\omega,
\label{wire_eom_omega2}
\end{equation}
from which we can find $\omega$ once $\varepsilon(r)$ is given.
Now we consider the exterior metric. Taking the ansatz \eqref{wire_metric_ansatz} and setting $\tensor{T}{^\mu _\nu} = 0$ in \eqref{wire_einsteq} results in the exterior solution
\begin{subequations}
\begin{gather}
\begin{align}
\chi &= m \, \ln{|r + K|} + C_1 \\ 
\psi &= m(m - 1) \, \ln{|r + K|} + C_2 \\
\omega &= (1 - m) \, \ln{|r + K|} + C_3, 
\end{align}
\end{gather}
\end{subequations}
where $K, C_i$ are constants to be determined. If we make $\bar{r} = r + K$ we get
\begin{subequations}
\begin{gather}
\begin{align}
e^\chi &= c \, \bar{r}^m \\
e^\psi &= b \, \bar{r}^{m(m-1)} \\
e^\omega &= a \, \bar{r}^{1-m},
\end{align}
\end{gather}
\end{subequations}
which results in the following exterior metric
\begin{equation}
ds_{out}^2 = c^2 \,\bar{r}^{2m} dt^2 - b^2 \,\bar{r}^{2m(m-1)} \,(dr^2 + dz^2) - a^2 \,\bar{r}^{2(1-m)}d\varphi^2.
\label{wire_extmetric1}
\end{equation}
Now we need to determine $a,b,c$ based on the continuity of the metric functions and their derivatives on the boundary $r_0$. Let us start with the continuity of derivatives. Inside the cylinder the functions $\chi$ and $\psi$ are constants hence continuity of derivative yields $m = 0$. Now taking $m = 0$ on the continuity of the functions $e^\chi$ and $e^\psi$ we see that $c^2 = \chi_{0}^2$ and $b^2 = \psi_{0}^2$, which means the scaling of the coordinates $t$ and $z$ are equal in the interior and exterior of the cylinder. Also invariance under boosts along the z-axis demands $c = b$. For these reasons we set $c = b =1$. Therefore, the exterior metric takes the following form
\begin{equation}
ds_{out}^2 = dt^2 - (dr^2 + dz^2) - a^2 \bar{r}^2 d\varphi^2, \quad r > r_0.
\label{wire_extmetric2}
\end{equation}
Since we have neither applied any condition on the size of the cylinder nor on the energy function $\varepsilon(r)$, we know that the spacetime outside a hard-wall cylindrically symmetric energy density \emph{has} to have the form \eqref{wire_extmetric2}. Nevertheless, notice that we still have to determine two constants, namely $a$ and $K$. When $r_0 \to 0$ we know, by regularity of the metric at the origin, the inside metric is Minkowskian, and continuity of metric at $r = r_0$ implies $K = r_0 (1 - a)$, hence $K$ vanishes when $r_0 \to 0$. For this reason from now on we use $\bar{r} = r$.\par
Now we intend to determine $a$. Imposing continuity of the derivative of $e^\omega$ across the boundary yields

\begin{equation}
e^\omega (r_0) \omega' = a.
\label{wire_boundcond_omega}
\end{equation}
Integrating \eqref{wire_eom_omega2} gives
\begin{equation}
e^\omega \omega' \Big{|}_0^{r_0} = - 8 \pi \int_{0}^{r_0}{\varepsilon(r) e^\omega dr},
\end{equation}
and employing the condition that the metric at $r = 0$ is Minkowskian, $e^\omega \xrightarrow{r \to 0} r$, we see that $e^\omega \omega' \xrightarrow{r \to 0} 1$ which means
\begin{equation}
e^\omega \omega'(r_0) = a = 1 - 8 \pi \int_{0}^{r_0}{\varepsilon(r) e^\omega} dr ,
\label{wire_a}
\end{equation}  
where \eqref{wire_boundcond_omega} was used. By now we have already determined the exterior metric in terms of the matter content of the string, but we can go even further. The mass per unit length of the string (or linear energy density), $\mu$, is given by

\begin{equation}
\mu = \int_{0}^{r_0}{\varepsilon(r) \sqrt{-g} \, dr d\varphi  } = 2 \pi \int_{0}^{r_0}{\varepsilon(r) e^\omega dr}.
\end{equation}
where $g$ is the determinant of the interior metric. This yields

\begin{equation}
ds_{out}^2 = dt^2 - (dr^2 + dz^2) - \left(1 - 4\mu \right)^2 r^2 d\varphi^2. 
\end{equation}

One interesting feature of this metric is that it does not depend on $r_0$, so it seems there is no meaning in applying the zero-thickness limit; we will get back to this point shortly. When $r_0 \to 0$ a reasonable approximation is that the energy distribution inside the tube is uniform, $\varepsilon(r) = \varepsilon_0$. Consider Equation \eqref{wire_eom_omega2} with $\varepsilon(r) = \varepsilon_0$. Employing regularity of the metric at the origin, the solution is 
\begin{equation}
e^\omega = r_* \sin(r/r_*),
\end{equation}
where $r_* = 1/\sqrt{8 \pi \varepsilon}$. With the metric functions in hand, we can calculate the mass per unit length explicitly
\begin{equation}
\mu = \int{\varepsilon r_* \sin(r/r_*)dr d\varphi}  = 2 \pi \varepsilon_0 r_{*}^2 \left[1 - \cos\left(\frac{r_0}{r_*} \right) \right].
\label{wire_mass}
\end{equation}
Now we see that when we apply the limit $r_0 \to 0$ we have to do it such that $\mu$, hence $r_0/r_*$, is kept constant. It implies that we should also have $\varepsilon_0 \to \infty$, turning the EM tensor into a delta-function. This procedure justifies that the metric \eqref{wire_extmetric2} is valid for all values of $r$, i.e., we extended the validity of \eqref{wire_extmetric2} to the whole space.\par

Consider the line-element outside the zero-thickness infinite straight string
\begin{equation}
ds^2 = dt^2 - dr^2 -(1 - 4\mu)^2 r^2 d\varphi^2 - dz^2, \quad r > 0.
\label{ds_string}
\end{equation}
This line-element represents a locally flat spacetime where the angular variable range from $0$ to $2 \pi (1 - 4\mu)$ or, equivalently, with a deficit angle $\Delta\varphi = 8\pi\mu$. Notice that $\Delta\varphi = 2 \pi$ is not physical, hence $\mu$ is constrained, $0 < \mu < \frac{1}{4}$. Since we are using natural units, $\mu = \frac{1}{4}$ means roughly $3.2 \times 10^{26}$ kg/m. At this density, a string with the mass of the Jupiter planet, roughly $2 \times 10^{27}$ kg, would be approximately $10$ meters long. If we consider the mass of the black hole at the center of the Milky way galaxy, Sagittarius A*, with mass of approximately $4 \times 10^{36}$ kg, the corresponding string would have $1.3 \times 10^{7}$ km.\\

One can think of this geometry as a flat spacetime where one has cut a slice and glued the edges, resulting in a conical structure. Imagine taking a flat sheet of paper, cutting a slice of it with a specific angle, and then gluing the edges. It is only possible if one turns the flat sheet into a cone. Now take another sheet on a table and draw a constant vector field (a set of arrows with the same size and same direction) all over the paper, cut a slice, and glue the edges again. One can see that the vector field on the two sides of the glued edge is \emph{not} in the same direction. The vector field changes the direction when it is parallel transported around the tip of the cone, meaning that there is curvature somewhere. However, the curvature can not be extended since we have not folded the paper, so it must be located on the tip.\par

\begin{figure}
\caption{The space around a cosmic string is conical. It can be seen as Minkowski space without a slice.}
\includegraphics[width=0.65\textwidth]{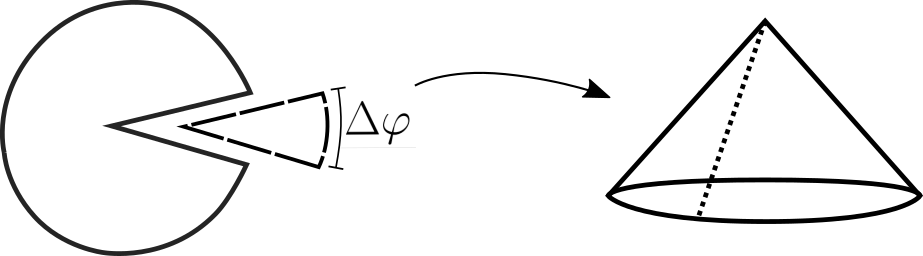}
\caption*{Source: The author (2021).}
\label{fig:cuttingandgluing}
\end{figure}

\begin{figure}[H]
\caption{A vector paralell transported around the string changes its direction, suggesting non-zero curvature.}
\includegraphics[width=0.55\textwidth]{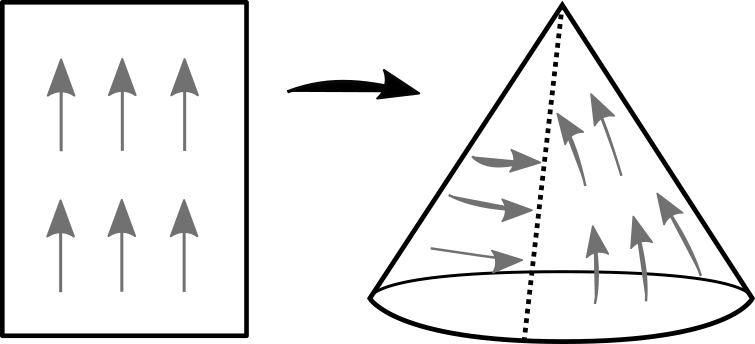}
\caption*{Source: The author (2021).}
\label{fig:conevector}
\end{figure}

We can formally demonstrate this idea. Taking the trace of EFE \eqref{wire_einsteq}
\begin{equation}
g^{ \mu \nu} \left(R_{\mu \nu} - \frac{1}{2} R g_{\mu \nu} \right) = R - 2R = -R = 8\pi T,
\end{equation}
and using the averaged energy-momentum tensor $\tensor{T}{^\mu _\nu} = \mu \delta(x) \delta(y) \text{diag}(1, 0, 0, 1)$. It results in
\begin{equation}
R = - \Delta\varphi \, \delta(x) \delta(y) = - \Delta\varphi \,\delta^{(2)}(r),
\label{wire_curvature}
\end{equation}
which means the discontinuity generated by the curvature is located precisely at the string axis. The spacetime is locally flat, i.e., it is everywhere flat except a small region. Here we will not address the controversy of using distributions\footnote{Calling it delta-function is an abuse of terminology. Though physicists commonly treat it as a function, mathematically $\delta (r)$ is a distribution.} as sources in EFE. A distributional formulation of the straight string can be found in \cite[Chapter~7]{anderson2015mathematical}.\par

The "cut and glue" process described above is one of Volterra processes, which are used to visualize the formation of crystalline defects in materials. By geometrizing a crystalline lattice, we can describe crystalline defects by geometrical properties such as curvature and/or torsion. For instance, defects called disclinations cause non-vanishing curvature while dislocations are related to non-zero torsion \cite{katanaev1992theory, puntigam1997volterra, kleinert1989gauge}. The cosmic string metric shown here is equivalent to a line-like crystalline defect called \emph{wedge disclination}, which is "created"\footnote{Volterra processes are just tools to visualize the creation of such defects.} using the process outlined before. However, this is not the only way to create such a line-like defect. Some works propose that the cosmic string can also be described by a torsion singularity, instead of a curvature one \cite{hammond2016static, fujishiro1993cosmic, de1990spin}. In this case the cosmic string condensed matter counterpart is called an \emph{edge dislocation}. \\

Another interesting property of \eqref{ds_string}, is that if we remove the z-axis from the conical line element \eqref{ds_string} the resulting line-element describes the spacetime around a point-particle with mass $\mu$ in (2+1)-dimensional gravity \cite{deser1989string, jackiw1985lower}.

Finally, a geometrical way of detecting a cosmic string is to enclose the string with a circular loop and measure its perimeter. In doing this one sees that the perimeter is $2\pi r (1 - 4\mu)$ and not just $2\pi r$. This is, however, a global measure. In the next section we will see that this global feature induces some important local physical effects.

\subsection{Topology induces physics}
\epigraph{\textit{In this terrifying world, all we have are the connections we make.}}{Bojack Horseman}
In the last section, we have seen that the curvature outside a zero-thickness string vanishes everywhere. Considering only this fact, we could wrongfully conclude that the string is gravitationally sterile. Nevertheless, we shall see that the conical structure is phenomenologically quite rich. \par

We start with analyzing the geodesics in conical space. The line element \eqref{wire_extmetric2} can be written in a Minkowski-form if we make $\theta = (1 - 4\mu)\varphi$
\begin{equation}
ds^2 = dt^2 - dr^2 - d\theta^2 - dz^2, \quad 0 \leq \theta \leq 2\pi(1 - 4\mu),
\end{equation}
which is more convenient since $\theta$ is what we measure outside the string. Besides that, we know the geodesics in Minkowski coordinates are straight lines. A displacement in the radial direction is given by
\begin{figure}[H]
\caption{Coordinates of the geodesic in conical flat space.}
\includegraphics[width=0.4\textwidth]{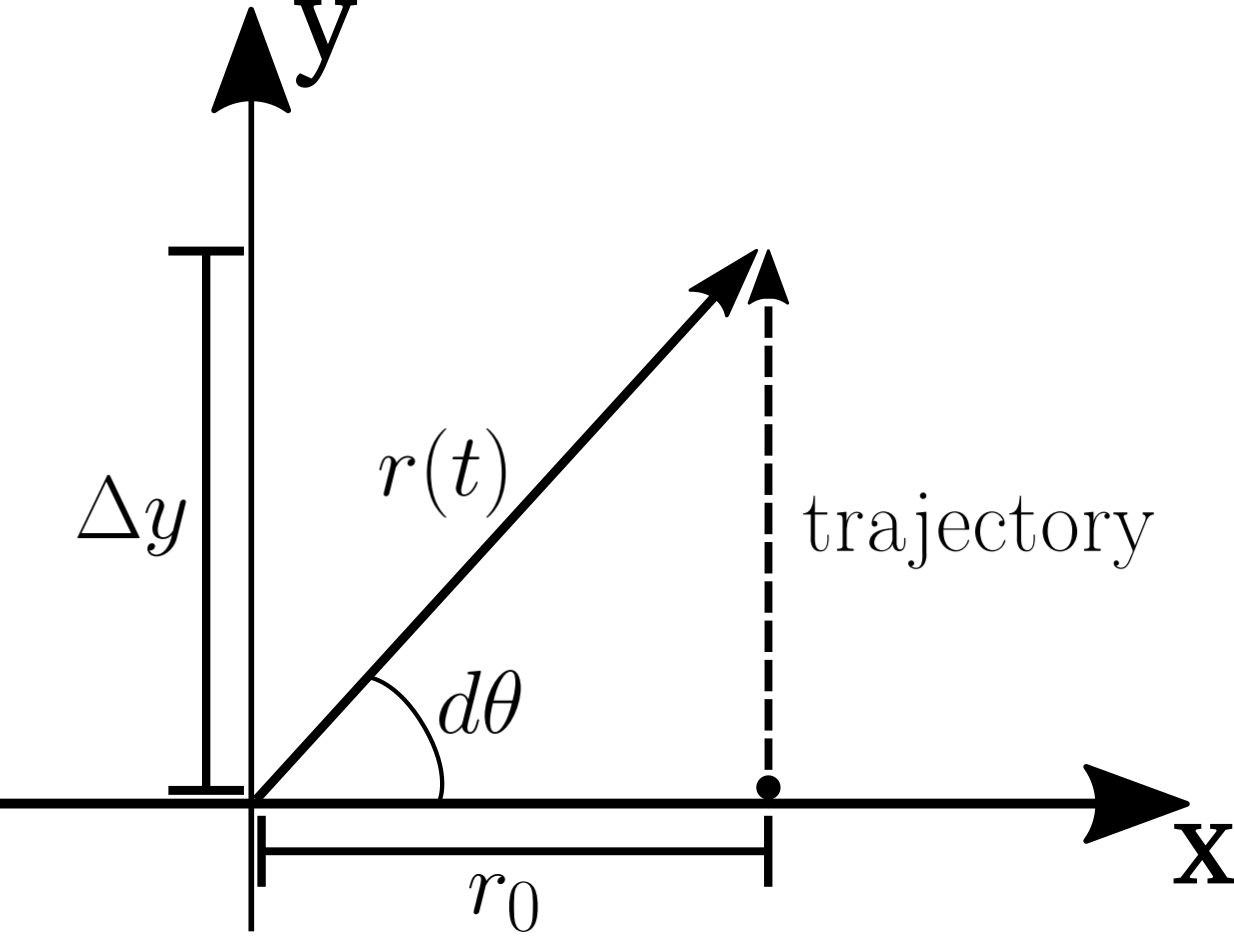}
\caption*{Source: The author (2021).}
\end{figure}
\begin{equation}
r^2(t) = r^2(0) + (dr)^2 = r^2(0) + v^2 t^2 = r^2(0) + \left(\frac{d\theta}{dt} r(0) t \right)^2.
\end{equation}
Now the displacement in the angular direction is
\begin{equation}
\begin{gathered}
\varphi = \varphi_0 + d\varphi = \varphi_0 + \frac{1}{1 - 4\mu} d\theta ,\\
d\theta = \tan^{-1} \left(\frac{\Delta y}{\Delta x} \right) = \tan^{-1}\left(\frac{d\theta}{dt}t\right) = \tan^{-1} \left[(1 - 4\mu)\omega t \right], 
\end{gathered}
\end{equation}
where $\omega = \frac{d\varphi}{dt}$. The final equations of motion are given by
\begin{gather}
\begin{aligned}
r(t) &= r(0) \left[\left(1 - 4\mu \right)^2 \omega^2 t^2 \right]^{1/2},\\
\varphi(t) &= \varphi(0) + \frac{1}{1 - 4\mu} \tan^{-1} \left[ (1 - 4\mu) \omega t \right].
\end{aligned}
\end{gather}

\begin{figure}[H]
\caption{Light rays are bent when passing near a cosmic string.}
\includegraphics[width=0.6\textwidth]{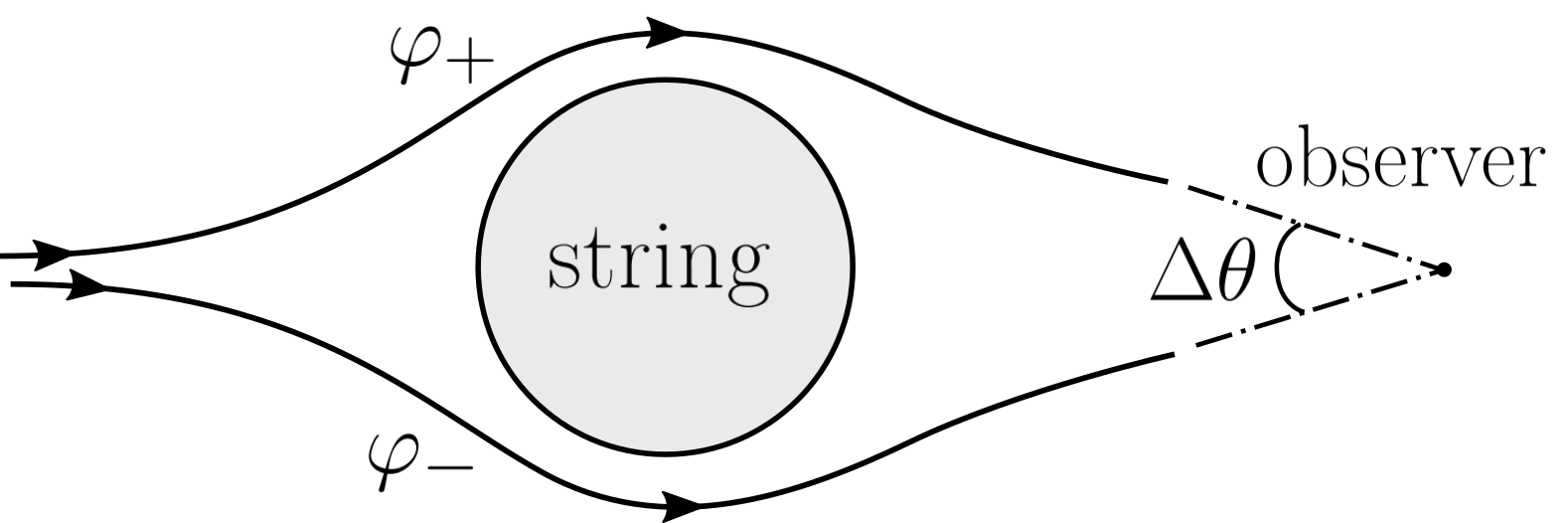}
\caption*{Source: The author (2021).}
\end{figure}

Now, suppose two light rays coming from the same star encounter a cosmic string on their paths, each ray traveling by one side of the string. It is important to find the difference, if there is any, in the angle between these two light rays when they reach a distant observer, like us here on earth. First, we adapt the formula for $d\theta$ shown before because we want it to reflect the fact that the light rays are coming from \emph{opposite} sides of the string, so 
\begin{equation}
d\theta = \frac{\pi}{2} + \tan^{-1} \left[ (1 - 4\mu) \omega t \right].
\end{equation}
We can express the $\varphi$-coordinate of both light rays as
\begin{gather}
\begin{aligned}
\varphi_+ (t) &= \varphi(0) + \frac{\pi/2}{1 - 4\mu} + \frac{1}{1 - 4\mu}\tan^{-1} \left[ (1 - 4\mu) \omega t \right], \\
\varphi_- (t) &= \varphi(0) - \frac{\pi/2}{1 - 4\mu} - \frac{1}{1 - 4\mu}\tan^{-1} \left[ (1 - 4\mu) \omega t \right]. \\
\end{aligned}
\end{gather}
When $t \to \infty$, they become
\begin{gather}
\begin{aligned}
\varphi_+ (t \to \infty) &= \varphi(0) + \frac{\pi}{1 - 4\mu},\\
\varphi_- (t \to \infty) &= \varphi(0) - \frac{\pi}{1 - 4\mu}. \\
\end{aligned}
\end{gather}
The angular difference between the two light rays are
\begin{equation}
\Delta\varphi_g = \frac{2\pi}{1 - 4\mu} - 2\pi = \frac{8\pi\mu}{1 - 4\mu} \Rightarrow \Delta\theta = 8\pi\mu.
\label{angulardif_idealstring}
\end{equation}
This is a rather interesting result. The measured angular difference between the two light rays, $\Delta\theta$, is precisely the deficit angle of the spacetime! Moreover, this is a purely geometrical effect that arises from the global conical structure, not from local spacetime curvature. Notice that we have not used any assumption of the particle's mass, so this result is also valid for massive particles. A careful study of the possible detection of cosmic strings via gravitational lensing can be found in \cite{gott1985gravitational}.

Now imagine we are asked to find the Newtonian gravitational force on a massive point particle near a cosmic string. At first, it might seem there should be no such force since the space is flat, though the non-vanishing mass of the string could correctly anticipate that a tidal force does exist. Suppose the particle has mass $m$ and is at rest at $(r, \theta, z) = (a, \theta_0, 0)$, where $\theta_0$ is an arbitrary angle. The equation for the gravitational field $\Phi$ is
\begin{equation}
\nabla^2 \Phi \propto \frac{m}{a} \delta(r - a)\delta(\theta - \theta_0) \delta(z),
\end{equation} 
which is to be solved under the boundary conditions
\begin{gather}
\begin{aligned}
\Phi(r, 0, z) &= \Phi\left(r, 2\pi(1 - 4\mu), z \right),\\
 \frac{\partial}{\partial \theta} \Phi(r, \theta, z) \Bigg{|}_{\theta = 0} &= \frac{\partial}{\partial \theta} \Phi\left(r, \theta, z \right)\Bigg{|}_{\theta = 2\pi(1 - 4\mu)}.
\end{aligned}
\end{gather}
These are \emph{not} the boundary conditions we face in flat Minkowski spacetime; hence the solution should be different, i.e., the deficit angle changes the solution of Poisson's equation by altering the boundary of the problem. In 1990 Dmitry Gal'tsov \cite{gal1990cosmic} solved this problem and found that the particle experiences an attractive self-force, $\vec{F}$, proportional to $m^2$ 
\begin{equation}
\vec{F} \propto - \frac{m^2\mu}{ a^2} f(\mu) \, \hat{r},
\end{equation}
where $f(\mu)$ is a monotonically increasing function of $\mu$
\begin{equation}
f(\mu) = \frac{1}{4\pi \mu}\int_{0}^{\infty}{\left\{ \frac{\sinh(\eta/B)}{B[\cosh(\eta/B) - 1]} - \frac{\sinh\eta}{\cosh\eta - 1}\right\} \frac{d\eta}{\sinh(\eta/2)}},
\end{equation}
with $B = (1 - 4\mu)$.

\begin{figure}[H]
\caption{The scaling factor for the force exerted on a particle by a cosmic string of mass per unit length $\mu$.}
\includegraphics[width=0.65\textwidth]{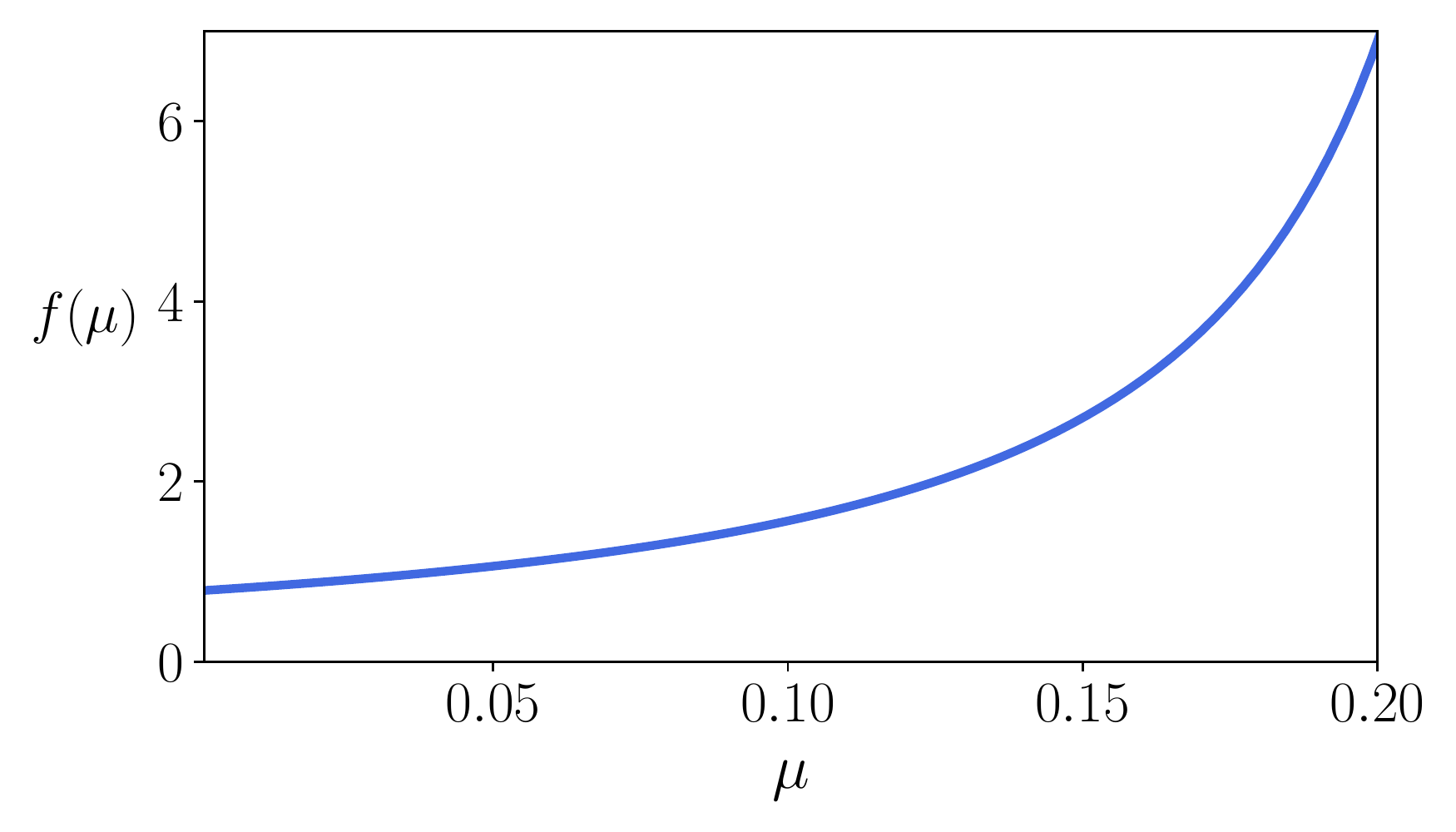}
\caption*{Source: The author (2021).}
\label{f(mu)}
\end{figure}

Four years before Gal'tsov, in 1986, Bernard Linet \cite{linet1986force} solved the problem of a charged particle in conical space, which is mathematically equivalent to the gravitational one, but with a sign change in Poisson equation, and found a repulsive interaction proportional to the charge squared. An intuitive explanation for arising of an electrostatic (gravitational) force is that the field lines "refract" on the string, causing the illusion that there is another particle on the opposite side. Also, because the charged (massive) particle accelerates, it radiates electromagnetic (gravitational) waves. Both of them can, in principle, be measured.\par

The appearance of both gravitational and electrostatic forces sheds light on the common cause of both phenomena: the non-trivial boundary conditions both fields have to satisfy are caused by the non-vanishing deficit angle. This is commonly mentioned as a topological feature since it comes from a \emph{global} property of space. \par

\section{Extended gravitating vortices}
\epigraph{La nature ne fait jamais des sauts \\ \textit{Nature does not make jumps}}{Gottfried Leibniz}

The zero-thickness cosmic string is a good approximation in a cosmological context, but the presence of a delta-function in the energy can be seen with distrust, especially when one realizes that the scalar curvature becomes a distribution in the limit $r_0 \to \infty$. Here we use the expression "extended vortex" to denote vortex solutions with internal structure, i.e., we take into account all components of the energy-momentum tensor. In fact, by taking a look at the energy-density profile of the Nielsen-Olesen vortex, one could anticipate that its effect on spacetime has a finite length and is highly dependent on the vortex parameters.\par 

However, the cosmic string zoo has many more characters. As we have seen in Chapter \ref{chap1}, any model with the Goldstone/Higgs potential $V(\phi) \propto (|\phi|^2 - \eta^2)^2$ admits a vortex excitation, hence it is of no surprise that many cosmic string solutions have been found through the years. Abelian-Higgs strings, described by the Nielsen-Olesen solution, were found in 1973 \cite{nielsen1973vortex}, then Vachaspati and Achucarro, in 1991, showed the existence of semilocal strings\footnote{They arise from a model that only breaks the local, not global, $U(1)$ symmetry.}\cite{vachaspati1991semilocal}, and in 1993 Vachaspati found Eletroweak strings in the Weinberg-Salam model \cite{vachaspati1993electroweak}, to cite a few. This section aims to see how the coupling of an extended vortex with gravity affects the spacetime at the location of and far from the string. \par

Gravitating cosmic strings with non-negligible thickness can be generated by coupling the Lagrangian of the matter with gravity and then solving Euler-Lagrange and Einstein field equations simultaneously. These equations are usually too hard to be solved in a closed form, and frequently one resorts to approximations or asymptotical limits in order to get an idea of what happens with the system before solving the equations numerically. \par

Given an action $S$ defined in Minkowski spacetime, we can minimally couple it with gravity by doing the following substitutions inside the action
\begin{equation}
\eta_{\mu \nu} \to g_{\mu \nu}, \quad \partial_\mu \to \nabla_\mu, \quad d^4x \to \sqrt{-g} d^4x,
\end{equation}
where $\eta_{\mu \nu}$ is the metric tensor in Minkowski spacetime, $g_{\mu \nu}$ is the metric tensor in a curved spacetime, $g$ is the determinant of the curved metric and $\nabla_\mu$ is the covariant derivative operator. When dealing with fermions we still need to substitute the gamma matrices, $\gamma^a$ with their curved-spacetime versions, $\gamma^\mu = \gamma^a \tensor{e}{_a ^\mu}$, where $\tensor{e}{_a ^\mu}$ is the a-th tetrad vector field, defined by $\tensor{e}{^a _\mu} e_{a\nu} = \eta_{\mu \nu}$. For example, consider the action of the free scalar field $\phi$ defined in Minkowski spacetime
\begin{equation}
S = \int{d^4x \, \frac{1}{2}\left( \eta^{\mu \nu}\partial_\nu \phi \partial_\mu \phi - M^2 |\phi|^2 \right)}.
\end{equation}
Coupling this action with gravity results in
\begin{equation}
S = \int{d^4x \sqrt{-g} \, \frac{1}{2} \left( g^{\mu \nu} \nabla_\nu \phi \nabla_\mu \phi - M^2 |\phi|^2 \right)},
\end{equation}
which means the Lagrangian of the free scalar field in curved spacetime is
\begin{equation}
\mathcal{L}_{KG} = \frac{1}{2}\left(\nabla_\mu \phi \nabla^\mu \phi - M^2 |\phi|^2 \right).
\end{equation}

As we are searching for cylindrically symmetric solutions that are invariant under boosts along the z-axis, the usual metric ansatz is
\begin{equation}
ds^2 = N^2(r)dt^2 - dr^2 - L^2(r)d\varphi^2 - N^2(r) dz^2,
\label{metric_cylindrical}
\end{equation}
where $N(r)$, $L(r)$ are functions of the radial coordinate only, and shall be determined by the matter content. Additionally, we impose spacetime boundary conditions 
\begin{gather}
\begin{aligned}
&\text{The metric should be regular at the origin, i.e: }N(r \to 0) = 1 \text{ and }L(r \to 0) = r. \\
&\text{Spacetime should be asymptotically flat, } R(r \to \infty) = 0.
\end{aligned}
\label{spread_metric_bc}
\end{gather}
This is the basic prescription to find gravitating cosmic string solutions. In 1981 Vilenkin \cite{vilenkin1981gravitational} found that, as we have seen in the last section, an infinite string using the wire approximation creates conical geometry outside it. This approach, however, neglects terms that might be present in the EM tensor, other than $T_{zz} \text{ and } T_{tt}$, which accounts for internal dynamics. In 1985 Garfinkle tackled this issue by showing that, under some weak conditions, any cosmic string solution of the gravitating abelian-Higgs model has to be asymptotically conical \cite{garfinkle1985general}. Finally, \textcite{christensen1999complete} showed the existence of at least four different classes of solutions to the gravitating abelian-Higgs model, including cosmic strings. The other three are Melvin, where the functions $N(r), L(r)$ are asymptotically powers of $x$, Kasner, and inverted cone solutions; the last two being in the category of closed solutions since the metric is terminated at some finite radial coordinate, suggesting that they do not represent an isolated system. In what follows, we deal only with cosmic string solutions.\par
\begin{figure}[H]
\caption{Conical parameters generated by the Nielsen-Olesen vortex with $n = 1$. Notice that for $\alpha = 1.0$, $a$ decreases with $\gamma$, while it is increasing for $\alpha = 3.0$. The parameter $b$ seems to always decrease with $\gamma$.}
\includegraphics[width=0.45\textwidth]{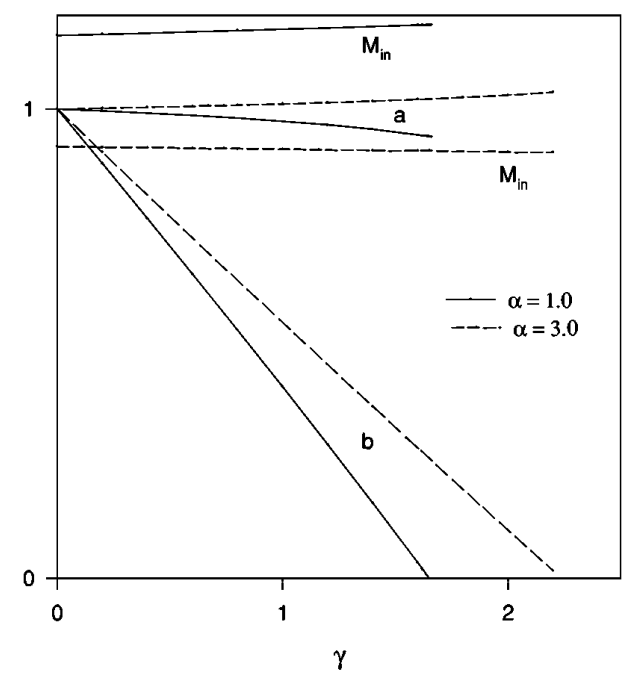}
\caption*{Source: \cite{brihaye2000classical}}
\label{fig:brihaye1}
\end{figure}

In \cite{christensen1999complete} the authors used the following standard ansatz
\begin{equation}
\phi = \eta f(r) e^{in \varphi}, \quad A^\varphi = \frac{n - H(r)}{er},
\label{notation_abelianhiggs_new}
\end{equation}
where $n$ is the winding number and $f(r)$, $H(r)$ are functions to be determined. Notice that this notation is different from the one we used in Chapter \ref{chap1}. Considering \eqref{notation_abelianhiggs_new}, the boundary conditions are
\begin{gather}
\begin{aligned}
r &\to 0: f(r) = 0, H(r) = 1, \\
r &\to \infty: f(r) = 1, H(r) = 0.
\end{aligned}
\label{spread_bc}
\end{gather}
In particular, in \cite{christensen1999complete} the authors used the parameters $\alpha, \gamma$, given by
\begin{equation}
\alpha = \frac{e^2}{\lambda} = \frac{1}{2 \beta}, \quad \gamma = 8 \pi \eta^2,
\end{equation}
to show that every solution in the cosmic string branch is related to a solution in the Melvin branch with the same position in the $\alpha-\gamma$ plane. Following the approach of \cite{christensen1999complete}, in 2000 Brihaye and Lubo \cite{brihaye2000classical} studied classical gravitating solutions of the abelian-Higgs model with $n = 1$ and explictly showed that the metric of cosmic string solutions become conical\footnote{Technically the conical strucuture only refers to the function $L(r)$, while the behavior of $N(r)$ accounts for blue/red shift of the time coordinate. Here we abuse the terminology and refer to \eqref{conical_limit} as conical limit.} far from the core, i.e.,
\begin{gather}
\begin{aligned}
N(r \to \infty) &= a\\
L(r \to \infty) &= br + c.
\end{aligned}
\label{conical_limit}
\end{gather}
In particular the authors showed the dependence of the conical parameters $a, b$ and the mass per unit length $M_{in}$, on $\alpha$ and $\gamma$. In Figure \ref{fig:brihaye1} we can see that while $M_{in}$ does not change much, the parameter $b$ decrease linearly with $\gamma$. This means the angular deficit is proportional to the scale of symmetry breaking of the vortex solution. There is however a critical value of $\gamma$, $\gamma_{cr}(\alpha)$, for which $b = 0$ and the solution becomes non-physical, since the deficit angle becomes $2 \pi$. This means there is a region in the $\alpha-\gamma$ plane where the metric solution stops making sense. Moreover, in Figure \ref{fig:brihaye1} we can see that when $\alpha$ changes from 1.0 to 3.0, $a$ changes from decreasing to increasing, which suggests that for some $\alpha = \alpha_0$, $ 1 < \alpha_0 < 3$, we have $a(\alpha_0) = 1$. In fact, another finding of \textcite{brihaye2000classical} is that when $\alpha = 2$ it is possible to find analytical solutions to $a, b, \text{ and } \gamma_{cr}$

\begin{equation}
a = 1, \quad b = 1 - \frac{\gamma}{2}, \quad \gamma_{cr}(2) = 2.
\end{equation}

For other interesting works in gravitating abelian-Higgs cosmic strings, we refer to \cite{laguna1989spacetime, dyer1995complete, linet1987vortex}. Although \textcite{garfinkle1985general} anticipated the asymptotical conical structure for the gravitating abelian-Higgs model, we shall see that the conical limit \eqref{conical_limit} is also present in gravitating non-abelian strings \cite{slagter1998self, de2015gravitating}. \par
In \cite{de2015gravitating} the authors studied the following gravity-coupled non-abelian-Higgs model
\begin{gather}
\begin{aligned}
S &= \int{d^4 x \sqrt{-g} \left( \frac{1}{16 \pi G}R + \mathcal{L}_{m} \right)}, \\
\mathcal{L}_m &= \frac{1}{2}(D_\mu \phi^a)^2 + \frac{1}{2}(D_\mu \chi^a)^2 - \frac{1}{4} F^{a}_{\mu \nu}F^{\mu \nu}_{a} - V(\phi^a, \chi^a), \quad a = 1, 2, 3
\end{aligned}
\label{padua_Lagrangian}
\end{gather}
which describes two bosonic fields, $\phi$ and $\chi$, interacting via a potential $V(\phi^a, \chi^a)$ given by
\begin{equation}
V(\phi^a, \chi^a) = 
\frac{\lambda_1}{4} \left[ (\phi^a)^2 - \eta^{2}_{1} \right]^2 + 
\frac{\lambda_2}{4} \left[ (\chi^a)^2 - \eta^{2}_{2} \right]^2 +
\frac{\lambda_3}{2} \left[ (\phi^a)^2 - \eta^{2}_{1} \right]  \left[ (\chi^a)^2 - \eta^{2}_{1} \right],
\end{equation}
in the presence of the $SU(2)$ gauge field $A^{a}_{\mu}$ that generates the field strength
\begin{equation}
F^{a}_{\mu \nu} = \partial_\mu A^{a}_{\nu} - \partial_\nu A^{a}_{\mu} + 
e \epsilon^{abc}A^{b}_{\mu}A^{c}_{\nu}.
\end{equation}
In the above expression, $e$ is the gauge coupling, and $\epsilon^{abc}$ is the antissymetric Levi-Civita symbol. The gauge-covariant derivatives have the form
\begin{gather}
\begin{aligned}
D_\mu \phi^a &= \partial_\mu \phi^a + e \epsilon^{abc}A^{b}_\mu \phi^c, \\
D_\mu \chi^a &= \partial_\mu \chi^a + e \epsilon^{abc}A^{b}_\mu \chi^c ,
\end{aligned}
\end{gather}
and just for the sake of clarity $(D_\mu \phi^a)^2 = D_\mu \phi^a (D^\mu \phi^a)^* $ and $(\phi^a)^2 = \phi^a \phi^*_a$.
Notice that each bosonic field has three complex components, and there is one vector field for each component of the scalar field. Although the situation seems much more complicated than the abelian-Higgs scenario, the prescription is the same, i.e., solve the E-L and EFEs that arise from \eqref{padua_Lagrangian} using the boundary conditions \eqref{spread_bc} and  \eqref{spread_metric_bc}. The authors considered the following ansatz
\begin{align}
\phi(r) &= f(r) 
\begin{pmatrix}
\cos{\theta} \\
\sin{\theta}  \\
0 
\end{pmatrix},
\label{padua_phi} \\
\chi(r) &= g(r) 
\begin{pmatrix}
\cos{\theta} \\
-\sin{\theta} \\
0
\end{pmatrix} ,
\label{padua_chi}\\
\vec{A^a}(r) &= \delta^{a, 3} \, \frac{1 - H(r)}{e r} \hat{\varphi} \label{padua_gauge} ,
\end{align}
which means the scalar fields are orthogonal in the field space, $\phi^a \chi_a = 0$, and only the third gauge field, $\vec{A^3}$, is non-vanishing. Notice that this model reduces to the abelian-Higgs one when $\lambda_2, \lambda_3, \text{and } \chi$ all vanish. \par
\begin{figure}[H]
\caption{Results of \textcite{de2015gravitating} for abelian strings. Left panel shows the field functions, right panel shows the metric functions. Parameters used are $\alpha = 1.0, \gamma = 0.6$.}
\includegraphics[width=0.9\textwidth]{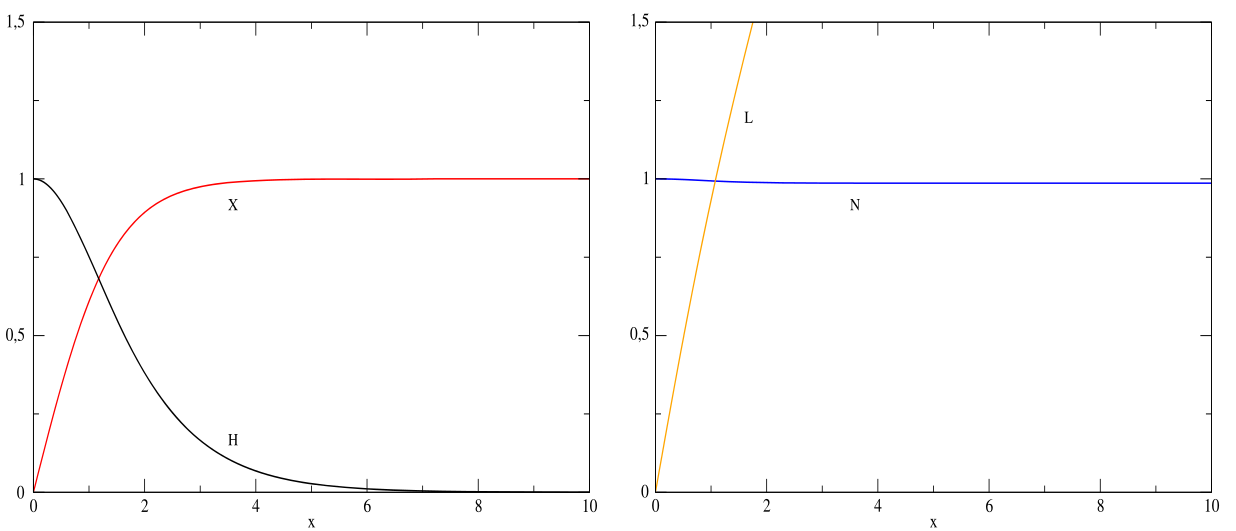}
\caption*{Source: \cite{de2015gravitating}}
\label{fig:padua_abelian}
\end{figure}
\begin{figure}[H]
\caption{Results of \textcite{de2015gravitating} for non-abelian strings. Left panel shows the field functions, right panel shows the metric functions. Parameters used are $\alpha = 1.0, \gamma = 0.6, \beta_2 = 2.0, \beta_3 = 1.0 \text{ and } q = 1.0$. }
\includegraphics[width=0.9\textwidth]{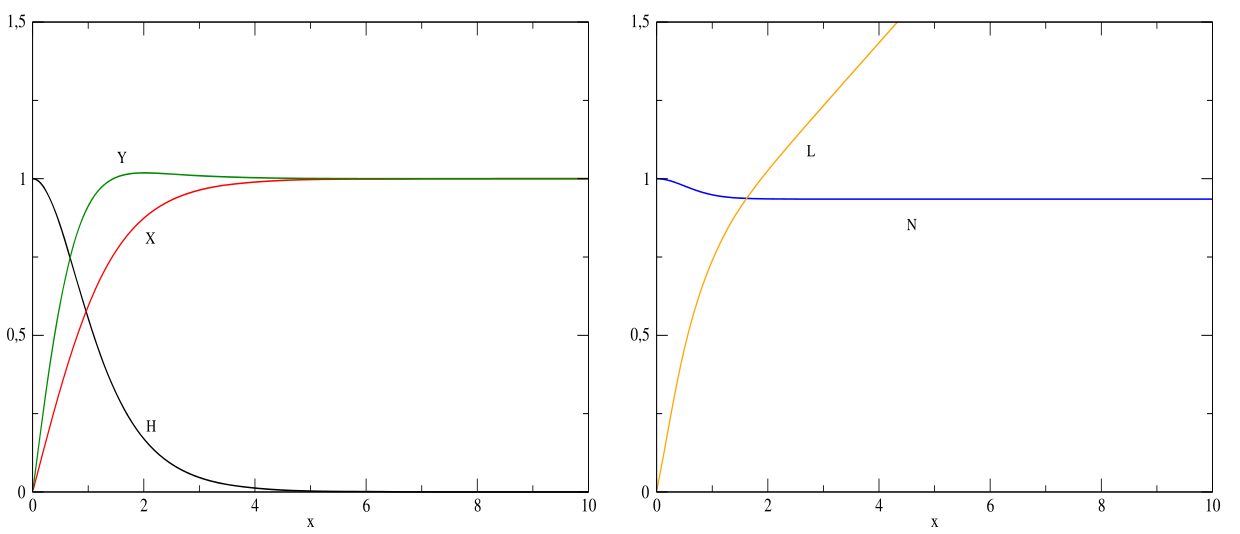}
\caption*{Source: \cite{de2015gravitating}}
\label{fig:padua_nonabelian}
\end{figure}
\begin{figure}[H]
\caption{Comparison between metric in both scenarios. The parameters used are the same as in Figures \ref{fig:padua_abelian} and \ref{fig:padua_nonabelian}}
\includegraphics[width=0.50\textwidth]{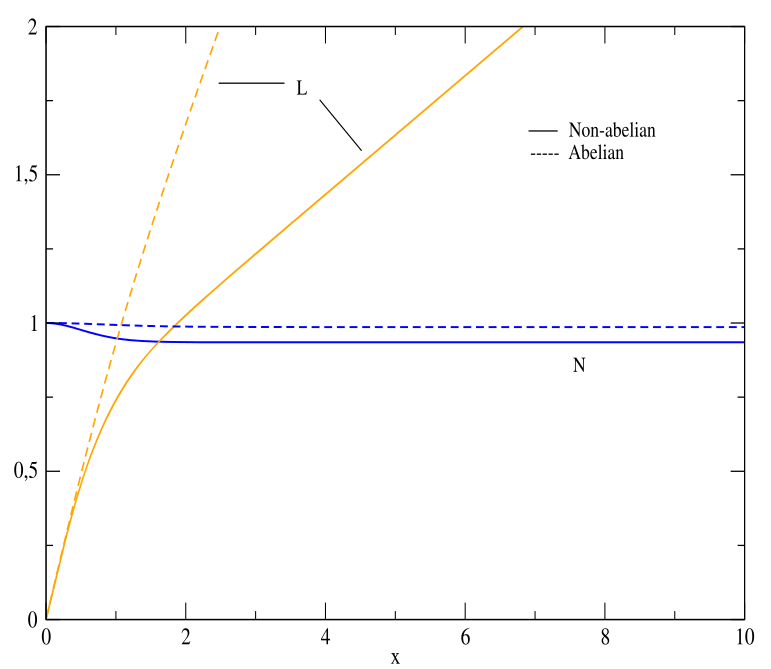}
\caption*{Source: \cite{de2015gravitating}}
\label{fig:padua_metric_comparison}
\end{figure}

For numerical analysis it is useful to define the following dimensionless functions
\begin{equation}
x = \sqrt{\lambda_1}\eta_1 r, \quad L(x) = \sqrt\lambda_1\eta_1 L(r),\quad f(r) = \eta_1 X(x), \quad g(r) = \eta_1 Y(x)
\end{equation}
which makes the Lagrangian depend only on the dimensionless parameters
\begin{equation}
\alpha = \frac{e^2}{\eta_1}, \quad q = \frac{\eta_1}{\eta_2}, \quad \beta_i = \frac{\lambda_i}{\lambda_1},\quad \gamma = 8\pi G \eta_1^2 .
\label{padua_parameters}
\end{equation}
Their results for the field and metric functions can be seen in Figures \ref{fig:padua_abelian} and \ref{fig:padua_nonabelian}. In Figure \ref{fig:padua_metric_comparison} one can compare the metric functions in the abelian and non-abelian scenarios.

\begin{figure}[H]
\caption{A possible shape for the curvature around a gravitating cosmic string.}
\includegraphics[width=0.55\textwidth]{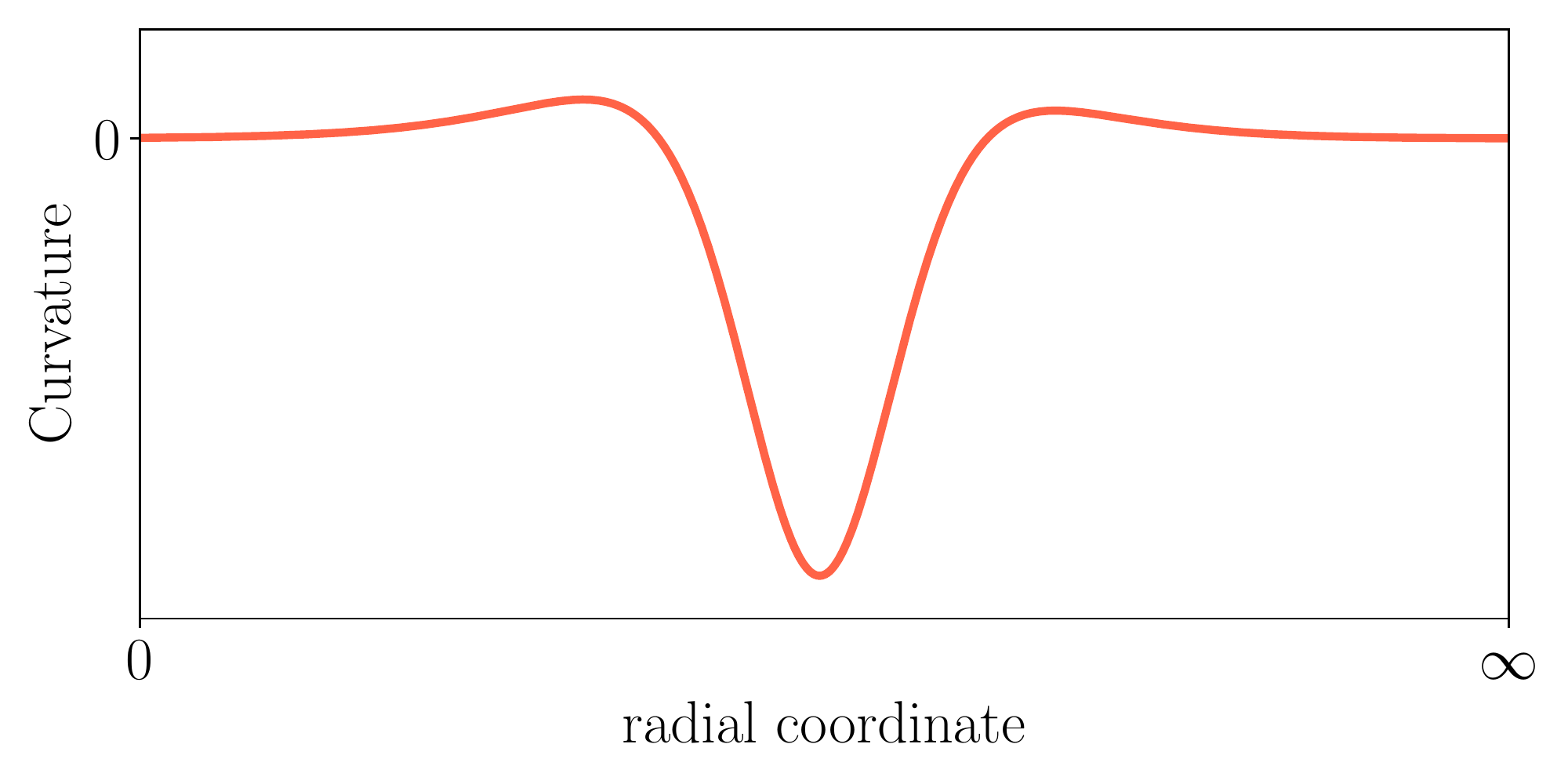}
\caption*{Source: The author (2020).}
\label{fig:spread_vortex_curvature}
\end{figure}

From Figures \ref{fig:padua_abelian}, \ref{fig:padua_nonabelian}, and \ref{fig:padua_metric_comparison}, one can conclude that non-abelian cosmic strings also generate asymptotical conical geometry, and this effect is more pronounced compared with the abelian case.
Moreover, the transition from Minkowski to conical is smooth in these spacetimes. They consist of a region with non-vanishing curvature, where the exact form is dependent on the vortex parameters (see Figure \ref{fig:spread_vortex_curvature}).

Here we end our discussion on gravitating extended vortices. It is worth mentioning that we did not try to exhaust the literature. There have been many studies in this line of research, and we did not include everything here. Instead, our goal was to give an introduction with some guiding references. Now we turn our attention to how the asymptotical conical structure of cosmic string spacetimes presents a difficulty for studying bosonic and fermionic scattering and our approach to deal with the problem.

\chapter{Scattering cross-section in gravitating cosmic string spacetimes}
\label{chap3}

\section{Bosonic cross-section}
\epigraph{\textit{O que é que trazemos de um dia como este? Nada além de uma quantidade de esboços. Ainda assim, trazemos outra coisa mais: um tranquilo desejo de trabalhar.}}{Vincent van Gogh}
In Chapter \ref{chap2} we have seen the general structure of the spacetime around a cosmic string. It consists of three distinct regions: at the center of the vortex, spacetime is flat, with Minkowskian coordinates, then at a finite distance from the center, it becomes curved and, far from the core, it approaches the conical limit.\par
In 1988 Deser and Jackiw \cite{deser1988classical} studied the quantum scattering in a pure conical background and found that the usual partial wave approach produces singularities in the scattering amplitude; they fixed this problem by changing the usual asymptotical ansatz of QM. In \cite{Silva_2021} we showed that although the spacetime of a cosmic string is much more complex than pure conical, the usual partial-wave approach also generates a divergent scattering amplitude. To avoid the singularity, we proposed a modification of the asymptotic ansatz in the partial-wave formalism and presented the corrections in the phase-shift, scattering amplitude, and cross-section. The first part of this chapter is devoted to these results.

After that, we apply the same formalism to the scattering of a massive fermionic field in the same background.

\subsection{Scalar field scattering}
The lagrangian of the free massive scalar field $\Phi$ in a curved spacetime is
\begin{equation}
\mathcal{L}_{KG} = \frac{1}{2}\left(\nabla_\mu \Phi \nabla^\mu \Phi - M^2 |\Phi|^2 \right),
\end{equation}
where $M$ is the mass of the field. Using the relations
\begin{equation}
\begin{aligned}
\frac{\partial \mathcal{L}}{\partial (\nabla^\mu \Phi^*)} &= \nabla_\mu \Phi \to  \nabla^\mu \left( \frac{\partial \mathcal{L}}{\partial (\nabla^\mu \Phi^*)} \right) = \nabla^\mu \nabla_\mu \Phi, \\
\frac{\partial \mathcal{L}}{\partial \Phi^*} &= - M^2 \Phi, 
\end{aligned}
\end{equation}
we get the following E-L equation
\begin{equation}
\nabla_\mu \nabla^\mu \Phi + M^2 \Phi = 0
\label{KGeq_curved}
\end{equation}
which is the Klein-Gordon equation minimally coupled with gravity. We can go further and add non-minimal terms inside \eqref{KGeq_curved}, for example
\begin{equation}
    (\Box + M^2 + \xi R)\Phi = 0,
    \label{scalar_kleingordon}
\end{equation}
where $\Box = \nabla_\mu \nabla^\mu$ is the D'Alembertian operator, $R$ is the Ricci scalar and $\xi$ is the non-minimal coupling. Taking $\Box \Phi =\frac{1}{\sqrt{-g}}\partial_\mu{(\sqrt{-g}g^{\mu\nu}\partial_\nu{\Phi})}$ \cite{carroll2019spacetime}, we obtain
\begin{equation}
    \Bigg{\{}\frac{1}{N^2L}\Bigg{[}L\partial_{t}^2-\Big{(}2NN'L+N^2L'\Big{)}\partial_{r}-N^2L\partial_{r}^2-\frac{N^2}{L}\partial_{\varphi}^2-L\partial_{z}^2\Bigg{]} + M^2 +\xi R\Bigg{\}}\Phi=0,
    \label{scalar_eom1}
\end{equation}
where prime denotes derivative with respect to $r$. Imposing conservation of energy, cylindrical symmetry and restricting the dynamics to the $r - \varphi$ plane (invariance under boosts along the z-axis) we can take the following ansatz
\begin{equation}
    \Phi = e^{\mp iEt} e^{ikz} \sum_{m=-\infty}^{\infty} a_m R_{m}(r) e^{im\varphi},
    \label{scalar_solution_ansatz}
\end{equation}
where $E$ is the energy of the field, $k$ the momentum in z-direction, $m = \pm 1, \pm 2, \pm 3,...$, is the mode of the field, and $a_m$ is a mode-dependent constant to be defined by the initial condition. Notice $m$ \emph{has to} be integer for the solution to be single-valued in $\varphi$. The summation over $m$ reflects the fact that for a linear equation, the sum of solutions is still a solution. Combining \eqref{scalar_solution_ansatz} with \eqref{scalar_eom1} and using $\lambda^2 \equiv E^2 - M^2 - k^2$, result in
\begin{equation}
    R_{m}''(r) + \left(\frac{2N'(r)}{N(r)}+\frac{L'(r)}{L(r)}\right)R_{m}'(r)+\left[\frac{\lambda^2}{N^2(r)}- M^2\left(1-\frac{1}{N^2(r)}\right)-\frac{m^2}{L^2(r)}+ \xi R\right]R_{m}(r)=0.
    \label{scalar_eom2}
\end{equation}
Now remember that near the origin we have $N(r) = 1$ and $L(r) = r$, hence $R_m(r \to 0) \propto J_m(\lambda r)$ where $J_m$ is the Bessel function of first kind of order $m$, and we absorb the proportionality constant in $a_m$. Here we set $a_m = i^m$ such that in the limit $r \to 0$ the field is a plane-wave in the $\hat{x}$-direction. The general solution in the limit $r \to \infty$ is
\begin{equation}
R_m(r \to \infty) = b_m J_{m^\prime}(\lambda^\prime w) + c_mY_{m^\prime}(\lambda^\prime w),
\end{equation}
where $m^\prime = m/b$, ${\lambda^\prime}^2 = \lambda^2/a^2 - M^2(1 - 1/a^2)$, $w = r + c/b$, $Y_{m'}$ is the Neumann function of order $m'$, and $b_m, c_m$ are mode-dependent constants. Notice that the momentum $\lambda$ changes to $\lambda'$ to account for the new parametrizations of time and distance in the z-direction, which is evident from $\lambda'^2 = (E/a)^2 - (k/a)^2 - M^2$. Asymptotically, Bessel and Neumann functions approach
\begin{gather}
\begin{aligned}
J_\nu(x) &\xrightarrow{x \to \infty} \sqrt{\frac{2}{\pi x}} \cos\left[x - \frac{\pi}{2} \left(\nu + \frac{1}{2}\right) \right], \\
Y_\nu(x) &\xrightarrow{x \to \infty} \sqrt{\frac{2}{\pi x}} \sin\left[x - \frac{\pi}{2} \left(\nu + \frac{1}{2}\right) \right],
\end{aligned}
\end{gather}
which together with the relations $b_m = C_m \cos d_m$ and $c_m = -C_m \sin d_m$ leads to
\begin{equation}
    R_m(r \rightarrow \infty) = C_m \sqrt{\frac{2}{\pi \lambda^{\prime} r}} \cos \left(\lambda^{\prime} r + \beta_{m^\prime}\right),
    \label{scalar_asymp_solution}
\end{equation}
In the above expression $\beta_{m^\prime} = \frac{\lambda^\prime c}{b} - \alpha_{m^\prime} + d_m(\lambda)$, $\alpha_m = \frac{\pi}{2}(m + 1/2)$ while $C_m(\lambda)$ and $d_m(\lambda)$ are model-dependent constants to be determined, usually numerically. Also notice that the phase shift of the m-th mode is given by $\delta_m(\lambda) = \beta_{m^\prime} + \alpha_m = \frac{\lambda^\prime c}{b} + \frac{m\pi}{2}\left(1 - \frac{1}{b} \right) + d_m(\lambda)$. It is interesting to take a look at the effect of conical structure in other observables of the scalar field. For instance, a temporal element $dt$ at the center of the vortex becomes $a \, dt$ far from the core, hence the frequency of oscillations appears higher for a local observer outside the string than it is measured by someone at the core. Although the energy of the field does not change, the hamiltonian operator, which is the local measuring scale, does change. Explicitly we have

\begin{equation}
\hat{H}_{in} = i \frac{\partial}{\partial t}, \quad  \hat{H}_{out} = i \frac{\partial}{\partial(at)} = \frac{1}{a} \hat{H}_{in}
\label{hamiltonian_operator}
\end{equation}
where $\hat{H}_{in}$ and $\hat{H}_{out}$ are the hamiltonian near and far from the core, respectively. Equation \eqref{hamiltonian_operator} implies $\hat{H}_{out} \Phi = E/a \Phi$, which makes sense if we compare the expression of $\lambda$ and $\lambda'$. Because $g_{tt} = -g_{zz}$ the linear momentum in the z direction is perceived as larger far from the core. Finally, angular momentum is also affected by the conical structure. The angular momentum operator is given by
\begin{equation}
(\hat{L}_z)_{in} = -i \frac{\partial}{\partial \varphi}, \quad (\hat{L}_z)_{out} = -i \frac{\partial}{\partial(b \varphi)} = \frac{1}{b} (\hat{L}_z)_{in}.
\label{angmom_operator}
\end{equation}
where $\hat{L}_{in}$ and $\hat{L}_{out}$ are the angular momentum operators near and far from the core, respectively. Equation \eqref{angmom_operator} suggests the angular momentum outside the vortex is measured to be a non-integer value $m' = m/b$. Thus, the angular deficit of spacetime affects the measured angular momentum of the field.\par

In addition, note that if the flat spacetime for $r \rightarrow \infty$ is not conical, i.e. $c=0$ and $b=1$, the phase shift becomes equal to $d_m(\lambda)$ determined by the gravitational potential in the region $0<r<\infty$. Moreover, one might realize that any local interaction does not change the general form of the solution at infinity. The effect of any such interaction shall be felt only by the parameters $C_m$ and $d_m$ to be measured at infinity. For example, suppose the scalar field also interacts with the gauge field creating the vortex. In this situation, the constants $C_m$ and $d_m$ are certainly modified by the local interaction when compared with the situation without the gauge field. This illustrates the fact that the constants $C_m$ and $d_m$ store information about \emph{any} local interaction in the region $0 < r < \infty$. Although cosmic string models are usually too hard to solve in closed form, we later devise a toy model and show the effects of the gauge-field coupling in the total cross-section of a scalar test field. \par

Now let us get back to the scattering. In usual partial wave approach we rewrite the cosine in \eqref{scalar_asymp_solution} in terms of plane waves, resulting in

\begin{equation}
    \Phi_{solution} = \frac{1}{\sqrt{2\pi\lambda^\prime r}} \left[e^{-i \lambda^\prime r}\left(\sum_m C_m i^m e^{im\varphi} e^{-i\beta_{m^\prime}} \right) + e^{i\lambda^\prime r} \left(\sum_m C_m i^m e^{im\varphi}e^{i\beta_{m^\prime}} \right) \right].
    \label{solution_expanded}
\end{equation}
The usual QM ansatz is given by
\begin{equation}
\Phi_{usual \, ansatz} = f(\varphi) \frac{e^{i \lambda r}}{\sqrt{r}} + e^{i\lambda r \cos\varphi},
\end{equation}
where $f(\varphi)$ is called scattering amplitude. This ansatz says the scattered wave is a cylindrical wave, regulated by $f(\varphi)$, plus a plane-wave accounting for the part of the incident wave that does not interact with the potential. \\
Usually in QM, the spacetime before and after the potential is the same. Hence the unscattered wave has momentum $\lambda$, the same as the incoming plane wave\footnote{Considering elastic scattering.}. However, this is not possible in the spacetime of a cosmic string since a change in the metric component $N(r)$ affects $\lambda$. Also, far from the core, the radial part of the solution to the Klein-Gordon equation, $R_m(r)$, is a Bessel function of non-integer order, making it impossible to take the pure plane-wave in the second part of the ansatz. Finally, it is clear that the regions before and after the scattering are \emph{not} equivalent, which motivates a change in the canonical approach. Later we show another reason to do such modification. We take the asymptotic ansatz in the form
\begin{equation}
    \Phi_{ansatz} = f(\varphi) \frac{e^{i \lambda^{\prime}w}}{\sqrt{r}} + \sum_{m=-\infty}^\infty A_m i^m J_{m^\prime} (\lambda^\prime w)e^{im\varphi} = f(\varphi) \frac{e^{i \lambda^{\prime}r}}{\sqrt{r}} + (e^{i\lambda^\prime r \cos\varphi})_{mod}.
    \label{scalar_ansatz}
\end{equation}
The reason for writing the second term in this form is that when $A_m = 1, a = 1, b = 1$ and $c = 0$ it takes the form of a plane wave with momentum $\lambda$ travelling in the x-direction. Here, however, we left $A_m$ free to be determined by the form of the solution at infinity. Expressing the ansatz in terms of plane waves, we get
\begin{align}
    \Phi_{ansatz} &= \frac{1}{\sqrt{2 \pi \lambda^\prime r }} \left[ 
    e^{-i\lambda^\prime w} \left(\sum_{m=-\infty}^\infty  A_m i^m e^{i\alpha_{m^\prime}}e^{im\varphi}  \right) +\right.\nonumber\\ 
   & \left. +e^{i\lambda^\prime w} \left( \sqrt{2 \pi \lambda^\prime} f(\varphi) + \sum_{m=-\infty}^\infty  A_m i^m e^{-i\alpha_{m^\prime}} e^{im\varphi}\right)   \right].
    \label{ansatz_expanded}
\end{align}
Now comparing the coefficients of $e^{-i\lambda'w}$, we obtain
\begin{equation}
  A_m e^{-i (\beta_{m^\prime} - d_m)} = C_m e^{-i \beta_{m^\prime}} \rightarrow A_m = C_m e^{-i d_m(\lambda)}.
\label{coefficient_ansatz}
\end{equation}
Comparing the coefficients of $e^{i \lambda^\prime w}$ and considering Equation (\ref{coefficient_ansatz}), results in
\begin{equation}
    f(\varphi) = \frac{1}{\sqrt{2\pi i \lambda^\prime}}
    \sum_{m=-\infty}^\infty A_m \left[e^{2id_m(\lambda)} - 1 \right]e^{im(\varphi - \delta \varphi)} = \sum_{m=-\infty}^\infty f_m(\varphi),
    \label{scalar_SA}
\end{equation}
where $\delta \varphi = \frac{\pi}{2} \left(\frac{1}{b} - 1 \right) \ge 0$ knowing $b \leq 1$. The solution (\ref{scalar_SA}) has the extra factors $A_m$ and $e^{- i m \delta \varphi}$ when compared with the result in QM. However, if the spacetime at $r \to \infty$ is not conical, i.e. $N(r) \to 1$ and  $L(r) \to r$, the extra factors disappear and our result matches the standard approach.

Now let us justify better the inclusion of the free parameter $A_m$ in \eqref{scalar_ansatz}. Suppose $A_m = 1$, as it is in the standard QM approach. The scattering amplitude reads 
\begin{gather}
\begin{aligned}
f(\varphi) &= \frac{1}{\sqrt{2 \pi i \lambda'}} \left( \sum_{m} e^{2id_m} e^{im (\varphi - \delta\varphi)} - \sum_{m} e^{im(\varphi - \delta\varphi)} \right), \\
&= \frac{1}{\sqrt{2 \pi i \lambda'}} \left( \sum_{m} e^{2 i d_m}e^{i m (\varphi - \delta\varphi)}  - \delta(\varphi - \delta\varphi) \right),
\end{aligned}
\end{gather}
i.e., the scattering amplitude has a delta-contribution coming from the deficit angle of spacetime. In our approach, with $A_m$ free to be determined by the field solution at infinity, the scattering amplitude becomes
\begin{equation}
f(\varphi) = \frac{1}{\sqrt{2 \pi i \lambda'}} \left ( \sum_{m} C_m e^{idm} - \sum_{m} C_m e^{-i d_m} e^{im(\varphi - \delta\varphi)} \right ),
\end{equation}
which does not have any delta contribution due to the nontrivial mode dependent constants $C_m$ and $d_m$. In conclusion, our formalism avoids the singularity of the scattering amplitude. \par
Let us pause and connect our approach with the one in \cite{deser1988classical} for the particle scattering in a conical geometry. In their scenario, the whole space is conical and analytically determined. Because of a non-vanishing deficit angle, the scattering amplitude $f(\varphi)$ is singular, taking the standard partial wave expansion ansatz in quantum mechanics. They circumvented this problem by modifying the second term in (\ref{scalar_ansatz}) to match the solution determined at $r \rightarrow \infty$. Our reasoning is similar. Due to the conical structure at $r \rightarrow \infty$, we need to leave an extra free parameter to be fixed with the asymptotic solution after the scattering.
In contrast with the result in \cite{deser1988classical}, one needs to find the parameters $C_m$ and $d_m$ numerically since our formalism deals with a class of spacetimes typically too complex to have a metric in a closed-form due to non-trivial matter-gravity interaction. Similar to \cite{deser1988classical}, here the optical theorem is not satisfied, although there is no problem with unitarity. It can be shown that the probability current $\vec{J}$ is
\begin{equation}
\vec{J} \propto Im\left( \sum_{m, n} a_n a^{*}_{m} e^{i (n - m)\varphi} R_n \frac{dR_m}{dr} \right) \hat{e}_r - |\phi|^2\left( \frac{m}{L(r)} \hat{e}_\varphi + \frac{k}{N(r)} \hat{e}_z \right).
\label{scalar_probcurrent}
\end{equation}
Restricting the dynamics to the x-y plane, calculating the outgoing flux from a circle of radius $r_0$ and setting $a_m = i^m$, results in
\begin{equation}
\int_{\partial S}{\vec{J} d\vec{A}} = r_0 \, L(r) \int{J^r d\varphi} = 0,
\end{equation}
which by Stokes theorem
\begin{equation}
\int_{S}{\nabla \cdot \vec{J} \, dV} = \int_{\partial S}{\vec{J} d\vec{A}}.
\end{equation}
It means that the probability current on the plane is divergenceless, hence conserved. We conclude that in this class of spacetimes the optical theorem is no longer suitable to determine particle or probability conservation.

Moreover, the appearance of an angular deficit in the scattering amplitude originates from the spacetime's deficit angle equal to $\delta=2\pi (1-b)$ \cite{de2015gravitating}. In fact, we noticed that the extra factor $\delta \varphi$ is proportional to the angular difference between geodesics in the ideal cosmic string spacetime \eqref{angulardif_idealstring}, $\Delta\varphi = \frac{8\pi \mu}{b}$ with $b=1-4 \mu$,
\begin{equation}
    \delta\varphi = \frac{1}{4} \Delta\varphi.
\end{equation}

Now we turn our attention to the phase shift. The phase shift of the scalar field scattered in this class of spacetimes is given by
\begin{equation}
    \delta_m(\lambda) = \beta_{m^\prime} + \alpha_m = 
    \lambda^\prime \frac{c}{b} + \frac{m\pi}{2}\left(1 - \frac{1}{b} \right) + d_m(\lambda),
    \label{scalar_phaseshift}
\end{equation}
and notice the first term accounts for the rescaling of the radial coordinate, the second term matches the result found in \cite{deser1988classical} and comes from conical structure, and the last one is due to any interaction in the midway, e.g. curvature and/or gauge field for example. In the low-energy regime, the isotropic mode, $m = 0$, holds the largest contribution to scattering \cite{le2011quantum} since any other mode has some part of the flux in the azimutal direction, as seen in \eqref{scalar_probcurrent}. In QM the scattering length is defined as
\begin{equation}
l_{sc} = \lim_{\lambda^\prime \rightarrow 0} \left | \frac{\delta_0(\lambda)}{\lambda^\prime} \right|,
\label{scat-length}
\end{equation}
such that, for instance, the potential of a hard sphere of radius $R_0$ yields $l_{sc} = R_0$. Knowing that for $\lambda' = 0$ there is no scattering, which means $d_0(\lambda^\prime = 0) = 0$, the scattering length of a cosmic string is given by
\begin{equation}
    l_{sc} = \frac{c}{b} + \frac{d(d_0)}{d\lambda^\prime}(\lambda^\prime = 0).
\end{equation}

Now we have all the ingredients to calculate the scattering cross-section. Ignoring the z-axis, due to the dynamics being restricted to the x-y plane, and knowing the incoming and outgoing momenta, $\lambda$ and $\lambda'$, are not the same in general, the differential cross-section is given by \cite{zettili2003quantum}
\begin{equation}
    \frac{d\sigma}{d\varphi} = \frac{\lambda^\prime}{\lambda} |f(\varphi)|^2
    \label{scalar_difcross},
\end{equation}
One can see that if we include the z-direction, the total cross-section diverges. Sustituting eq. (\ref{scalar_SA}) into eq. (\ref{scalar_difcross}) and then integrating (\ref{scalar_difcross}) in $\varphi$ results in
\begin{equation}
    \sigma = \frac{4}{\lambda} \sum_{m=-\infty}^\infty |C_m|^2 \sin^2(d_m).
\label{scalar_totalcross}
\end{equation}
This result is very similar to the standard expression from QM, except for the factor $|C_m|^2$. However, the extra factor becomes $1$ in the limit where the asymptotical spacetime has the same parametrization as the origin. Clearly, the convergence of this formula depends on the convergence of the amplitudes $C_m$, which, as we shall see, does converge for both a general toy model and a realistic scenario.

\subsection{Toy model}
\label{sec6}

To illustrate our formalism in a concrete example, we develop an analytical model similar to a cosmic string spacetime, although somewhat simplified, in order to calculate the factors $C_m$ and $d_m$ analytically. We have seen in Chapter \ref{chap2} that the metric outside a hard-wall cylindrically symmetric energy density of radius $r_0$ is conical. We then use the following metric as a first approximation to the cosmic string spacetime
\begin{gather}
\begin{aligned}
r < r_0 &: \quad N(r) = 1, \quad L(r) = r \\
r > r_0 &: \quad N(r) = a, \quad L(r) = br + c ,
\end{aligned}
\label{toymod_metric}
\end{gather}
which has conical geometry outside, as expected. Continuity of $L(r)$ at $r = r_0$ results in
\begin{equation}
r_0 = \frac{c}{1 - b},
\label{toymod_size}
\end{equation}
which gives a connection between the conical parameters $b, c$ and the size $r_0$ of the vortex. In what follows we consider $a = 1$, to avoid any delta singularities in the field equation\footnote{The derivative of $N(r)$ is a delta-function. Look at \eqref{toymod_metric_heaviside} later in the text}, and $b$ to be close to 1, which means $r_0 \gg c/b = \mathcal{O}(1)$. Later we relax both conditions. \par 

\begin{figure}[H]
\caption{Metric functions \eqref{toymod_metric_smooth} using $a = 0.98, b = 0.64, c = 0.39$ and $p = 3$. Dashed lines show the Minkowski counterparts.}
\includegraphics[width=0.9\textwidth]{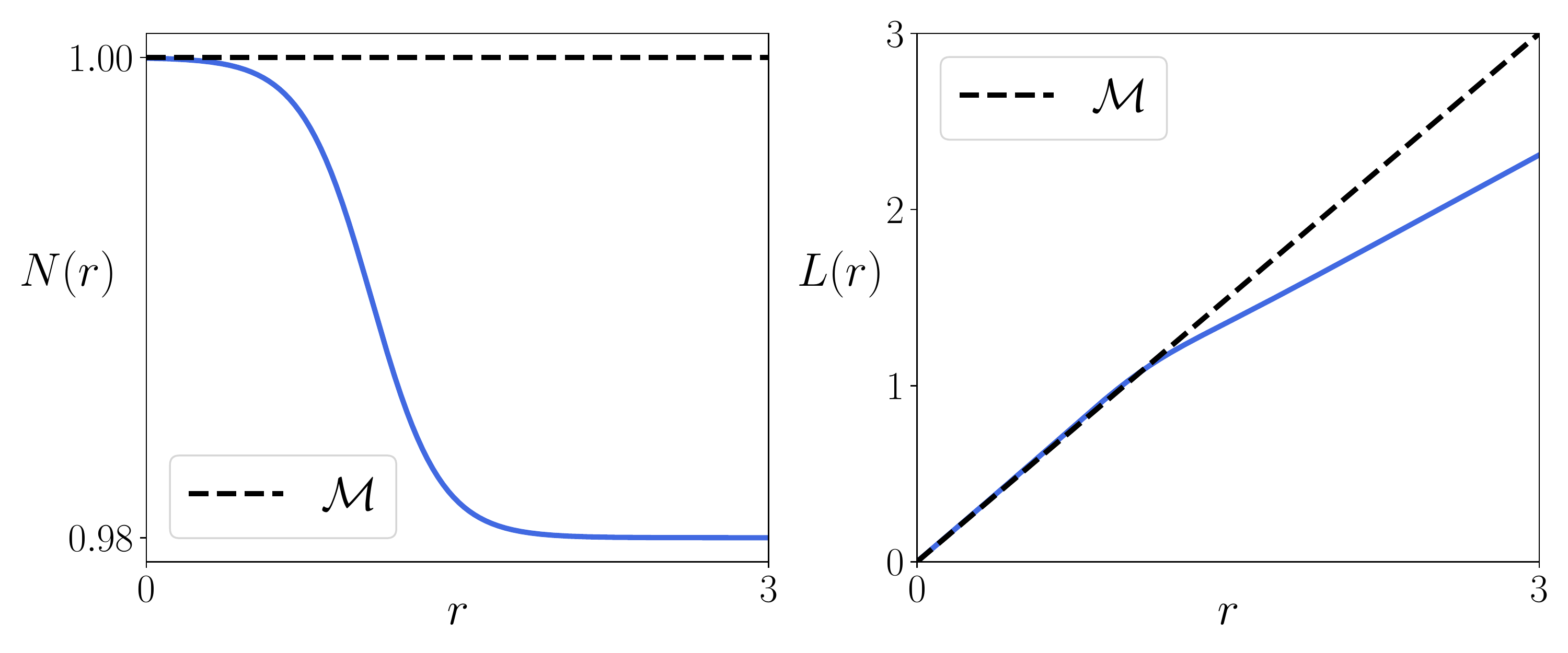}
\caption*{Source: The author (2021).}
\label{fig:toymod_metric}
\end{figure}

The scalar field solutions are
\begin{gather}
\begin{aligned}
r < r_0 &: \phi = e^{-iEt} e^{ikz} \sum_{m} i^m J_m(\lambda r) e^{im\varphi} \\
r > r_0 &: \phi = e^{-iEt} e^{ikz} \sum_{m} i^m C_m J_{m'}(\lambda w) e^{im\varphi}.
\end{aligned}
\end{gather}
Taking the asymptotic form of the solution outside ($r_0 \gg c/b$) and imposing continuity of $\phi$ at $r = r_0$ yields 
\begin{equation}
C_m \sqrt{\frac{2}{\pi \lambda r_0}} \cos(\lambda'r_0 + \beta_{m'}) = J_m(\lambda r_0).
\label{toymod_cond1}
\end{equation}
Imposing continuity of the gradient of $\phi$ at $r = r_0$ gives 
\begin{equation}
\frac{\lambda}{2} \Delta J_m(\lambda r_0) = C_m \sqrt{\frac{2}{\pi \lambda r_o}} \cos(\lambda r_0 + \beta_{m'}) \left[ \frac{1}{2r_0} + \lambda \tan(\lambda r_0 + \beta_{m'} )\right],
\label{toymod_cond2}
\end{equation}
where $\Delta J_m(x) = J_{m+1}(x) - J_{m-1}(x)$. Using \eqref{toymod_cond1} in \eqref{toymod_cond2} gives
\begin{equation}
\tan(\lambda r_0 + \beta_{m'}) = \frac{1}{2} \left( \frac{\Delta J_m(\lambda r_0)}{J_m(\lambda r_0)} - \frac{1}{\lambda r_0} \right),
\label{toymod_cond3}
\end{equation}
resulting in 
\begin{equation}
d_m(\lambda) = \alpha_{m'} - \lambda\left(r_0 + \frac{c}{b} \right) + \tan^{-1}\left[ \frac{1}{2} \left( \frac{\lambda}{\lambda} \frac{\Delta J_m(\lambda r_0)}{J_m(\lambda r_0)} - \frac{1}{\lambda r_0} \right) \right].
\label{toymod_d}
\end{equation}
Combining \eqref{toymod_cond3} and \eqref{toymod_cond1} yields the amplitude of the scattered field
\begin{equation}
C_m = \sqrt{\frac{\pi \lambda r_0}{2}} J_m(\lambda r_0) \left \{ 1 + \frac{1}{4} \left[ \frac{\Delta J_m(\lambda r_0)}{J_m(\lambda r_0)} - \frac{1}{\lambda r_0} \right]^2 \right \}^{1/2}.
\label{toymod_C}
\end{equation}
It is worth mentioning, however, that the cross-section that comes from \eqref{toymod_d} and \eqref{toymod_C} converges too slowly, which may happen because this toy model has serious limitations. The curvature scalar that comes from \eqref{toymod_metric}\footnote{The use of Levi-Civita connection is implicit.} diverges at $r = r_0$, so this is a singular spacetime. To circumvent this problem, we design a smooth version of \eqref{toymod_metric}. The metric \eqref{toymod_metric} can be expressed using the Heaviside step function $\Theta(x)$
\begin{gather}
\begin{aligned}
N(r) &= \Theta(r_0 - r) + a \Theta(r - r_0), \\
L(r) &= r \Theta(r_0 - r) + (br + c) \Theta(r - r_0).
\end{aligned}
\label{toymod_metric_heaviside}
\end{gather}
One might have noticed that \eqref{toymod_metric_heaviside} presents no transition between the interior and the conical spacetime, which is precisely the reason why the delta-curvature appears. We avoid this problem by employing a smooth transition via an analytical approximation of the step function, $H(x)$, defined by
\begin{equation}
H(x) = \frac{1}{2} \left(1 + \tanh(px) \right), \quad H(x) \xrightarrow{p \to \infty} \Theta(x).
\label{toymod_Hfunc}
\end{equation}
Now substituting $\Theta(x)$ by $H(x)$ in \eqref{toymod_metric_heaviside} gives
\begin{gather}
\begin{aligned}
N(r) &= \frac{1}{2}\left \{ (a + 1) + (a - 1)\tanh\left[p(r - r_0)\right] \right \}, \\
L(r) &= \frac{1}{2} \left \{ \left( (b + 1)r + c \right) + \left( (b - 1)r + c \right) \tanh \left[p (r - r_0) \right] \right \},
\end{aligned}
\label{toymod_metric_smooth}
\end{gather}
which imitates the spacetime around a vortex, as can be seen in Figure \ref{fig:toymod_metric}. One might realize that $r_0$ still defines a characteristic radius of the vortex since $r_0$ is the center of the transition between Minkowski and conical. In this situation the curvature is no longer divergent and size of the curvature well is controlled by the parameter $p$.

\begin{figure}[H]
\caption{Comparison between the curvature generated by \eqref{toymod_metric_smooth} with two different values of $p$. We can see that $p$ regulates the depth of the curvature well. Conical parameters $a, b$ and $c$ are the same as in Figure \ref{fig:toymod_metric}.}
\includegraphics[width=0.6\textwidth]{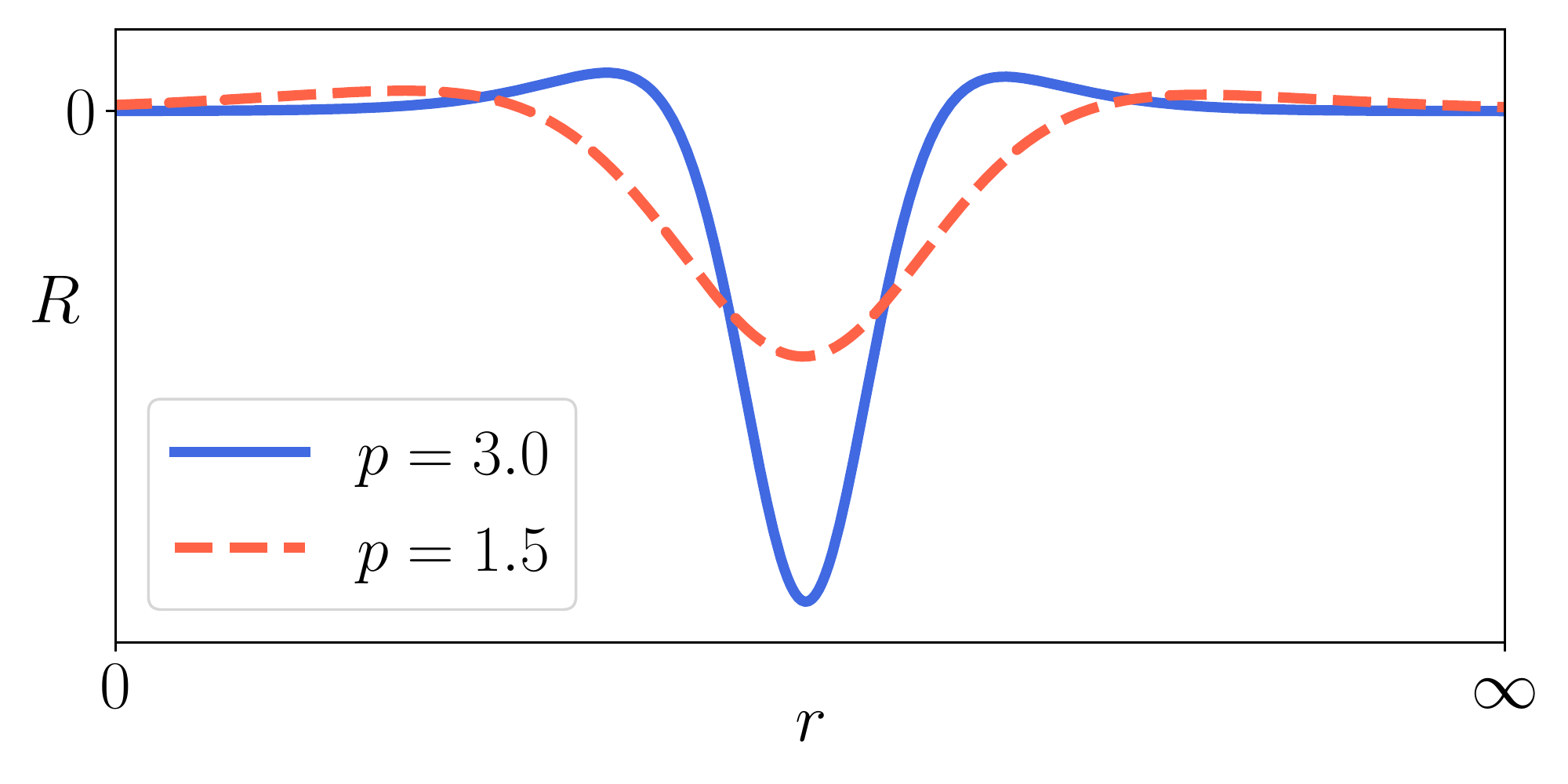}
\caption*{Source: The author (2021).}
\label{fig:toymod_curvature}
\end{figure}

\begin{figure}[H]
\caption{The solid (dashed) line shows the total scattering cross-section of scalar field with $M = 1.0$ in the absence (presence) of the gauge field.}
\includegraphics[width=1.0\textwidth]{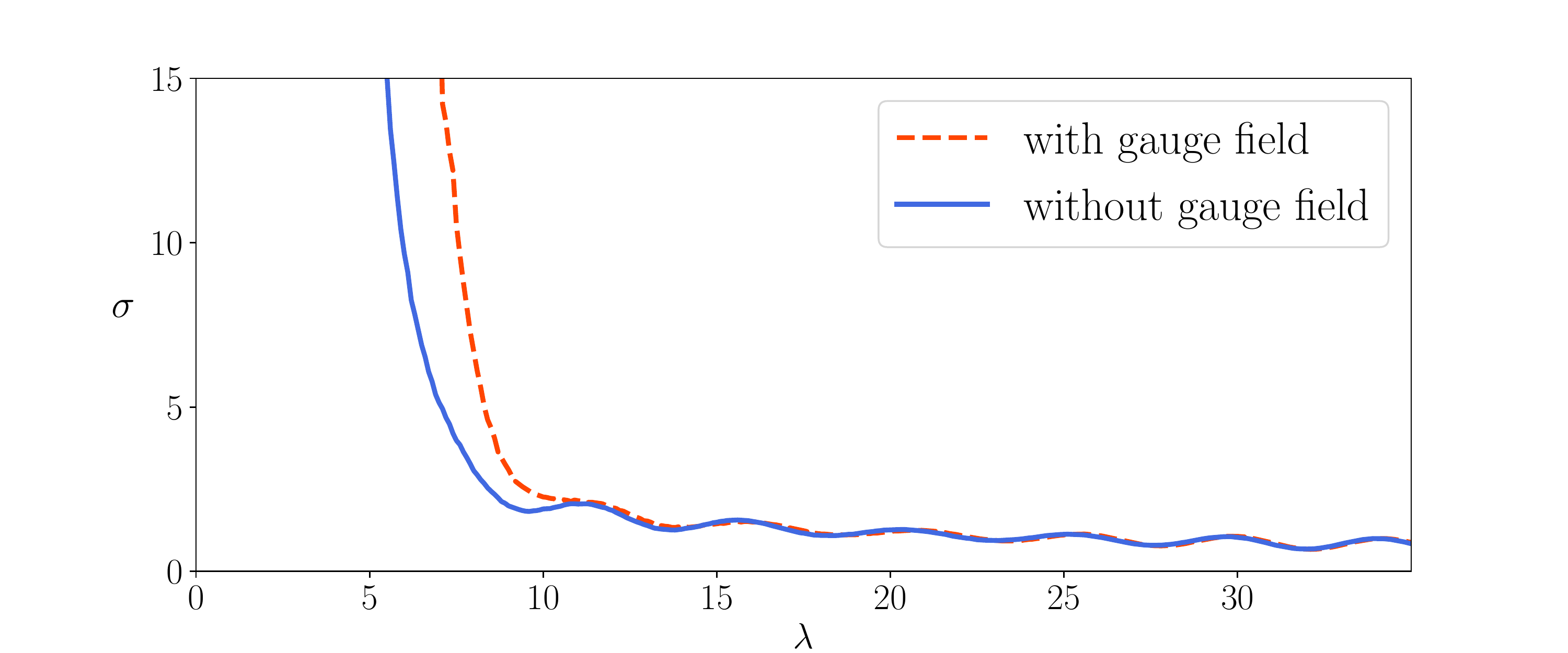}
\caption*{Source: The author (2021).}
\label{fig:toymod_sigma}
\end{figure}

With a faithful toy model in hand, all we need to do is to insert \eqref{toymod_metric_smooth} in \eqref{scalar_eom2}, extract constants $C_m$, $d_m$ from the numerical solution of the equation of motion and then use \eqref{scalar_totalcross}. Now we consider the scenario where the scalar test field also interacts with the gauge field of the vortex. To simulate this situation, we take the gauge field solution of the abelian-Higgs vortex
\begin{equation}
A^\varphi = \frac{n}{e r} \alpha(r) \hat{\varphi},
\end{equation}
where $\alpha(r)$ is coming from the Nielsen-Olesen solution found in Chapter \ref{chap1}. For the abelian-Higgs vortex with $n = 1$ and $\beta = 0.5$, \textcite{de2015gravitating} found conical parameters approximately $a = 0.98, b = 0.64, c = 0.39$. We take the same spacetime and field parameters, together with $p = 3.0$. In Figure \ref{fig:toymod_sigma}, we show the scattering cross-section of the scalar field with and without the gauge field interaction. It is clear that the gauge field has most influence in large wavelength (small momentum) particles, while it becomes irrelevant for small wavelength (large momentum) particles. As the momentum increases, all local interactions tend to become irrelevant and the cross-section approaches zero.  

\section{Fermionic cross-section}
\epigraph{\textit{Science, being necessarily performed with passion of Hope, is poetical.}}{Samuel Taylor Coleridge}

Now it is fairly straightforward to extend the formalism to analyze scattering of spin-$\frac{1}{2}$ fields. Each component of the spinorial field can be considered to be similar to a scalar field with its own associated differential cross-section. The average value of all 4 cross-sections is the total cross-section of the fermionic field \cite{bazeia2018dirac}.\par

We start with the Dirac equation in a curved spacetime
\
\begin{equation}
(\gamma^\mu \nabla_\mu + iM)\Psi = 0,
\label{Dirac-eq}
\end{equation}
where $ \{ \gamma^\mu \}$ is the set of gamma matrices in a curved spacetime, $M$ is the mass of the fermion field, and $\nabla_\mu$ is the spinorial covariant derivative. In order to convert the known gamma matrices, defined in flat spacetime, to their curved counterparts we have to make use of tetrads, which are objects that parametrize the geometry of spacetime. The tetrad field in 4-dimensional spacetime $\{ (e_a)^\mu \}$ is a set of 4 vector fields labelled by the latin index $a$, each one with, in general, four components labelled by the index $\mu$. These objects are related to the metric via
\begin{gather}
\begin{aligned}
g_{\mu \nu} = \tensor{e}{_a _\mu} \tensor{e}{^a _\nu} =  \eta_{ab} \tensor{e}{^b _\mu} \tensor{e}{^a _\nu} ,\\
\eta_{ab} = \tensor{e}{_a _\nu} \tensor{e}{_b ^\nu} = g_{\mu \nu} \tensor{e}{_a ^\mu} \tensor{e}{_b ^\nu},
\end{aligned}
\end{gather}
where the latin indices\footnote{Here latin indices always denote quantities in flat spacetime, while  greek indices represent quantities in curved spacetime.} are raised and lowered by the Minkowski metric $\eta_{ab} = \text{diag}(1, -1, -1, -1)$ and the greek indices by the general curved spacetime metric $g_{\mu \nu} = \text{diag}(N^2, -1, -L^2, -N^2)$. The sets $\{ (e_a)^{\mu} \}$ and $\{ (e^a)_{\mu} \}$ are the components of vectors $\hat{e}_a$, and 1-forms $\hat{\theta}^a$ defined by
\begin{subequations}
\begin{align}
\hat{e}_a &= \tensor{e}{_a ^\mu} \partial_\mu,
\label{verbein_basis1} \\
\theta^a &= \tensor{e}{^a _\mu} dx^\mu.
\label{vierbein_basis2}
\end{align}
\end{subequations}
Inverting \eqref{vierbein_basis2} and using it in $ds^2 = g_{\mu \nu} dx^\mu dx^\nu$ yields
\begin{equation}
ds^2 = (\theta^0)^2 - (\theta^1)^2  - (\theta^2)^2  - (\theta^3)^2,
\label{tetrad_}
\end{equation}
which provides a straightforward prescription to construct matrices $(e^a)_\mu$. With a tetrad field in hands, the curved gamma matrices are calculated using
\begin{equation}
\gamma^\mu = (e_a)^\mu \gamma^a.
\end{equation} 
The covariant derivative is defined as
\begin{gather}
\begin{aligned}
\nabla_\mu &= \partial_\mu + \Gamma_\mu, \\
\Gamma_\mu &= \frac{i}{2} (\omega_\mu)_{ab} \Sigma^{ab},
\end{aligned}
\label{covderiv_spinor}
\end{gather}
where $\Gamma_\mu$ is called the spin coefficient, $(\omega_\mu)_{ab}$ the spin connection and the set ${\Sigma^{ab}}$, here called $\Sigma$ matrices, is the set of generators of Lorentz transformations for spinors. Spin connection can be expressed in terms of the tetrad field and Christofell symbols $\{ \tensor{\Gamma}{^\lambda _\mu _\nu} \}$\cite{kleinert1989gauge, buchbinder2017effective, parker2009quantum, shapiro2016covariant} via 
\begin{gather}
\begin{aligned}
\omega_{\mu ab} = &\frac{1}{2} \left( \tensor{e}{_b ^\nu} \partial_\mu \tensor{e}{_a _\nu} - \tensor{e}{_a ^\nu} \partial_\mu \tensor{e}{_b _\nu} \right) \\
+ &\frac{1}{2} \tensor{\Gamma}{^\lambda _\mu _\nu} \left(\tensor{e}{_b _\lambda} \tensor{e}{_a ^\nu} - \tensor{e}{_a _\lambda} \tensor{e}{_b ^\nu} \right),
\end{aligned}
\label{spin_connection}
\end{gather}
while the Lorentz generators $\Sigma^{ab}$ are constructed with flat gamma matrices
\begin{equation}
\Sigma^{ab} = \frac{i}{4} [\gamma^a, \gamma^b] = \frac{i}{4} \left( \gamma^a \gamma^b - \gamma^b \gamma^a \right).
\label{lorentz_gen}
\end{equation}
Now all we have to do is to choose a convenient tetrad field, calculate the spin coefficients and find the equations of motion for the fermionic field. \par
The simplest choice of the tetrad field is given by
\begin{equation}
(e^a)_\mu = 
\begin{pmatrix}
N & 0 & 0 & 0 \\
0 & 1 & 0 & 0 \\
0 & 0 & L & 0 \\
0 & 0 & 0 & N \\
\end{pmatrix},
\hfill
(e_a)^\mu = 
\begin{pmatrix}
N^{-1} & 0 & 0 & 0 \\
0 & 1 & 0 & 0 \\
0 & 0 & L^{-1} & 0 \\
0 & 0 & 0 & N^{-1}
\end{pmatrix}.
\end{equation}
Such a choice, however, does not reflect the symmetries of spacetime, which yields a set of equations of motion that are not separable. A more suitable choice of the tetrad field is
\begin{gather}
\begin{aligned}
dt &= \frac{\theta^0}{N}, \\
dr &= \theta^1 \cos\varphi - \theta^2 \sin\varphi, \\
d\varphi &= \frac{1}{L}(\theta^1 \sin\varphi + \theta^2 \cos\varphi), \\
dz &= \frac{\theta^3}{N},
\end{aligned}
\end{gather}
which results in

\begin{equation}
\tensor{e}{_a ^\mu} = 
\begin{pmatrix}
1/N & 0 & 0 & 0 \\
0 & \cos\varphi & -\sin\varphi & 0 \\
0 & \sin\varphi/L & \cos\varphi/L & 0   \\
0 & 0 & 0 & 1/N     
\end{pmatrix}, \quad
\hfill
\tensor{e}{^a _\mu} = 
\begin{pmatrix}
N & 0 & 0 & 0 \\
0 & \cos\varphi & L\sin\varphi & 0 \\
0 & -\sin\varphi & L\cos\varphi & 0   \\
0 & 0 & 0 & 1/N     
\end{pmatrix},
\label{tetrad_matrix}
\end{equation}
and one can verify that indeed $(e^a)_\mu (e_b)^\mu = \delta^a_b$. From now on, when writing components explictly, latin indices are replaced by numbers, $a, b, c... \to 0, 1, 2, 3$, and greek indices by letters representing the coordinates, $\mu, \nu, \lambda... \to {t, r, \varphi, z}$. In addition notice that, when reading the tetrad matrices, lower index runs on the columns while upper index runs on the rows of the matrix, so, for example, $\tensor{e}{^2 _\varphi} = \cos\varphi / L$ and $\tensor{e}{_2 ^\varphi} = L \cos\varphi$.\par
The curved gamma matrices read
\begin{equation}
\begin{subequations}
\begin{aligned}
\gamma^t &= \frac{1}{N} \gamma^0, \\
\gamma^r &= \cos\varphi \gamma^1 - \sin\varphi \gamma^2 ,\\
\gamma^\varphi &= \frac{1}{L} \left( \sin\varphi \gamma^1 + \cos\varphi \gamma^2 \right) ,\\
\gamma^z &= \frac{1}{N} \gamma^3 .
\end{aligned}
\end{subequations}
\end{equation}
In what follows we use the following representation of the flat gamma matrices
\begin{equation}
\gamma^0 =
\begin{pmatrix}
\mathbf{1} & 0 \\
0 & -\bold{1} \\
\end{pmatrix},
\quad
\gamma^i = 
\begin{pmatrix}
0 & \sigma^i \\
-\sigma^i & 0 \\
\end{pmatrix},
\end{equation}
which gives the $\Sigma$ matrices
\begin{equation}
\Sigma^{0i} = \frac{i}{4}
\begin{pmatrix}
0 & \sigma^i \\
\sigma^i & 0 \\
\end{pmatrix},
\quad
\Sigma^{ij} = \frac{1}{2} \epsilon_{ijk} 
\begin{pmatrix}
\sigma^k & 0 \\
0 & \sigma^k \\
\end{pmatrix}.
\end{equation}
The non-vanishing Christofell symbols are
\begin{gather}
\begin{aligned}
\tensor{\Gamma}{^t _t _r} &= \frac{N'}{N}, \quad \tensor{\Gamma}{^r _t _t} = N N', \quad \tensor{\Gamma}{^\varphi _\varphi _r} = \frac{L'}{L}, \\
\tensor{\Gamma}{^r _\varphi _\varphi} &= -L L', \quad \tensor{\Gamma}{^z _z _r} = \frac{N'}{N}, \quad \tensor{\Gamma}{^r _z _z} = -N N' \, .
\end{aligned}
\end{gather}
To calculate the spin connection we consider each term separately. So, we split $(\omega_\mu)_{ab}$ in part A and part B
\begin{gather}
\begin{aligned}
\omega_{\mu ab} = &\underbrace{\frac{1}{2} \left( \tensor{e}{_b ^\nu} \partial_\mu \tensor{e}{_a _\nu} - \tensor{e}{_a ^\nu} \partial_\mu \tensor{e}{_b _\nu} \right)}_{(\omega^A_\mu)_{ab}} \\
+ &\underbrace{\frac{1}{2} \tensor{\Gamma}{^\lambda _\mu _\nu} \left(\tensor{e}{_b _\lambda} \tensor{e}{_a ^\nu} - \tensor{e}{_a _\lambda} \tensor{e}{_b ^\nu} \right)}_{(\omega^B_\mu)_{ab}}.
\end{aligned}
\end{gather}
The only non-vanishing components of the spin connection are
\begin{equation}
\begin{aligned}
(\omega^A_\varphi)_{12} &= -1, \\
(\omega^B_\varphi)_{12} &= -L' \\ 
(\omega^B_t)_{01} &= -N' \cos\varphi, \;(\omega^B_t)_{02} = N' \sin\varphi, \\
(\omega^B_z)_{13} &= -N' \cos\varphi, \; (\omega^B_z)_{23} = N' \sin\varphi,
\end{aligned}
\end{equation}
which when put together with the $\Sigma$ matrices inside the spin coefficients yield
\begin{equation}
\begin{aligned}
\Gamma_t =& \frac{1}{2} N' ( \cos\varphi \sigma^1 - \sin\varphi \sigma^2 )
\begin{pmatrix}
0 & \bold{1} \\
\bold{1} & 0 \\
\end{pmatrix} , \\
\Gamma_r =&  \; 0 , \\
\Gamma_\varphi =& -\frac{i}{2} (1 + L') \sigma^3
\begin{pmatrix}
\bold{1} & 0 \\
0 & \bold{1} \\
\end{pmatrix} ,\\
\Gamma_z =& \frac{i N'}{2} (\cos\sigma^2 + \sin\varphi \sigma^1)
\begin{pmatrix}
\bold{1} & 0 \\
0 & \bold{1} \\
\end{pmatrix}.
\end{aligned}
\end{equation}
The covariant derivative contracted with the gamma matrices gives
\begin{equation}
\begin{aligned}
\gamma^t \nabla_t &= \gamma^t \partial_t + \frac{1}{2} \frac{N'}{N} \gamma^r \\
\gamma^r \nabla_r &= \gamma^r \partial_r \\
\gamma^\varphi \nabla_\varphi &= \gamma^\varphi \partial_\varphi + \frac{1 + L'}{2L} \gamma^r \\
\gamma^z \nabla_z &= \gamma^z \partial_z + \frac{1}{2} \frac{N'}{N} \gamma^r.
\end{aligned}
\end{equation}

Now, in order to make calculations easier we split the spinor $\Psi$ in two parts
\begin{equation}
\Psi = 
\begin{pmatrix}
\phi \\
\chi \\
\end{pmatrix},
\end{equation}
which renders the final form of Dirac's equation \eqref{Dirac-eq}
\begin{equation}
\begin{aligned}
(\gamma^\mu \nabla_\mu + iM)\Psi &= \frac{1}{N}
\begin{pmatrix}
\partial_t \phi + \sigma^3 \partial_z \chi\\
-\partial_t \chi - \sigma^3 \partial_z \phi
\end{pmatrix} \\
&+ \cos\varphi
\begin{pmatrix}
\sigma^1 \partial_r \chi + \sigma^2 \partial_\varphi \chi / L \\
-\sigma^1 \partial_r \phi - \sigma^2 \partial_\varphi \phi / L
\end{pmatrix}
- \sin\varphi
\begin{pmatrix}
\sigma^2 \partial_r \chi - \sigma^1 \partial_\varphi \chi / L \\
-\sigma^1 \partial_r \chi + \sigma^1 \partial_\varphi \phi / L
\end{pmatrix} \\
&+ \left[ \frac{N'}{N} + \frac{(1 + L')}{2L} \right] \left[ \cos\varphi \,\sigma^1 - \sin\varphi \, \sigma^2 \right]
\begin{pmatrix}
\chi \\
-\phi
\end{pmatrix}
= \bold{0}
\end{aligned}.
\label{Dirac_eq_open}
\end{equation}
We employ the following ansatz for the spinor field $\Psi$
\begin{equation}
\Psi(t,\rho,\phi,z)=e^{-iEt}e^{ikz}\sum_{j = -\infty}^{\infty}a_j\psi_j(\rho,\varphi)e^{ij\varphi},
\end{equation}
where $j = \pm 1/2, \pm 3/2, \pm 5/2 ...$, and
\begin{equation}
\psi_j(\rho, \varphi) = 
\begin{pmatrix}
\psi^{(0)}(\rho) e^{+i \varphi/2} \\
\psi^{(1)}(\rho) e^{-i \varphi/2} \\
\psi^{(2)}(\rho) e^{+i \varphi/2} \\
\psi^{(3)}(\rho) e^{-i \varphi/2} 
\end{pmatrix}
,
\label{fermion_ansatz2}
\end{equation}
For clarity, we have dropped the index $j$ from the components of the spinor $\psi_j$. The angular dependence of \eqref{fermion_ansatz2} may seem as an imposition, but one can perform the calculations with general angular phases and the separation of variables force them to appear this way. Substituting the ansatz \eqref{fermion_ansatz2} in  \eqref{Dirac_eq_open} leads to the following equations of motion
\begin{equation}
\begin{pmatrix}
(i/N) \big{(} - E \psi^{(0)} +  k \psi^{(2)} \big{)} + i M_f \psi^{(0)} +  [ \partial_x + (j+1/2)/L + \xi(x) ]\psi^{(3)}\\
(i/N) \big{(} - E \psi^{(1)} - k \psi^{(3)} \big{)} + i M_f \psi^{(1)} + [ \partial_x - (j-1/2)/L+ \xi(x)]\psi^{(2)} \\
(i/N) \big{(} +E\psi^{(2)} -  k \psi^{(0)} \big{)} + i M_f \psi^{(2)} - [ \partial_x + (j+1/2)/L + \xi(x)]\psi^{(1)} \\
(i/N) \big{(}  +E \psi^{(3)} + k \psi^{(1)} \big{)} + i M_f \psi^{(3)} - [ \partial_x - (j - 1/2)/L + \xi(x)]\psi^{(0)} \\
\end{pmatrix}
= \boldsymbol{0}.
\label{fermion_eom1}
\end{equation}
We have denoted $\xi(x)\equiv \big{[} N'/N + (1/2L)(L'-1) \big{]}$, that vanishes in Minkowski spacetime, but not in conical one. All the above $\psi^{(i)}$ are complex and only dependent on $r$. By cylindrical symmetry, the solution does not change if we make $\varphi \to -\varphi$, and since we are summing over all modes from $-\infty$ to $+\infty$, we do not have to worry about the sign of $j$. Now in order to apply the partial wave approach, we need to know the solution of \eqref{fermion_eom1} near and far from the origin. In the limit $r \to 0$ we have \cite{mohammadi2015finite}
\begin{equation}
\Psi_j(t, r \rightarrow 0, \varphi, z) = a_j e^{-iEt} e^{ikz} e^{ij\varphi}
\begin{pmatrix}
J_{\beta_j}(\lambda x)e^{-i\varphi/2} \\
J_{\beta_j + \epsilon_j}(\lambda x)e^{i\varphi/2} \\
\frac{k -i \epsilon_j\lambda}{E + M} J_{\beta_j}(\lambda x)e^{-i\varphi/2} \\
-\frac{k -i \epsilon_j\lambda}{E + M} J_{\beta_j + \epsilon_j}(\lambda x)e^{i\varphi/2}\\
\end{pmatrix},
\label{fermion_originsolution}
\end{equation}
where $j = \pm 1/2, \pm 3/2, ...$, $\epsilon_j = sgn(j)$, $s=\pm 1$, $\beta_j = |j| - \epsilon_j/2$ and $\lambda^2 = E^2 - k^2 - M^2$ as before. Here there is a sublety. The constant $a_j$ has to be set the same way for all components, which means we cannot make \emph{all} of them plane waves. Here we choose $a_j = i^{j - \frac{1}{2}}$ such that the component $\Psi^{(0)}$ is a plane wave in the x-direction near the origin. The field solution in the limit $r \to \infty$ is given by

\begin{equation}
\psi^\pm_j(t, r \rightarrow \infty, \varphi, z) = C_j e^{\mp iEt} e^{ikz} e^{ij\varphi}\sqrt{\frac{2}{\pi \lambda' r}}
\begin{pmatrix}
\cos(\lambda^\prime w -\alpha_{\beta_{j^\prime}}+ d_{j}^{0}(\lambda)) e^{-i\varphi/2} \\

s \cos(\lambda^\prime w - \alpha_{\beta_{j^\prime} + \epsilon_{j^\prime}}+ d_{j}^{1}(\lambda)) e^{i\varphi/2} \\

\pm \frac{k' - i s \epsilon_{j^\prime }\lambda'}{E' \pm M} \cos(\lambda^\prime w - \alpha_{\beta_{j^\prime}} + d_{j}^{2}(\lambda)) e^{-i \varphi/2} \\

\mp \frac{k' - i s \epsilon_{j^\prime}\lambda'}{E' \pm M} \cos(\lambda^\prime w - \alpha_{\beta_{j^\prime} + \epsilon_{j^\prime}} + d_{j}^{3}(\lambda)) e^{i \varphi/2}\\

\end{pmatrix},
\label{fermion_infinitysolution}
\end{equation}
where $k' = k/a, E' = E/a$, $\lambda'^2 = E'^2 - k'^2 - M^2$ and $C_j(\lambda)$, $d_{j}^i (\lambda)$ are model-dependent constants. Notice that here we have four different phase-shifts, $d^i_j$, and only one global amplitude $C_j$. We express the asymptotic ansatz as
\begin{equation}
\psi_{ansatz}^i = f^i(\varphi) \frac{e^{i\lambda' w}}{\sqrt{r}} + (e^{i\lambda'r \cos\varphi})_{mod}^i
\label{fermion_ansatz}
\end{equation} 
where the index $i = 0, 1, 2,3$ labels the component of the spinor field. Here we need to define the second term of \eqref{fermion_ansatz} such that in the limit $r \to \infty$ it reduces to a term similar to the solution near the origin but with the free parameter $A^i_j$ and $j \to j', r \to w, \lambda \to \lambda'$. The ansatz for the 0-th component becomes
\begin{equation}
\Psi_{ansatz}^0 = f^0(\varphi) \frac{e^{i \lambda' w}}{\sqrt{r}} + \sum_{j = - \infty}^{\infty} A^0_j i^{j - 1/2} J_{\beta_{j'}}(\lambda' w) e^{i (j - 1/2) \varphi},
\end{equation}
and notice that
\begin{equation}
\beta_j = |j| - \frac{\text{sgn}(j)}{2} = \left \{
\begin{aligned}
j &- \frac{1}{2}, \text{ if } j > 0 , \\
-j &+ \frac{1}{2} = - \left(j - \frac{1}{2} \right), \text{ if } j < 0.
\end{aligned}
\right.
\end{equation}
Instead of summing over $j$, we can sum over $n = j - 1/2$. The ansatz becomes
\begin{equation}
\Psi_{ansatz}^0 = f^0(\varphi) \frac{e^{i \lambda' w}}{\sqrt{r}} + \sum_{n = - \infty}^{\infty} A^0_n i^n J_{n'}(\lambda' w) e^{in\varphi}
\label{fermion_comp0_ansatz}
\end{equation}
where $n' = j' - 1/2$, and $\beta_{j'} = n'$. The asymptotic form of the 0-th component of the actual solution is given by
\begin{equation}
\Psi^0_{solution} = \sum_{n} C_n i^n \sqrt{\frac{2}{\pi \lambda' r}}\cos \left( \lambda'w - \alpha_{n'} + d_n^0 \right) e^{i n \varphi},
\label{fermion_comp0_sol}
\end{equation}
which when compared with \eqref{fermion_comp0_ansatz} gives
\begin{gather}
\begin{aligned}
\quad A_n^0 &= C_n e^{-i d_n^0} , \\
f^0(\varphi) &= \frac{1}{\sqrt{2\pi \lambda' r}} \sum_{n} A_n^0 \left(e^{2i d_n^0} - 1 \right) e^{in (\varphi - \delta \varphi)} ,
\end{aligned}
\end{gather}
with $\delta \varphi = \frac{\pi}{2}\left(\frac{n'}{n} - 1 \right)$. The cross-section for the 0-th fermionic component is
\begin{equation}
\sigma^0 = \frac{4}{\lambda} \sum_{n = -\infty}^{ \infty} |C_n|^2 \sin^2(d^0_n).
\end{equation}
Calculating $\sigma^2$ is completely analogous since the coefficients of the Bessel function and the angular exponential are the same. We can adapt this result for $\sigma^2$ by just making the change $C_n \to C_n \frac{k' - i s \epsilon_n \lambda'}{E' + M}$. Therefore, the cross-section of the second component is as follows 
\begin{equation}
\sigma^2 = \frac{4}{\lambda} \frac{k'^2 + \lambda'^2}{(E' + M)^2} \sum_{n} |C_n|^2 \sin^2(d^2_n). 
\end{equation}
The computation of $\sigma^1$ and $\sigma^3$ are tedious but straightforward. The ansatz for the first component is
\begin{equation}
\Psi^1_{ansatz} = f^1(\varphi) \frac{e^{i\lambda'w}}{\sqrt{r}} + \sum_{j} A_j^1 i^{j - 1/2} J_{\beta_{j'} + \epsilon_{j'}} e^{i ( j + 1/2) \varphi}.
\end{equation}
However, notice that
\begin{equation}
\beta_j + \epsilon_j = |j| + \frac{\text{sgn}(j)}{2} = \left \{
\begin{aligned}
j &+ \frac{1}{2}, \text{ if } j > 0 \\
-j &- \frac{1}{2} = - \left(j + \frac{1}{2} \right), \text{ if } j < 0
\end{aligned}
\right.
\end{equation}
which suggests that we can sum over $m = n + 1 = j + 1/2$, instead of $j$. The asymptotic ansatz for $\Psi^1$ becomes
\begin{equation}
\Psi_{ansatz}^1 = f^1(\varphi) \frac{e^{i\lambda'w}}{\sqrt{r}} + \sum_{m} \frac{A_m^1}{i} i^m J_{m'}(\lambda' w) e^{i m \varphi},
\label{fermion_comp1_ansatz}
\end{equation}
where $m' = n' + 1 = j' + 1/2$ and $\beta_{j'} + \epsilon_{j'} = m'$. The asymptotic form of the actual solution is
\begin{equation}
\Psi^1_{solution} = \sum_{m} \frac{C_m}{i} \sqrt{\frac{2}{\pi \lambda' r}} \cos\left(\lambda' w - \alpha_{m'} + d^1_j \right) e^{im\varphi},
\end{equation}
which when compared with \eqref{fermion_comp1_ansatz} gives
\begin{equation}
\begin{aligned}
A_m^1 &= C_m e^{-i d_m^1}, \\
f^1(\varphi) &= \frac{1}{\sqrt{2 \pi \lambda' i}} \sum_{m} \frac{A_m^1}{i} \left(e^{2i d_m^1} - 1\right) e^{i m(\varphi - \delta\varphi)},
\end{aligned}
\end{equation}
with $\delta \varphi = \frac{\pi}{2} \left(\frac{m'}{m} - 1 \right)$. Adapting this result to $\sigma^3$ gives
\begin{gather}
\begin{aligned}
\sigma^1 &= \frac{4}{\lambda} \sum_{n} |C_n|^2 \sin^2(d^1_n), \\
\sigma^3 &= \frac{4}{\lambda} \frac{k'^2 + \lambda'^2}{(E' + M)^2} \sum_{n} |C_n|^2 \sin^2(d^3_n).
\end{aligned}
\end{gather}

Now it is easy to compute the total cross-section of the fermionic field. The average of all $\{\sigma^i\}$ is

\begin{equation}
\begin{aligned}
\sigma &= \frac{1}{4} \left( \sigma^0 + \sigma^1 + \sigma^2 + \sigma^3 \right) \\
 &= \frac{1}{\lambda} \left\{ \sum_{j = - \infty}^{\infty} |C_j|^2 \left[ \sin^2(d^0_j) + \sin^2(d^1_j) + \frac{E'^2 - M^2}{(E' + M)^2} \left( \sin^2(d^2_j) + \sin^2(d^3_j) \right) \right] \right \},
 \end{aligned}
\label{fermion_totalcross}
\end{equation}
which is analogous to the scalar case \eqref{scalar_totalcross}. In the same spirit, we expect the total cross-section to behave similarly to the scalar case, with damped oscillations caused by the spacetime's asymptotical structure. One might notice that as the energy of the field increases, $E \gg M$, the contribution from the second and third components becomes just the sinusoidal terms, similar to the 0-th and first components.\par 

Here we end our discussion on how to calculate the cross-section of fermionic and scalar fields in the spacetime of a cosmic string. It is worth mentioning that, although in the fermionic case, the computational work is considerably harder, the whole discussion about local non-minimal interactions we had in the scalar case is equivalent here, i.e., their effects shall be felt only by the constants $C_j, d^i_j$. In the next chapter, we finally apply this formalism to a realistic gravitating cosmic string and analyze the results.

\chapter{Bosonic and fermionic scattering in a gravitating cosmic string spacetime}
\label{chap4}

\section{Bosonic scattering}
\epigraph{\textit{Acredito que se pense muito mais corretamente quando as ideias surgem do contato direto com as coisas, do que quando se olham as coisas com o objetivo de encontrar esta ou aquela ideia.}}{Vincent van Gogh \cite{van1995cartas}}

In the final part of this thesis, we apply the method developed in Chapter \ref{chap3} to the gravitating cosmic string found by \textcite{de2015gravitating}. Our goal is to find the dependence of the cross-section of the scalar field on the mass $M$ and momentum $\lambda$ of the field. To do that, we need to solve the Klein-Gordon equation in this spacetime, find the parameters $C_m, d_m$, and use formula \eqref{scalar_totalcross} to compute the total cross-section. Later we apply the same procedure to the fermionic field.

Remember that the model consists of two scalar fields $\phi$ and $\chi$ coupled with a $SU(2)$ gauge field. The complete model is given by
\begin{gather}
\begin{aligned}
S &= \int{d^4 x \sqrt{-g} \left( \frac{1}{16 \pi G}R + \mathcal{L}_{m} \right)}, \\
\mathcal{L}_m &= \frac{1}{2}(D_\mu \phi^a)^2 + \frac{1}{2}(D_\mu \chi^a)^2 - \frac{1}{4} F^{a}_{\mu \nu}F^{\mu \nu}_{a} - V(\phi^a, \chi^a), \quad a = 1, 2, 3 \, , \\
\text{with} \quad 
V(\phi^a, \chi^a) &= 
\frac{\lambda_1}{4} \left[ (\phi^a)^2 - \eta^{2}_{1} \right]^2 + 
\frac{\lambda_2}{4} \left[ (\chi^a)^2 - \eta^{2}_{2} \right]^2 +
\frac{\lambda_3}{2} \left[ (\phi^a)^2 - \eta^{2}_{1} \right]  \left[ (\chi^a)^2 - \eta^{2}_{1} \right],
\end{aligned}
\label{padua_Lagrangian_chap4}
\end{gather}
which becomes the abelian-Higgs model for $\lambda_2 = \lambda_3 = 0$ and $\chi = \mathbf{0}$.
The authors used dimensionless functions
\begin{equation}
x = \sqrt{\lambda_1}\eta_1 r, \quad L(x) = \sqrt\lambda_1\eta_1 L(r),\quad f(r) = \eta_1 X(x), \quad g(r) = \eta_1 Y(x)
\end{equation}
and the dimensionless parameters
\begin{equation}
\alpha = \frac{e^2}{\eta_1}, \quad q = \frac{\eta_1}{\eta_2}, \quad \beta_i = \frac{\lambda_i}{\lambda_1},\quad \gamma = 8\pi G \eta_1^2.
\end{equation}

\begin{figure}
\caption{The metric components in abelian (solid) and non-abelian (dashed) cases. We set $\alpha = 1.0$, $\gamma = 0.6$ for both abelian and non-abelian cases and $q = 1.0$, $\beta_2 = 2.0$, and $\beta_3 = 1.0$ for the non-abelian case.}
\includegraphics[width=0.8\textwidth]{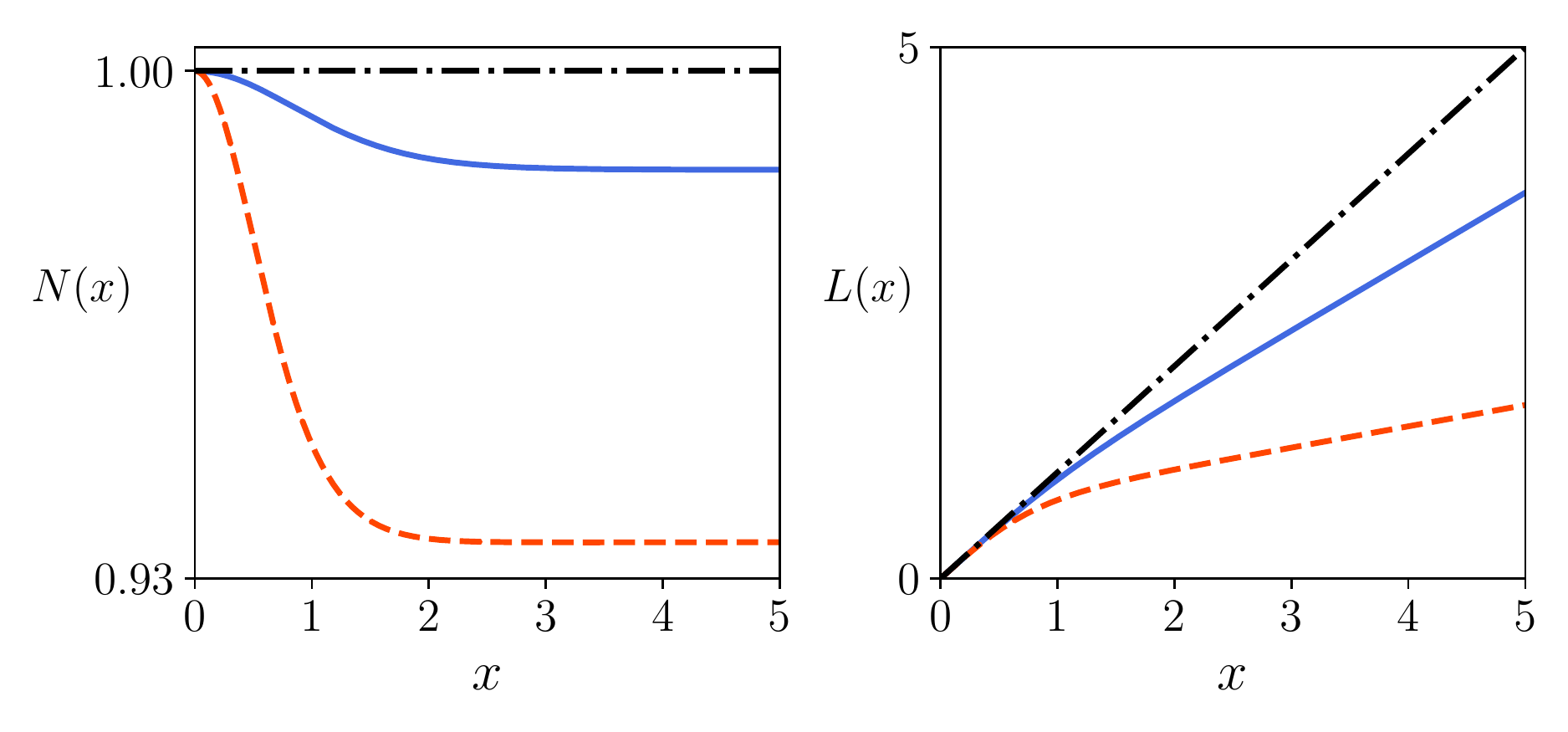}
\caption*{Source: The author (2021).}
\label{fig:padua_metric}
\end{figure}

From now on we set the dimensionless parameters to be $\alpha = 1.0$, $\gamma = 0.6$ for both abelian and non-abelian scenarios and $q = 1.0$, $\beta_2 = 2.0$, and $\beta_3 = 1.0$ in the non-abelian case. The associated metric components for both abelian and non-abelian cases are shown in Figure \ref{fig:padua_metric}.
For solving the equations of motion \eqref{scalar_eom2} numerically, we employed the Runge-Kutta method of eighth order. From each solution with specific mode $m$ and momentum $\lambda$ we extracted the parameters $C_m(\lambda)$ and $d_m(\lambda)$, which are needed to construct the cross-section. We also verified that the cross-section indeed converges for a finite number of modes. \par

\begin{figure}[htbp]
\caption{$|C_m|$ and $\sin^2(d_m)$ with respect to the incident momentum $\lambda$ for a scalar field with $M = 1.0$. Solid (blue) lines represent the abelian case and dashed (orange) the non-abelian one.}
\includegraphics[width=1.0\textwidth]{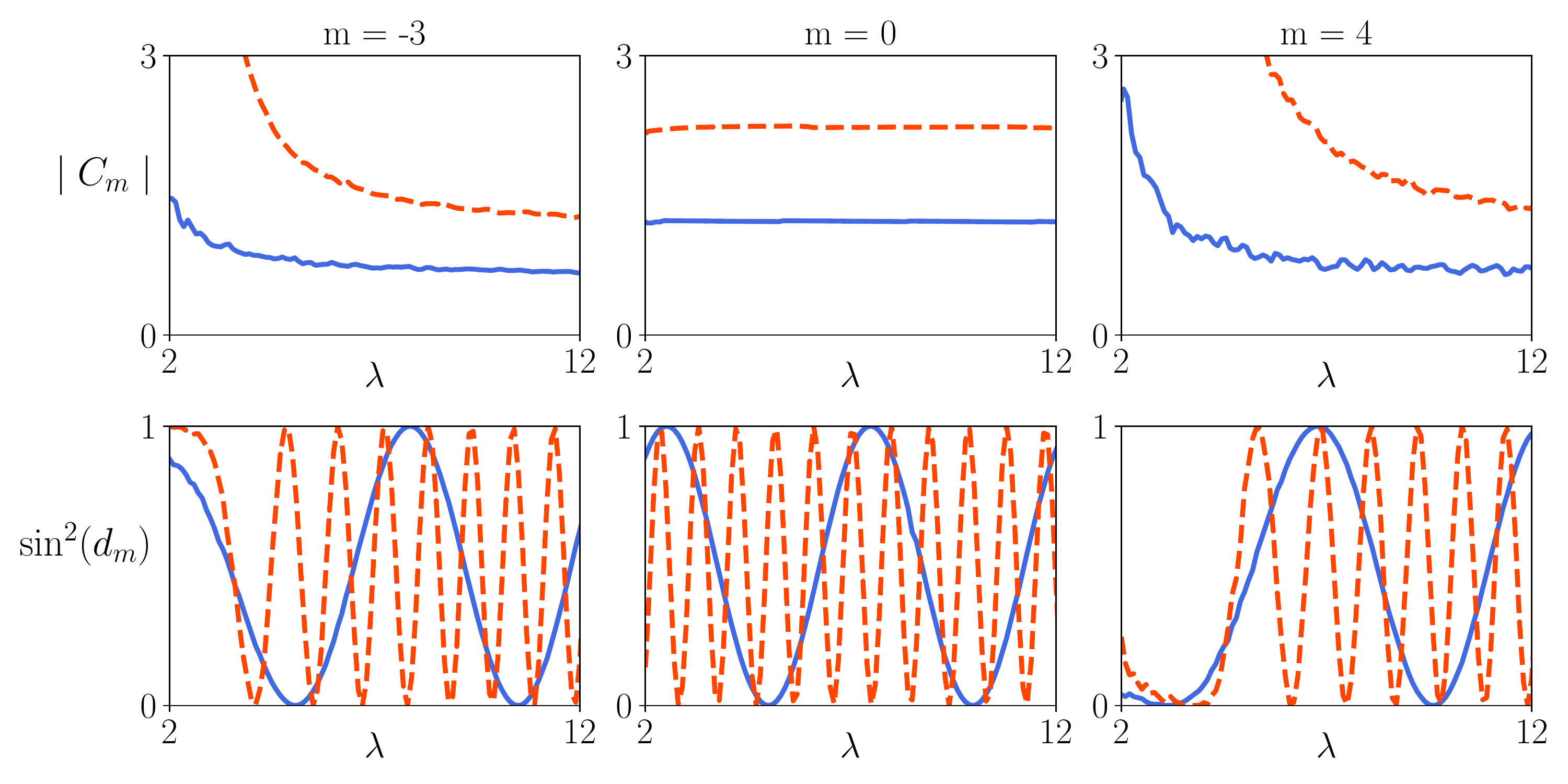}
\caption*{Source: The author (2021).}
\label{fig:Cmvsdm}
\end{figure}

In Figure \ref{fig:Cmvsdm} we present $ \lvert C_m \rvert$ and $\sin^2(d_m)$ for three different values of $m$ for a massive scalar field with $M = 1.0$. We noticed $|C_m|$, unlike $d_m$, is symmetric under $m \to -m$ and is always larger in the non-abelian scenario. It originates from the larger deviation of the conical parameters from the Minkowski counterparts. We also noticed that larger mass delays $\sin^2(d_m)$.

\begin{figure}
\caption{Total cross-section of the massive scalar field in the abelian case.}
\includegraphics[width=0.9\textwidth]{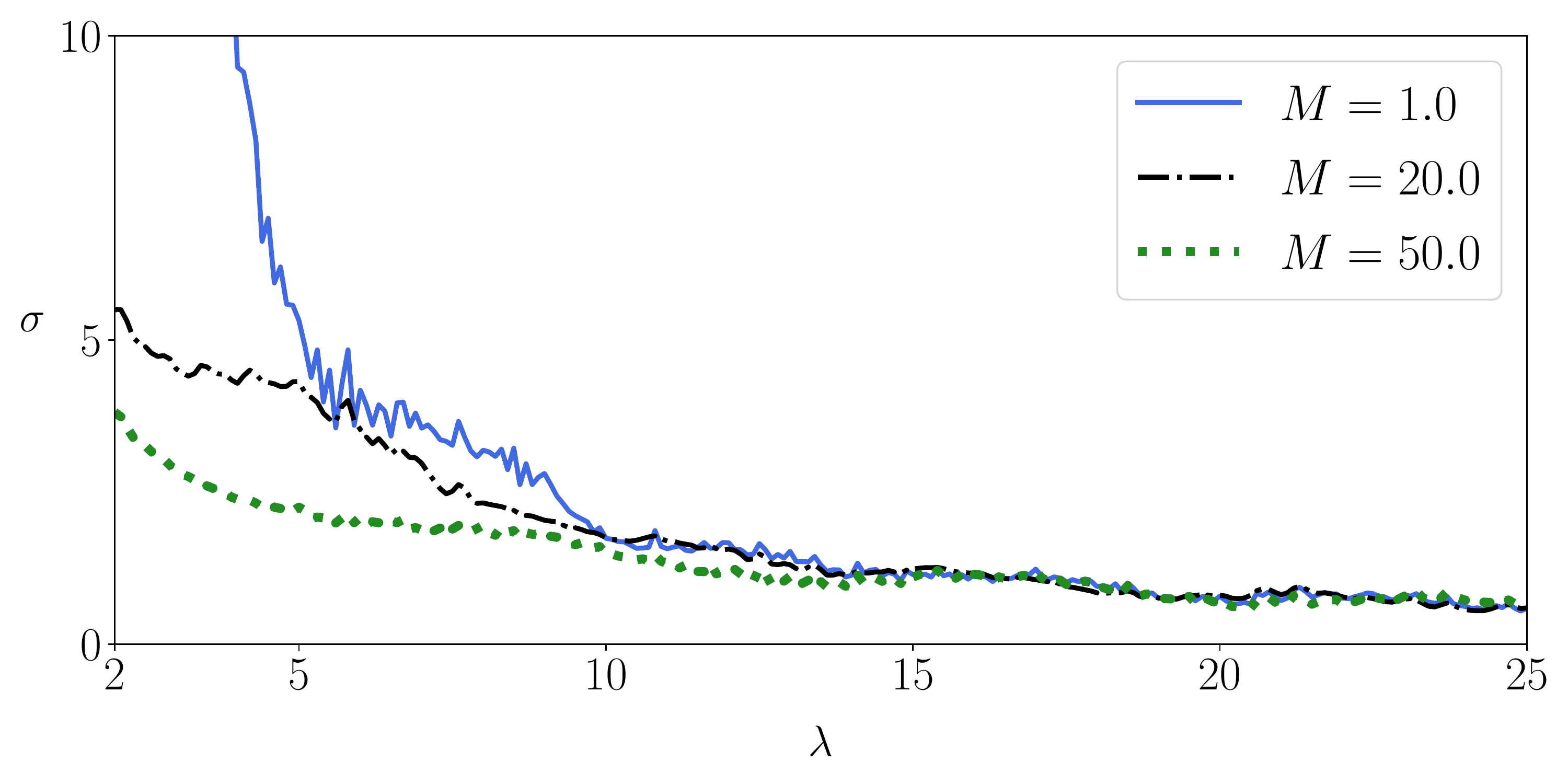}
\caption*{Source: The author (2021).}
\label{fig:sigma_scalar_ab}
\end{figure}

\begin{figure}
\caption{Total cross-section of the massive scalar field in the non-abelian case.}
\includegraphics[width=0.9\textwidth]{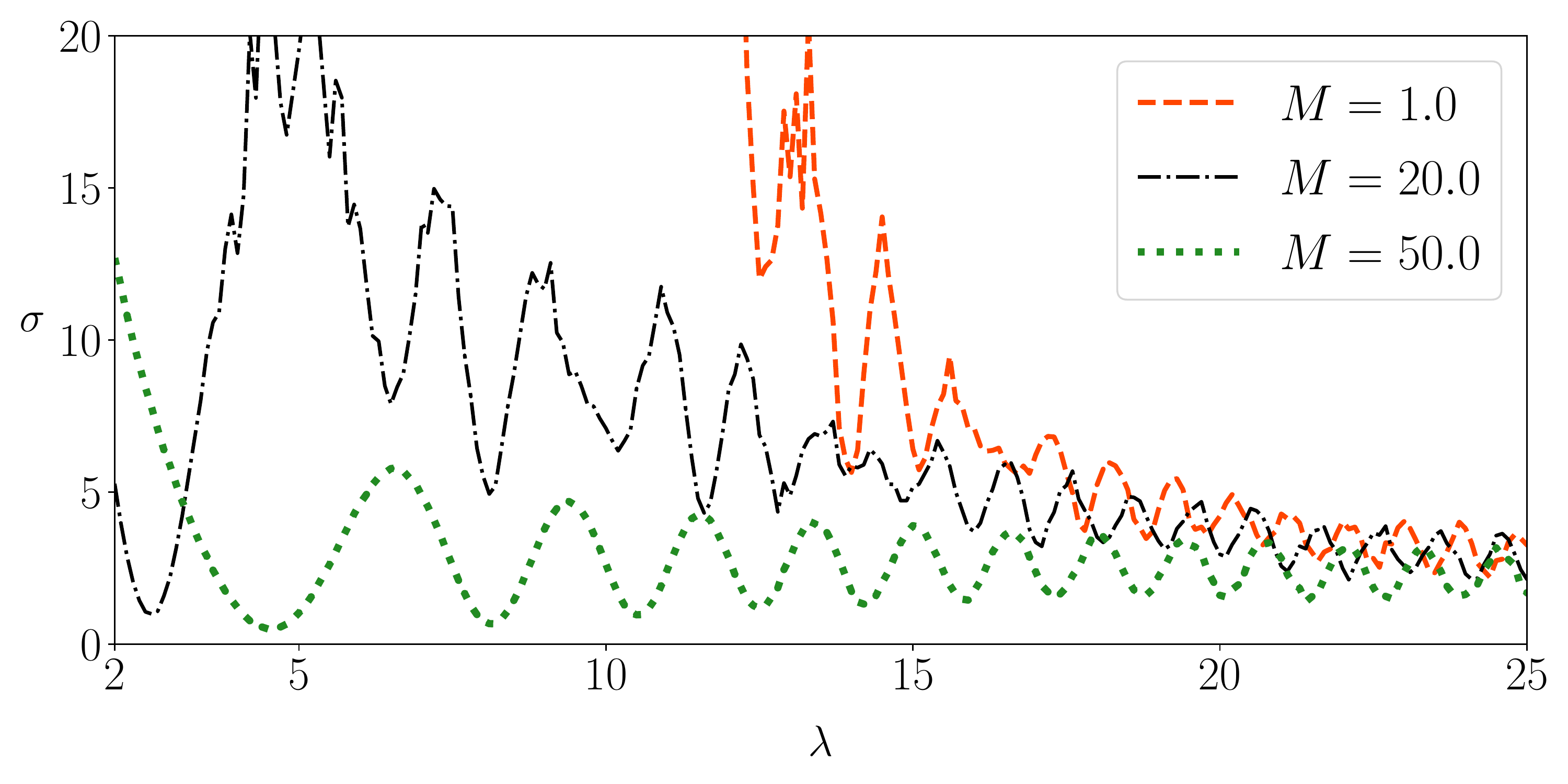}
\caption*{Source: The author (2021).}
\label{fig:sigma_scalar_na}
\end{figure}

In Figure \ref{fig:sigma_scalar_ab} and \ref{fig:sigma_scalar_na}, we show the total cross-section in the abelian and non-abelian scenarios, respectively, for three values of mass. We can see that in both cases, the mass dampens the cross-section in the small-momentum regime, but as $\lambda$ increases, the difference between the cross-sections becomes smaller. Eventually, the mass becomes irrelevant in the regime $\lambda \to \infty$.

\begin{figure}
\caption{Scattering cross-section in both abelian and non-abelian cases for a massless scalar field. We have also shown the mass effect in the non-abelian scenario in the zoomed region.}
\includegraphics[width=1.0\textwidth]{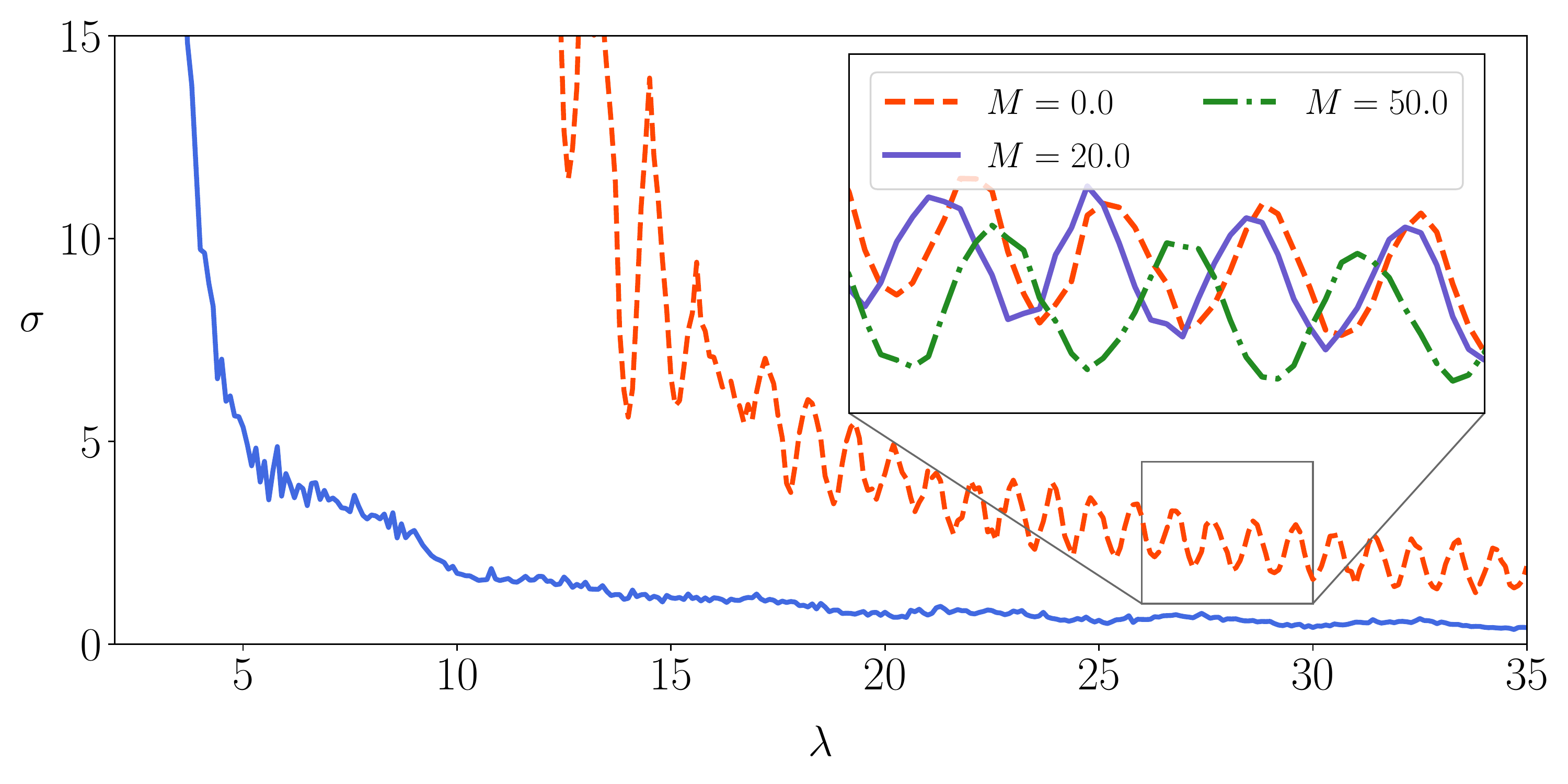}
\caption*{Source: The author (2021).}
\label{fig:sigma_scalar_abvsna}
\end{figure}

In Figure \ref{fig:sigma_scalar_abvsna} we compare the massless cross-section in abelian and non-abelian cases and show, in the zoomed region, how mass affects the region of large momentum in the non-abelian scenario. For large momentum, we see that the presence of mass dampens $\sigma$ but also delays the signal. This was expected since $\sin^2(d_m)$ is also delayed.  

We can see that the cross-section diverges as the incident momentum approaches zero and tends to zero as $\lambda \rightarrow \infty$, which is expected. However, in both scenarios, the total cross-section presents an unusual oscillatory behavior that becomes more evident as $\lambda$ increases. Since we have not detected any oscillatory pattern in the values of $|C_m|$, we are led to think the oscillation comes from an interference between the sinusoidal terms of equation \eqref{scalar_totalcross}. As we shall see, this particular interference pattern seems to be related to the asymptotical form of the spacetime metric.\par
\begin{figure}
\caption{Curvature for the toy model \eqref{toymod_metric_smooth} using $a = 1$.}
\includegraphics[width=0.8\textwidth]{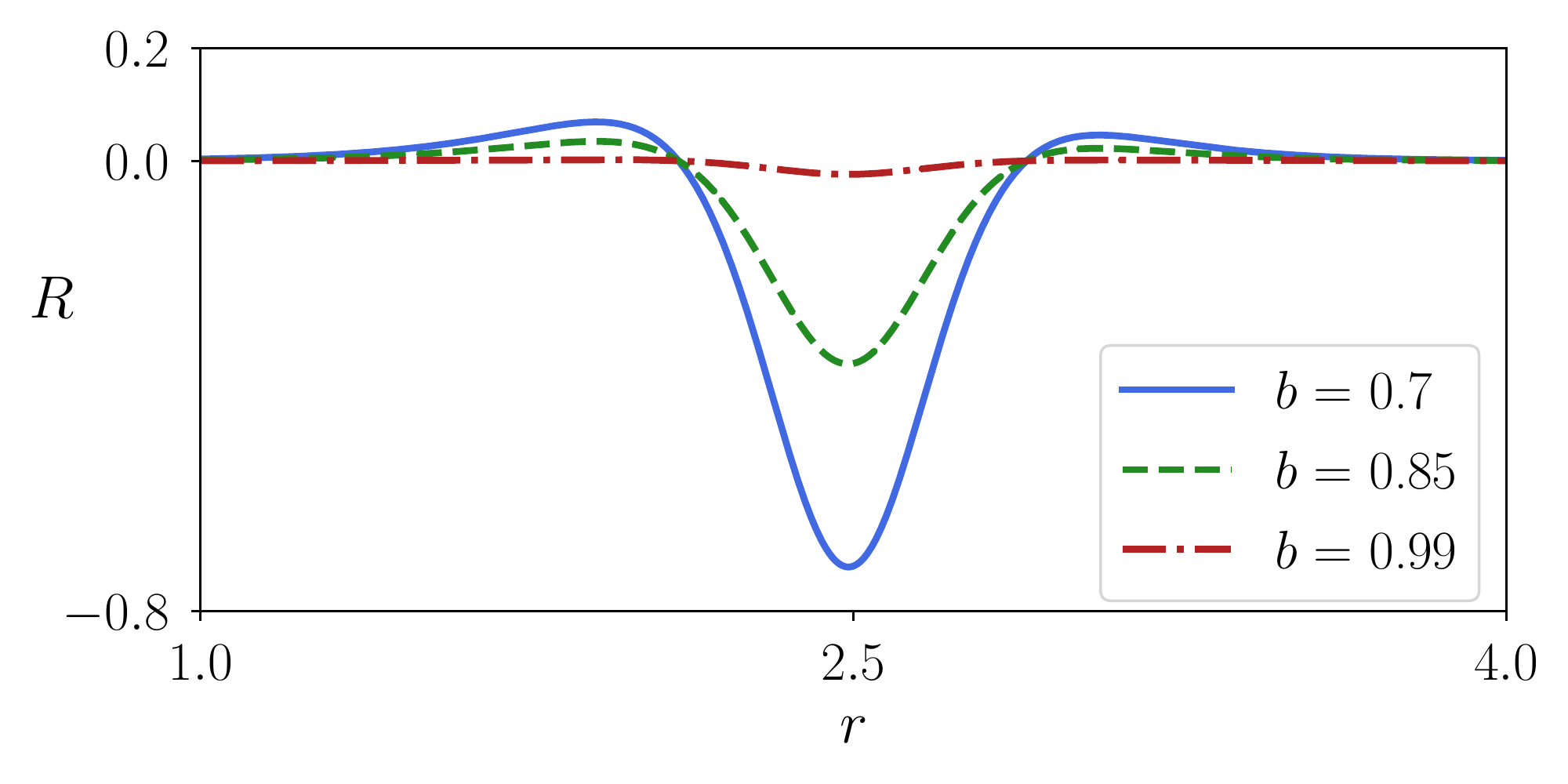}
\caption*{Source: The author (2021).}
\label{fig:toymod_a=1_curvature}
\end{figure}

\begin{figure}
\caption{Effect of $b$ on the oscillations of the total cross-section.}
\includegraphics[width=0.9\textwidth]{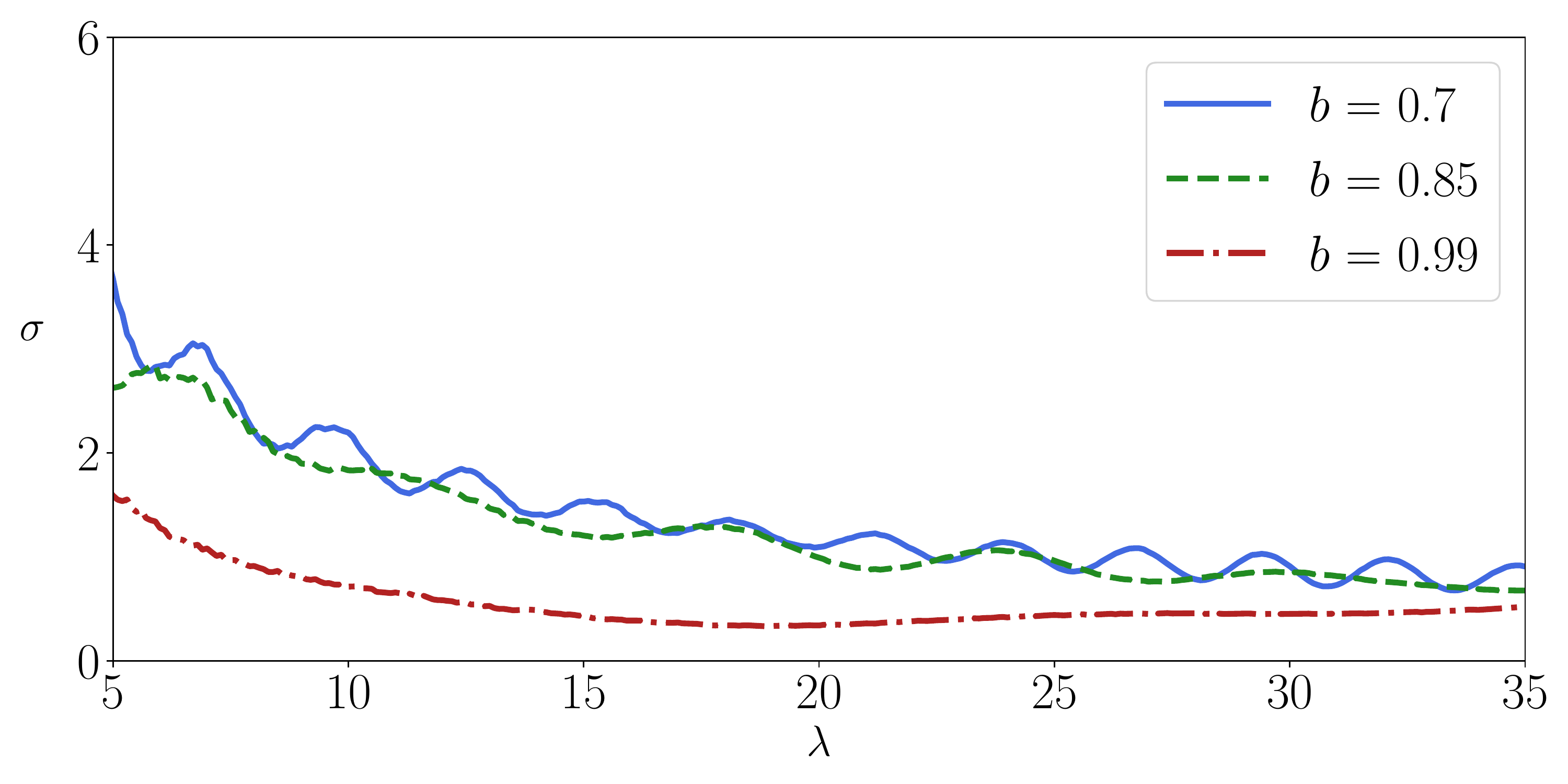}
\caption*{Source: The author (2021).}
\label{fig:toymod_a=1_sigma}
\end{figure}

 
The high-energy tail oscillation is not entirely new. \textcite{burt1975oscillations} already showed that non-linear persistent self-interactions lead to damped oscillations in the total cross-section of baryon-antibaryion scattering. In their case, the non-linear interaction appears as a potential in the lagrangian density, while in our case, it manifests itself as the asymptotical conical configuration of spacetime, which affects the lagrangian through the covariant derivative.
In fact, we can use the toy model developed in Chapter \ref{chap3} to investigate this hypothesis. We start considering the metric toy model \eqref{toymod_metric_smooth} with $a = 1, r_0 = 2.5, p = 3.0$ and analyze the scalar field total cross-section with respect to the conical parameter $b$. For consistency $c$ is always calculated according to $c = r_0 (1 - b)$. The curvature profile for the three chosen values of $b$ is shown in Figure \ref{fig:toymod_a=1_curvature}. We then calculated the total cross-section in these spacetimes. In Figure \ref{fig:toymod_a=1_sigma} we see that the frequency of oscillations in the total cross-section is proportional to the deficit angle of the asymptotic spacetime and it tends to zero as $b \to 1$.

\begin{figure}
\caption{Curvature profile from the metric \eqref{toymod_metric_smooth} with $b = 0.85$ and $r_0 = 2.5$.}
\includegraphics[width=0.8\textwidth]{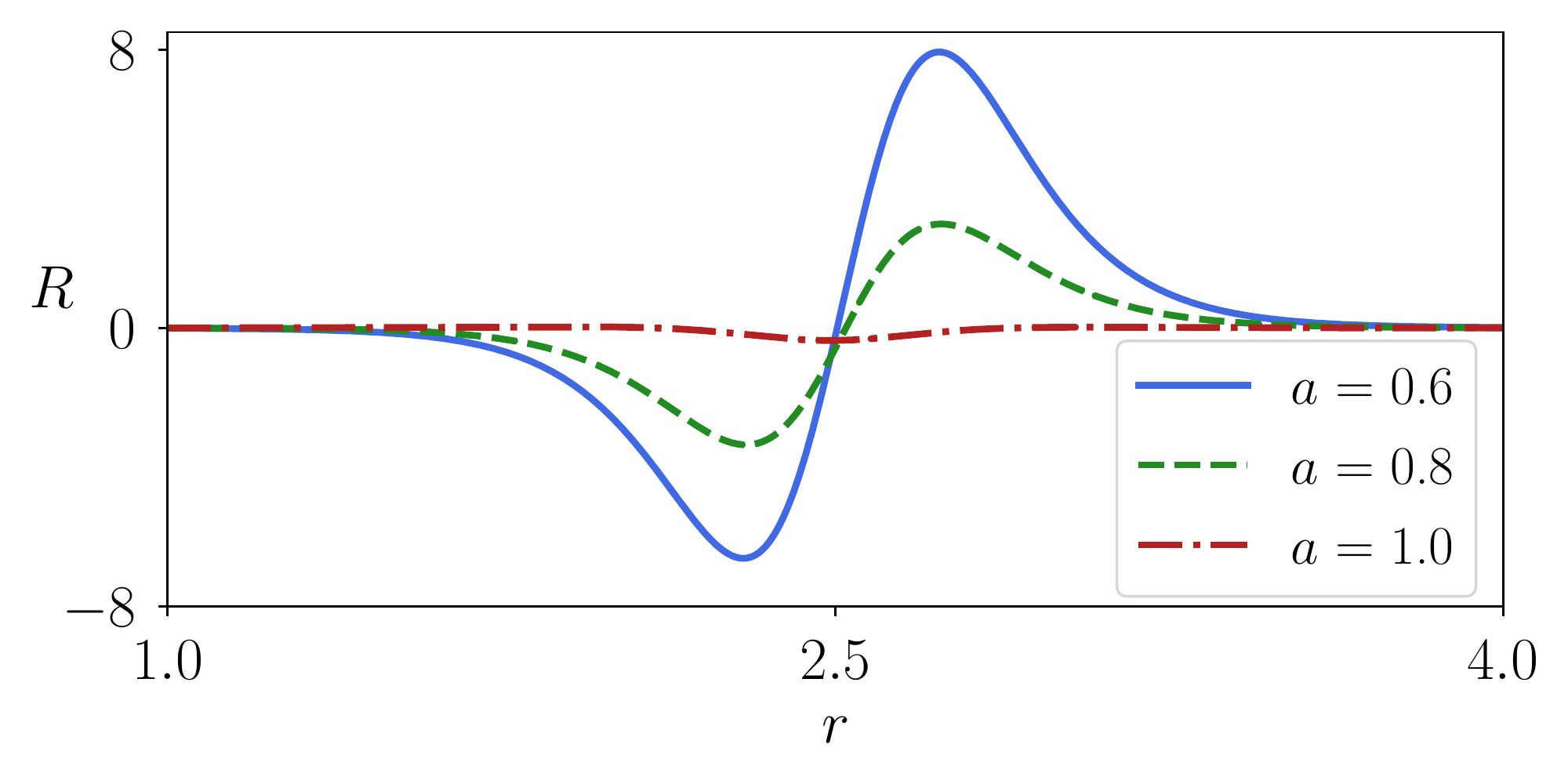}
\caption*{Source: The author (2021).}
\label{fig:toymod_b=0.85_curvature}
\end{figure}

\begin{figure}
\caption{Scalar field total cross-section in the spacetime \eqref{toymod_metric_smooth} with $b = 0.85$ and $r_0 = 2.5$.}
\includegraphics[width=0.9\textwidth]{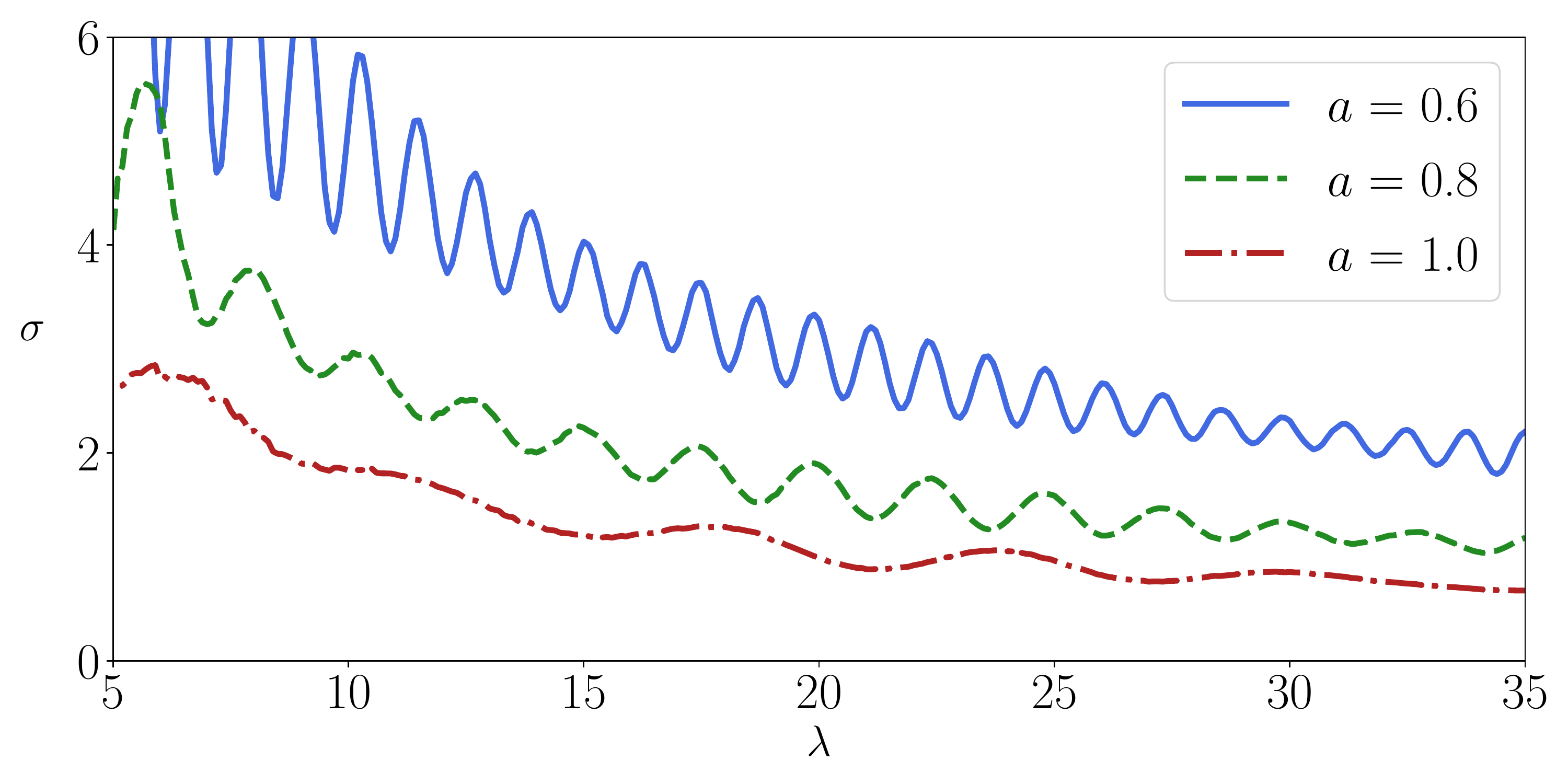}
\caption*{Source: The author (2021).}
\label{fig:toymod_b=0.85_sigma}
\end{figure}

However, that is not the end of the story. We set $b = 0.85, r_0 = 2.5, p = 3.0$ and analyzed how the cross-section changes with the parameter $a$, which represents a blue-shift of the time coordinate\footnote{The frequency of a light beam is perceived as higher outside the string than to an observer at the core.}. The curvature profile for the three chosen values of $a$ is shown in Figure \ref{fig:toymod_b=0.85_curvature}. The corresponding total cross-section is shown in Figure \ref{fig:toymod_b=0.85_sigma}. One can see that the parameter $a$ also has a strong influence on the frequency of oscillations and the average amplitude of $\sigma$. It is worth mentioning that the oscillations tend to disappear as the metric coefficients tend to pure Minkowski ones, namely $a = 1, b = 1, c = 0$. Therefore, we conclude that the damped oscillations in the total cross-section are caused by the persistent interaction of the field with the spacetime geometry, represented by the asymptotical structure of spacetime or, equivalently, by the parameters $a, b$ and $c$. \par 

\section{Fermionic scattering}
\epigraph{\textit{It should be written on every school chalkboard, "Life is a playground- or nothing."}}{Nemo Nobody}
In this section, we analyze the scattering of a fermionic field in the same spacetime. In Chapter \ref{chap3} We have shown that the total scattering cross-section of the fermionic field is given by the formula \eqref{fermion_totalcross}. Again we need to extract the parameters $C_j(\lambda), d^i_j(\lambda)$, and for doing so, we used the Runge-Kutta method of eighth order to solve the equations \eqref{fermion_eom1} numerically. The numerical procedure is similar to the one in the last section for the bosonic case.\par
\begin{figure}[H]
\caption{Quantities used to compute the cross-section of a fermionic field with $M = 1.0$. The other plots of $\sin^2(d^i_j)$ do not show any particularly different behavior.}
\includegraphics[width=1.0\textwidth]{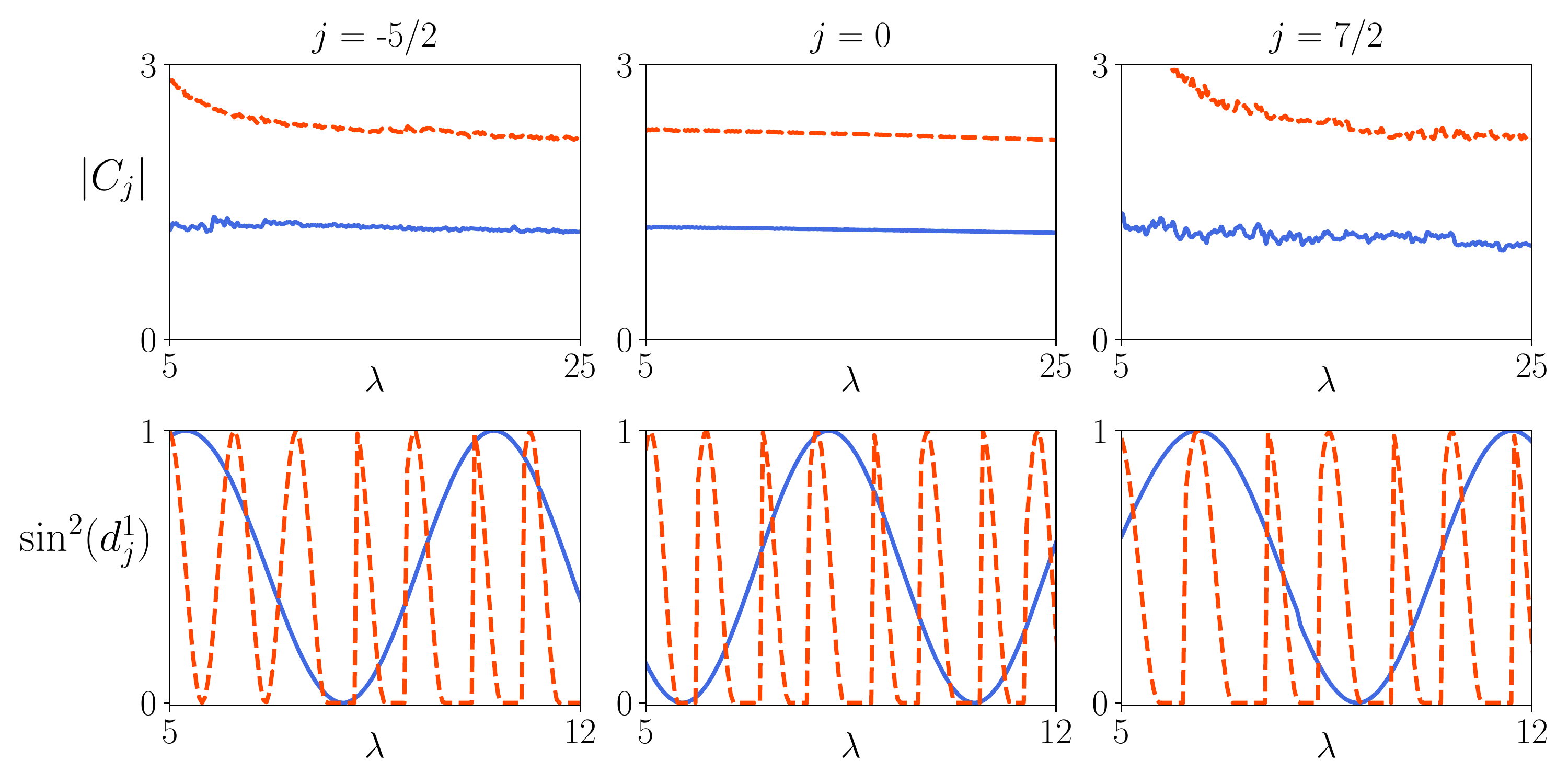}
\caption*{Source: The author (2021).}
\label{fig:fermion_CmvsdeltaM1}
\end{figure}

 
In Figure \ref{fig:fermion_CmvsdeltaM1} we show the amplitude $C_j$ and the phase $d^1_j$ of the component $\psi_1$. Unlike the phase of the scalar field, here $\sin^2(d_j^i)$ is symmetric under $j \rightarrow -j$. We noticed the presence of some plateaus in $\sin^2(d^2_j)$, which means that for each mode, there are windows, in $\lambda$, for which that mode has no contribution to the total cross-section. Though these windows are seen in both scenarios, in the abelian case, they are wider and were detected only for large momentum, i.e., $\lambda > 15$. In addition, the amplitude $|C_m|$ behaves similarly to the scalar counterpart. It is symmetric under $j \to -j$ and vanishes after a finite number of modes.

\begin{figure}[H]
\caption{Effect of mass on the total cross-section in the abelian case.}
\includegraphics[width=0.9\textwidth]{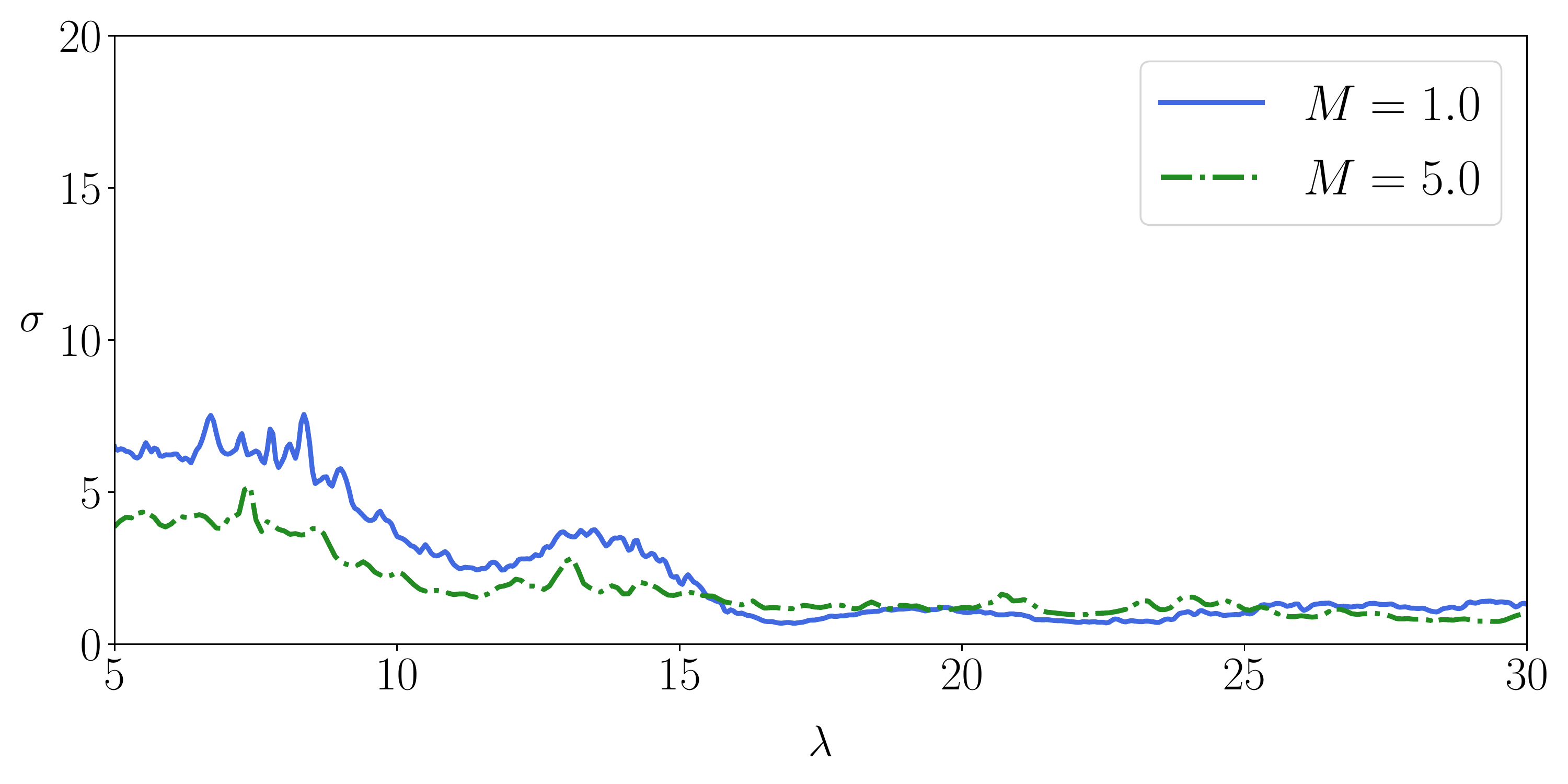}
\caption*{Source: The author (2021).}
\label{fig:fermion_sigma_ab}
\end{figure}

\begin{figure}[H]
\caption{Effect of mass on the total cross-section in the non-abelian case.}
\includegraphics[width=0.9\textwidth]{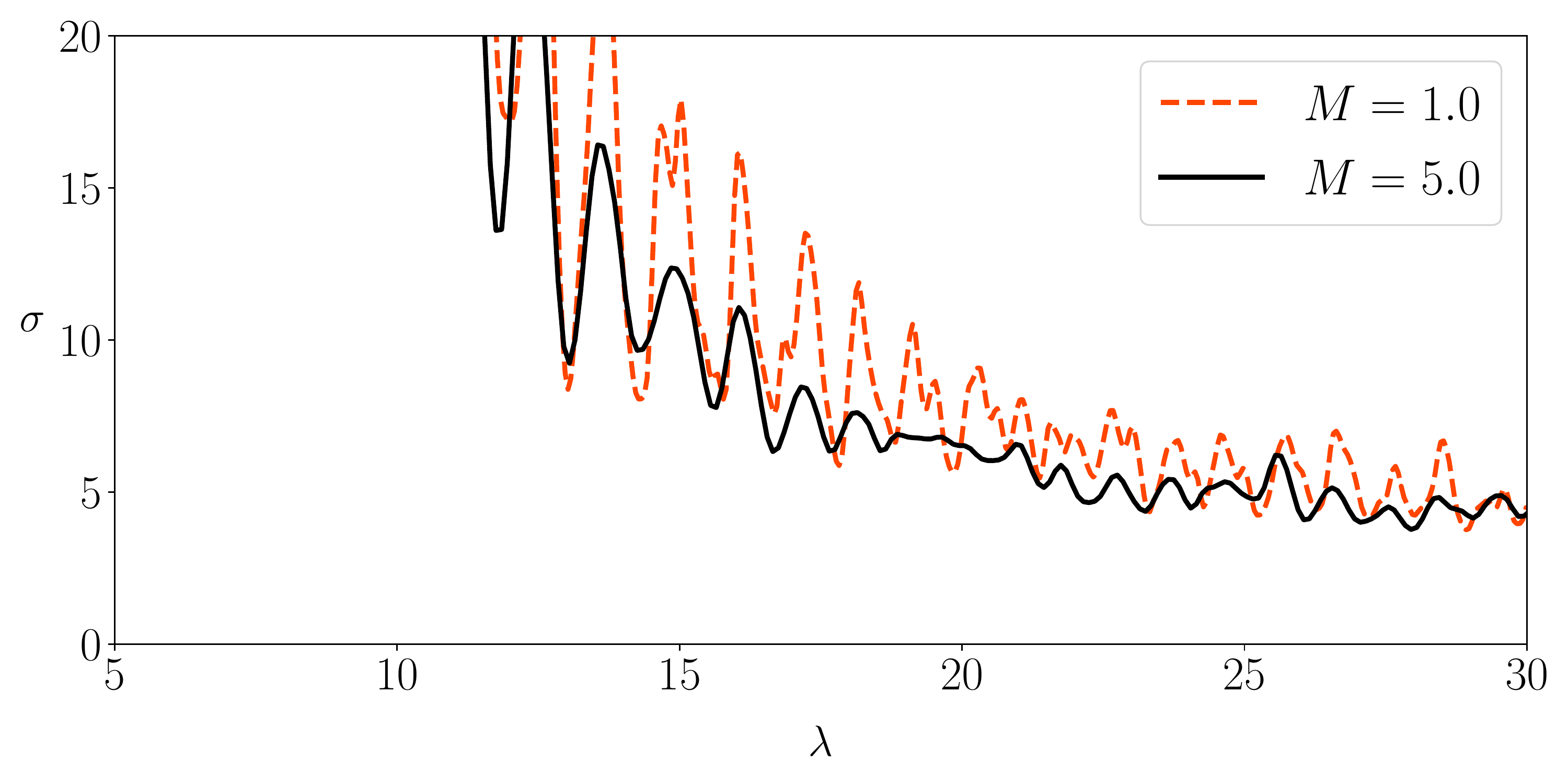}
\caption*{Source: The author (2021).}
\label{fig:fermion_sigma_na}
\end{figure}

In Figures \ref{fig:fermion_sigma_ab} and \ref{fig:fermion_sigma_na}, we show the influence of mass on the total cross-section in both abelian and non-abelian scenarios, respectively. We see that the effect is similar to the scalar case, i.e., larger mass reduces the average value of $\sigma$. Comparing the fermionic and bosonic results, one can conclude that the mass effect in the fermionic case is larger than the bosonic one.

\begin{figure}[H]
\caption{Total cross-section of the fermionic field with M = 1.0.}
\includegraphics[width=1.0\textwidth]{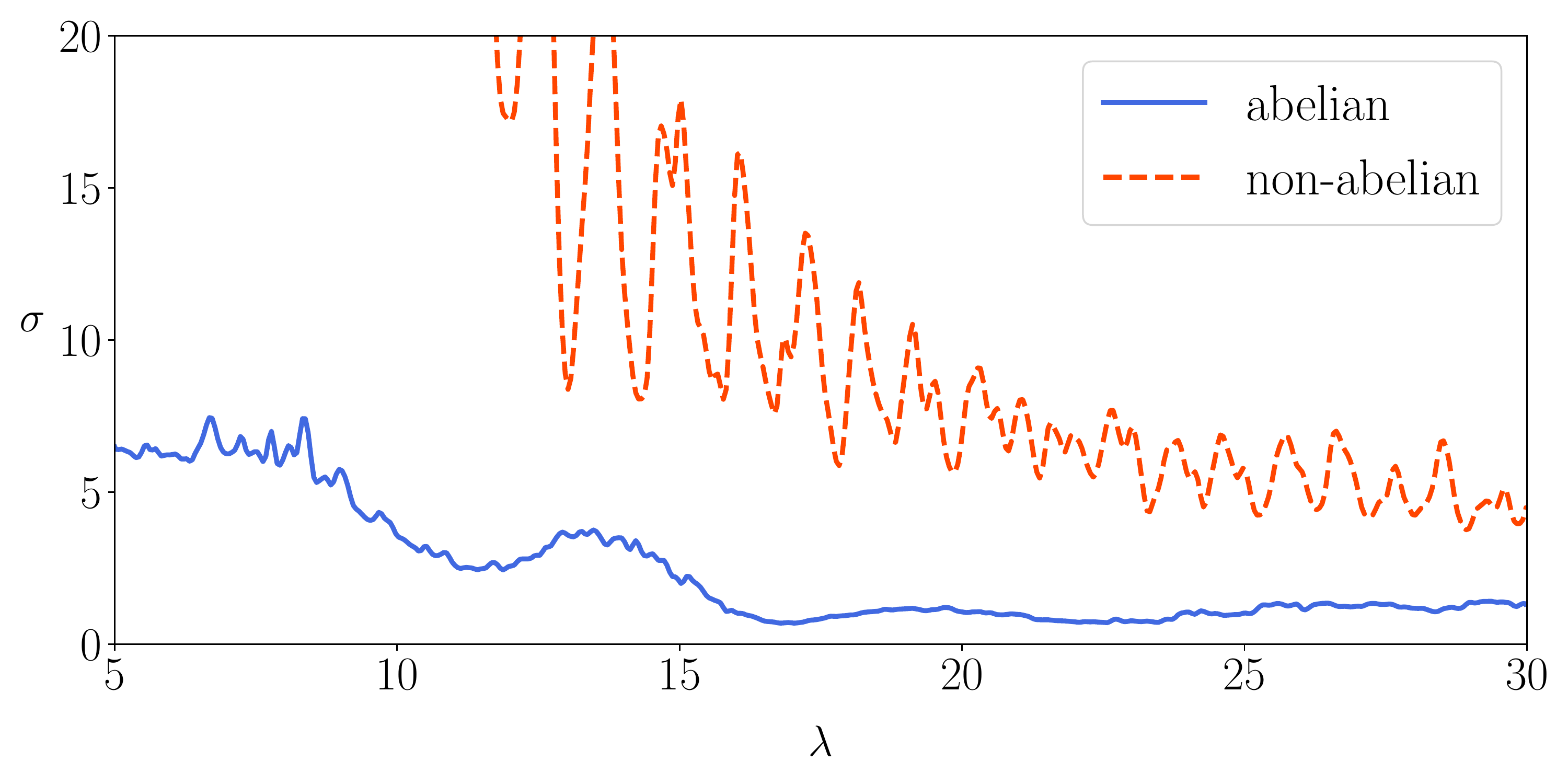}
\caption*{Source: The author (2021).}
\label{fig:fermion_sigma_abvsnaM1}
\end{figure}

Finally, it is clear that the fermionic cross-section also presents damped oscillations, which can be seen as evidence that damped oscillations in the total cross-section are a fundamental property of scattering in asymptotically conical spacetimes. In Figure \ref{fig:fermion_sigma_abvsnaM1} we show the cross-section for a massive fermionic field in both abelian and non-abelian scenarios.

One might have noticed that the cross-section of both scalar and fermionic fields is more significant in the non-abelian scenario compared with the abelian one. A crude estimation could have anticipated this result. Remember that the cross-section is proportional to the ratio outgoing flux/incident flux
\begin{equation}
\sigma =\frac{F_{out}}{F_{in}} = \frac{\frac{v_{out}}{A_{out}}}{\frac{v_{in}}{A_{in}}},
\end{equation} 
and that for a slice of a cylinder of height $\Delta z$, the effective area for the outgoing particles is proportional to $ab \Delta z$, hence when the deficit angle is larger (b is smaller), the area of the outgoing particles is smaller, hence the flux is larger. Of course, there is also the contribution from the increase in momentum/velocity of the particle since $\lambda^\prime > \lambda$.

\chapter{Conclusion}

This thesis has studied the general theory of cosmic strings and their gravitational interaction with nearby matter fields. In Chapter \ref{chap1}, we developed the basic theory of cosmic strings, discussed the two general types, global and local strings, and their topological stability. Then we presented the Nielsen-Olesen vortex, the first local string solution, and discussed some of its properties. We have seen that for some region of the Bogomolny'i parameter $\beta$ the Nielsen-Olesen vortex is unstable to unwinding, i.e. every string with winding number $n > 1$ eventually decays to $n$ strings each with $n = 1$. At the end of Chapter \ref{chap1}, we outlined how cosmic strings are expected to have formed in the early universe via the Kibble mechanism. \par 

In Chapter \ref{chap2} we proceeded to study gravitational properties of cosmic strings in the wire approximation and have seen that the conical structure presents non-trivial phenomenology. It was observed that gravitational (eletromagnetic) interaction is changed due to the non-trivial boundary conditions the conical spacetime imposes on the gravitational (eletromagnetic) potential. Still, in Chapter \ref{chap2} we investigated the literature on extended cosmic strings, i.e., vortex solutions with non-negligible internal structure, and have seen that the spacetime generated by a gravitating extended vortex presents a conical structure far from the core. \textcite{garfinkle1985general} showed this feature to be quite general in the abelian-Higgs model, and we presented a recent non-abelian model that possesses this feature as well. \par 

In Chapter \ref{chap3} we showed that the asymptotically conical structure of gravitating cosmic string spacetimes creates a divergent scattering amplitude if we follow the standard ansatz of the partial-wave formalism. \textcite{deser1988classical} had already investigated the simpler case of scalar field scattering on a cone and found a similar divergence. We avoided the singularity by modifying the asymptotical ansatz in the partial-wave approach, which leads to corrections in the phase-shift and total cross-section. The correction on the phase-shift is the addition of two terms induced by the conical structure, while the correction on the total cross-section is a multiplicative constant on each term of the cross-section series. This correction accounts for the idea that we need information about the field at infinity in order to construct the asymptotical ansatz, which is exactly what removes the singularity on the scattering amplitude. This happens because the spacetime before and after the scattering are not equivalent and it is impossible to construct the usual plane-wave solution at infinity. The essential conclusion is that we also need information about the amplitude, besides the phase, of the scattered field to be able to construct the total cross-section. We then developed an analytical toy model for the spacetime metric of an extended vortex, which is based on an analytical approximation of the Heaviside step function, and showed how the cross-section changes when considering the interaction with the string gauge field. In the second part of Chapter \ref{chap3}, we applied the same formalism to a Dirac field and found the explicit formula for the fermionic total cross-section. We have seen that defining the asymptotical ansatz for the fermionic field is tricky since we cannot make all components plane-waves before the scattering. The cross-section formula, however, is not dependent on how you define the initial condition, as expected. In addition, it was shown that in the high-energy limit all components of the fermionic field contribute to the cross-section in a symmetric way. \par 

In Chapter \ref{chap4}, we applied this formalism to an abelian and a nonabelian gravitating cosmic string model found by \textcite{de2015gravitating} and compared the cross-sections of both fields for two sets of the vortex parameters. We have seen that the spacetime with an asymptotically larger deficit angle has a larger cross-section for both fields which is related to a smaller effective area for the outgoing particles. In addition, all cross-sections present damped oscillations that, with the aid of our toy model, were shown to be caused by the particular spacetime structure, including the conicity, far from the core of the string. We also showed how each parameter of the asymptotical spacetime contributes to the curvature profile and oscillations in the total cross-section. We have found that both parameters, the conicicty $b$ and the blue-shift $a$, contribute to the cross-section oscillations. \par

The natural next step of our work is to apply this formalism to the scattering of gauge fields and see if it presents the same damped oscillations seen in the scalar and fermionic cases. We also look forward to explore our toy model and see if it could reveal new features of this class of spacetimes without the hard computational work usually required to obtain the metric components. Moreover, in the near future, we plan to study a cosmic string model from scratch in order to have control over all parameters and see how the scattering of classical and quantum particles is affected when we change the string configuration and when we consider interaction with the fields generating the vortex. Besides that, we plan to study other types of defects, topological and nontopological, and see if their interaction with nearby matter fields reveals novel or similar features.

\printbibliography[title={References}]

\end{document}